\def\USEACHEMSO{0} %

\if\USEACHEMSO1
\documentclass[journal=jctcce,manuscript=article,%
]{achemso}
\else
\documentclass[aip,jcp,notitlepage,
tightenlines,
reprint,%
floatfix]{revtex4-2}
\fi

\usepackage{chemformula}
\usepackage{siunitx}%
\usepackage[utf8]{inputenc}
\usepackage[T1]{fontenc}
\usepackage{stix}
\usepackage[version=3]{mhchem}
\usepackage{url}
\usepackage{hyperref}
\usepackage{numprint}
\usepackage[american]{babel}
\nprounddigits{2}
\npdecimalsign{\ensuremath{.}}  %
\usepackage[normalem]{ulem}

\usepackage{morefloats}

\usepackage{xifthen}  %
\usepackage{mathtools}
\usepackage{tabularray}
\UseTblrLibrary{booktabs}
\usepackage{booktabs}
\usepackage{dcolumn}%
\usepackage{caption}
\usepackage{layout}
\usepackage{multirow}
\usepackage{derivative}
\usepackage{array}
\newcolumntype{L}[1]{>{\raggedright\let\newline\\\arraybackslash\hspace{0pt}}m{#1}}
\newcolumntype{C}[1]{>{\centering\let\newline\\\arraybackslash\hspace{0pt}}m{#1}}
\newcolumntype{R}[1]{>{\raggedleft\let\newline\\\arraybackslash\hspace{0pt}}m{#1}}
\newcolumntype{P}[1]{>{\centering\arraybackslash}p{#1}}
\usepackage{enumitem}
\setenumerate[1]{label=(\arabic*)}

\usepackage{adjustbox}
\usepackage{lipsum}
\usepackage[dvipsnames]{xcolor}
\usepackage{soul}
\setstcolor{red}
\setul{}{0.28ex}
\definecolor{myred}{rgb}{1, 0.123, 0.404}
\definecolor{mypink}{rgb}{1, 0.2, 0.745}
\definecolor{mycyan}{rgb}{0.090, 0.667, 0.553}
\definecolor{freshgreen}{rgb}{0.051, 0.796, 0.733}
\definecolor{mygreen}{rgb}{0.184, 0.792, 0.0}
\definecolor{myviolet}{rgb}{0.256, 0.207, 1.00}
\definecolor{myorange}{rgb}{0.9, 0.44, 0.0}
\definecolor{mypink2}{rgb}{0.898, 0.0, 0.451}
\definecolor{mypurple}{rgb}{0.659, 0.251, 1.000}

\def\COLORIZE{1} %
\if\COLORIZE1

\else

\fi
\def\IMGSMALL{1} %
\if\IMGSMALL1
    \newcommand{\figdir}{figures}
\else
    \newcommand{\figdir}{figures}
\fi
\def\REDLINE{0} %
\if\REDLINE1
    
    \newcommand{\rlremove}[1]{\st{#1}}
\else
    
    \newcommand{\rlremove}[1]{\unskip}
\fi

\usepackage{xspace}
\usepackage{verbatim}
\let\oldtheequation\theequation
\makeatletter
\def\tagform@#1{\maketag@@@{\ignorespaces#1\unskip\@@italiccorr}}
\renewcommand{\theequation}{(\oldtheequation)}
\makeatother

\if\USEACHEMSO1
\fi

\newcommand{\myunit}[2][]{%
  \ifthenelse{\isempty{#1}}%
    {\ensuremath{#2}}%
    {\ensuremath{{#1}\:{#2}}}%
}

\newcommand{\matrgreek}[1]{\ensuremath{\pmb{#1}}} %

\usepackage{listings}
\if\USEACHEMSO1
    
\else
    
\fi
\newcommand{\figfull}{1\textwidth}

\usepackage[
textsize=small,
]
{todonotes}
\newcommand{\clace}{chloroacetylene\xspace}

\newcommand{\clbut}{chlorobutadiyne\xspace}

\newcommand{\Pheace}{Phenylacetaldehyde\xspace}
\newcommand{\pheace}{phenylacetaldehyde\xspace}
\newcommand{\furf}{furfural\xspace}
\newcommand{\Furf}{Furfural\xspace}
\newcommand{\butfa}{3-fluoro-2-methylpropanal\xspace}  %
\newcommand{\isobutfa}{sc-3-fluoro-2-methylpropanal\xspace}  %
\newcommand{\isobutfb}{sp-3-fluoro-2-methylpropanal\xspace}  %
\newcommand{\isobutfaB}{SC\xspace}  %
\newcommand{\isobutfbB}{SP\xspace}  %
\newcommand{\mydot}{}
\newcommand{\nmydot}{.}

\newcommand{\matr}[1]{\ensuremath{\mathbf{#1}}}
\newcommand{\braket}[2]{\ensuremath{ \langle #1 | \, #2  \rangle }}

\newcommand{\ketbra}[2]{\ensuremath{  | {#1} \rangle \langle {#2} |}}

\newcommand{\ket}[1]{\ensuremath{  | {#1} \rangle}}

\newcommand{\matrixe}[3]{\ensuremath{ \langle{#1} | {#2} | {#3} \rangle }}

\newcommand{\dd}{\ensuremath{\mathrm{d}}}
\newcommand{\ii}{\ensuremath{\mathrm{i}}}

\newcommand{\lit}[1]{Ref.~\mbox{[\!\!\citenum{#1}]}\xspace}
\newcommand{\lits}[1]{Refs.~\mbox{[\!\!\citenum{#1}]}\xspace}
\newcommand{\compdetails}{Appendix \ref{app:comp_details}\xspace}

\definecolor{CBdblue}{RGB}{5,113,176}

\newcommand{\sfigref}[2]{\autoref{#1}(#2)\xspace}
\newcommand{\sfigrref}[3]{Figs.~\ref{#1}(#2)--(#3)\xspace}
\newcommand{\siref}{Supplementary Information\xspace}
\newcommand{\sifig}[1]{Supplementary Fig.~S{#1}\xspace}

\newcommand{\sisec}[1]{Supplementary Section~S{#1}\xspace}

\if\USEACHEMSO0
\begin{document}
\fi

\title{Localized intrinsic bond orbitals decode correlated charge migration dynamics}
\author{Imam S. Wahyutama}
\author{Madhumita Rano}
\author{Henrik R.~Larsson}
\affiliation{Department of Chemistry and Biochemistry, University of California, Merced, CA 95343, USA}
\email{p_cm26 [a t] larsson-research . de}

\if\USEACHEMSO1
\begin{document}
\fi

\begin{abstract}

For decades, scientists have studied the intricate charge migration dynamics, where after ionization a localized charge distribution (``hole'') migrates across the molecule on a femtosecond timescale.
This  has the potential for %
controlling 
electrons %
in molecules,
yet a comprehensive understanding of the many aspects of charge migration is still missing.
In this work, we analyze charge migration using an extension of localized intrinsic bond orbitals (IBOs). 
These orbitals lead to a compact representation of the dynamics and map the complex, correlated many-electron charge migration to chemical concepts such as curly arrows and orbital-orbital interactions.
By analyzing multiple challenging scenarios, we show how IBOs %
enable us to identify key mechanisms in charge migration. For example, we show that different mechanisms are responsible for converting %
a $\pi$-shaped hole to a $\sigma$-shaped hole
and \textit{vice versa}. We explain these in terms of hyperconjugation interactions and configurations that couple orbitals with different symmetries.
We further demonstrate how IBOs can be used to find molecules with high charge migration efficiency.
We carry out all simulations using an efficient set up of the time-dependent density matrix renormalization group (TDDMRG), correlating as many as 45 electrons in 50 orbitals. 
We believe that our results will be useful to design future experiments.  The proposed IBO analysis is applicable to other types of real-time electron dynamics and spectroscopy.
 \end{abstract}

\maketitle

\section{Introduction\label{sec:intro}}

After a removal of an electron, how do the remaining electrons in a molecule respond, and how can this response be tuned and exploited?
Answering this is central to the recently established field of attochemistry,\cite{atto-el-dyn-mol-2017,atto-mol-dyn-2018,quantum-chem-atto-2020,attochemistry-future-2021}
and is useful to understand light-induced effects in DNA\cite{dna-ionize-2017,dna-radiation-dmg-2006} %
and photofragmentation,\cite{charge-transfer-weinkauf-1996, levine-select-react-1997, charge-direct-react-1998, control-select-reaction-1999,
  peptide-fragment-pathways-2005,
  cm-peptide-2007, uv-photogragment-2009} 
among many others.\cite{cm-select-reaction-covion-1999,reaction-control-kling-2013,what-observe-real-time-2014,cm-control-laser-2015,
Correlation2016larsson-2016,control-smooth-pulse-2017,Selfprobing2017azoury-2017}
One particular response after an electron removal is 
charge migration, in which 
an often localized ``charge cloud'' or electron ``hole'' migrates through a molecule.\cite{levine-select-react-1997, cm-cederbaum-1999, control-select-reaction-1999}
Charge migration is initiated within attoseconds\cite{cederbaum-atto-response-2005, timescale-chem-2006}%
and leads to the initial hole migrating to the opposite end of small molecules such as aminoacids
in just a few femtoseconds.\cite{cm-amino-acid-2012, cm-phenylalanine-2014,cm-amino-acid-2015,cm-iodo-2015,hole-survive-nucl-2015}
So far, a plethora of simulations have investigated many properties and effects of charge migration, such as
the controllability of the dynamics using ultrashort laser pulses,\cite{cm-control-laser-2015, control-smooth-pulse-2017}
nuclei-induced decoherence 
effect,\cite{cm-quantum-nucl-2015, atto-migrate-benzene-2015, decohere-full-quantum-2017, cm-quantum-nucl-2017,cm-propiol-full-quantum-2018, de-re-pyrene-2023, mol-long-cm-2023}
and
molecular symmetry\cite{cm-adc-chordiya-2023} or lack thereof,\cite{cm-break-symm-2020, 5-heterocyclic-cm-2023, symmetry-reduction-probe-2023} 
among others.\cite{cm-phenylethyl-2008, cm-oligopeptides-2013, cm-amino-acid-2015, cm-tddft3-2017, qcontrol-elflux-2017, cm-nucl-glycine-2017,%
cm-particle-2022,cm-fmatch-2024, corr-cm-mpi-2025}

To systematically understand charge migration, we must identify universal patterns and rules that allow us to predict charge migration starting from different orbitals in different molecules.\cite{cm-cederbaum-1999,cm-breidbach-2003, cm-breidbach-num-2007, cm-molmode-2021, cm-attochem-2023, cm-core-hole-shape-2025}
As the nonequilibrium, many-body, real-time electron dynamics of charge migration can be very complex and require the inclusion of electron-correlation effects,\cite{cm-cederbaum-1999}
many tools have been used to analyze charge migration.
Among others, these
include 
various time-frequency analyses,\cite{cm-cederbaum-1999,cm-breidbach-2003, cm-breidbach-num-2007,cm-localion-2025}
density matrices (``hole density,'' which we will define later)
and their eigenstates (``natural charge orbitals''),\cite{cm-breidbach-2003}
partial charges,
\cite{elnuc-cm-ct-cone-2013,cm-caffeine-2021,wahyutama-tddmrg-2024}
electron fluxes,\cite{qcontrol-elflux-2017, analyze-n-dyn-2017, probe-flux-trxrs-2020, ultrafast-cm-iccnh-2023, 5-heterocyclic-cm-2023, cm-coherence-increase-2025}
decompositions of the time-dependent wavefunction in terms of hole configurations,\cite{cm-breidbach-2003}
and
the time-dependent electron localization function.\cite{cm-tddft3-2017}
Despite these tools and both a large number of simulations and experiments, many features of charge migration still remain elusive.\cite{
  cm-ct-review-2017,
  cm-molmode-2021,
  cm-init-corr-bands-2022,  
  decohere-revival-attocm-2022,
  quantum-chr-glycine-2022,
  attochem-questions-2023,
  capture-chiral-dyn-2024,
  atto-chiral-triatom-2026}

Next to the aforementioned ways to  understand charge migration,
quantum chemistry offers many more tools 
to gain insights
by %
mapping the complicated many-body wavefunction to simple chemical concepts such as curly arrows\cite{partial-valence-interp-1922} 
and atomic charges. Examples of such tools include 
the quantum theory of atoms in molecules,\cite{bader1990atoms-1990,efficient-qtaim-2013}
energy decompositions,
\cite{kitaura-morokuma-eda-1976,neda-1996,lmoeda-2009,gkseda-2014,almoeda-2016,sobeda-2023}
wave function tilings,\cite{elstru-benzene-tiling-2020, hitchhiker-wave-function-2022}
entanglement patterns,\cite{entangle-single-multiref-2012,bond-corr-theory-2017,entangle-loc-orb-2023,meaos-2026} 
and localized orbitals.\cite{boys_loc-1960,er_loc-1963,pm_loc-1989,polar-atom-orb-scf-1997,polar-atom-orb-2000,fast_local_vorb-2005,intrnc-minbas-wvn-2011,ibo-2013,molintrsc_analysis-i-2013,meaos-2026} %
Particularly, intrinsic bond orbitals (IBOs)\cite{ibo-2013} have been shown to be very useful, e.g.,
to verify empirical laws,\cite{ibo-2013,electron-flow-2015}
to understand oxidation states,\cite{ibo-2013,stable-in-gold-2015,alkane-activate-2024}
and 
to gain an understanding of chemical reactions in terms of curly arrow  mechanisms, \cite{electron-flow-2015, cpcet-vs-hat-2018, alkane-activate-2024, reduction-CO2-2025}
e.g., for gold catalysts,\cite{stable-in-gold-2015} electron transfer in proteins,\cite{cpcet-vs-hat-2018} 
and electron flow in chemical reactions.\cite{electron-flow-2015,methane-to-methanol-2023}
However, most of these
use cases involve ground state electronic structures, 
with some exceptions covering excited eigenstates.\cite{livvo-core-excite-2017, loc-bond-o2-2020}

In this work, we extend the IBO formalism to describe real-time, correlated, nonequilibrium charge migration. 
Using  various challenging scenarios in complex molecules,
we show that IBOs work well even for complex charge migration where the commonly assumed molecular orbital picture breaks down. In addition, their localized and time-independent character makes IBOs easier to analyze than frequently used tools such as 
time-dependent natural charge orbitals.
We also show that IBOs can aid the interpretation of
ionization spectra without requiring real-time dynamics simulations.
Importantly,
we connect established concepts such as hole configurations to IBO-derived
quantities and interpret charge migration using
local symmetries, curly-arrow
mechanisms based on Lewis structures, as well as
intermolecular and 
through-space orbital interactions based on  hyperconjugation.\cite{hyperconjugation-2019}
We introduce the latter
as new, important concept for understanding charge migration.

Simulating correlated, nonequilibrium real-time electron dynamics extends
electronic structure methods beyond their usual domain.\cite{td-elstru-rev-2020}
We use a recently developed efficient pipeline\cite{wahyutama-tddmrg-2024} for the
time-dependent quantum-chemistry density matrix renormalization group
(TDDMRG).\cite{dmrg-study-polynom-2002, tdvp-tensor_train-2013,
tdvp-time_integration-2015, tdvp-unify-2016, PAECKEL_tddmrg_review-2019,
tddmrg-pfannkuche-2019, tddmrg-baiardi-2021, low-comm-dmrg-2021,
block2-2023, hrl-mctdh-rev-2024} As an inherently multireference method,
TDDMRG is well suited to charge migration involving superpositions of highly
excited configurations.
We simulate fully correlated dynamics of molecules with up to $45$ electrons in $50$ active orbitals, which is, to our knowledge, the largest TDDMRG simulation to date and
prohibitively expensive for many other multireference methods.
Together with our IBO analysis framework, the TDDMRG may also support
studies of other ionization and excitation spectra,
\cite{nto-2003,orb-to-obs-2020,visualize-tddft-excite-2024,exc-relax-diffdens-2026}
nonequilibrium dynamics of excited molecules,
\cite{charge-sep-holmob-2011,visible-optical-excitation-2022,rel-rtddft-valcor-atas-2023,chiral-magnet-from-4curr-2026}
and core-ionized molecules.\cite{probe-chr-core-vacant-2025}

The remainder of this paper is organized as follows. 
In \autoref{sec:theory} we  overview theories and tools employed in the present work, including an extension to the IBO concept. We present the results of our simulations and analyze them in \autoref{sec:results}. We conclude our findings and give our perspective on future use cases  in \autoref{sec:conclusions}.

\section{Theory\label{sec:theory}}

This Section is outlined as follows.
In \autoref{sec:tevo}, we present an overview of the TDDMRG to simulate real-time electron dynamics. We can gain additional insights into such dynamics from ionization spectra, which we discuss in \autoref{sec:gf}.
In  \autoref{sec:cm_analysis} we review charge migration and common tools to analyze it.
As a new tool to analyze charge migration, we give an overview of IBOs and their extensions in \autoref{sec:ibo}. %
Lastly, in  \autoref{sec:symmetry} we discuss global and local symmetries. Throughout, we use atomic units  unless specified otherwise.

\subsection{Time evolution using the density matrix renormalization group\label{sec:tevo}}

We simulate electron dynamics by approximately solving the time-dependent Schrödinger equation using the TDDMRG. In this method, the time-dependent state is expressed as a \textit{matrix product state} (MPS),
\begin{gather}
    \ket{\Psi(t)} = \sum_{\{\sigma\}} \left(\prod_{i=1}^K \mathbf M^{\sigma_i}(t) \right)
    \ket{\sigma_1 \ldots \sigma_K},
    \label{eq:mps_form}
\end{gather}
where $K$ is the number of active spatial orbitals, $\ket{\sigma_i}$ denotes one of the four possible spin configurations in the $i$-th orbital, %
and 
$\mathbf M^{\sigma_i}(t)$ are matrices of size $D_{i-1}\times D_{i}$. The largest of the $D_i$ is called \textit{bond dimension}, $D$. The larger $D$, the better the approximation to the exact full configuration interaction (CI) state.
By using the \textit{ansatz} of \autoref{eq:mps_form} to find the extremum of the time-dependent variational principle, one obtains a set of coupled equations to solve for $\mathbf M^{\sigma_i}(t)$. 
A Lie-Trotter splitting uncouples these equations, which leads to an algorithm that, essentially, sequentially propagates one matrix $\mathbf M^{\sigma_i}(t)$ at a time. For more information, we refer to \lits{tdvp-time_integration-2015,tdvp-unify-2016,PAECKEL_tddmrg_review-2019,hrl-mctdh-rev-2024},
\compdetails
and to our recently developed simulation pipeline.\cite{wahyutama-tddmrg-2024}

\subsection{Ionization spectra \label{sec:gf}}

While charge migration focuses on the time-dependent dynamics after ionization, analyzing it in terms of a time-independent view through ionization spectra provides an additional, helpful perspective, as it allows scientists to identify the relevant eigenstates that are contributing to charge migration.\cite{cm-cederbaum-1999,cm-breidbach-2003} 
Through linear response theory, the spectrum can be obtained from the dynamics via the  autocorrelation function $A(t)$,\cite{mol-correl-func-1968,tannor-book-2007}
  \begin{equation}
    A(t) = \braket{\Psi(t=0)}{\Psi(t)}.
  \end{equation}
A Fourier transform of $A(t)$ leads to the photo-absorption spectrum $P$ as a function of excitation energy $E$,
\begin{equation}
  P(E) = \lim_{\eta\to 0} \frac{1}{\pi} \Re \int_0^\infty \exp(\ii E t)  \exp(-\ii \eta t) A(t) \dd t . \label{eq:spectrum_td}
\end{equation}
To describe the experimental energy resolution,
a finite value of $\eta$ is used, which leads to a convolution of $P(E)$ with a Lorentzian.

As an alternative to the time-dependent approach,
the spectrum can also be obtained by computing a Green's function $G(E)$ as 
\begin{equation}
  P(E) = - \Im \matrixe{\Psi(t=0)}{\hat G(E)}{\Psi(t=0)}, \label{eq:spectrum_gf}
\end{equation}
with
\begin{equation}
    \hat G(E) =\frac{1}{\pi} \lim_{\eta \to 0}\frac{1}{E+ \ii \eta - \hat H}.
\end{equation}
In practice,
the time-dependent approach, \autoref{eq:spectrum_td} allows us to obtain the total spectrum with a finite resolution determined by the propagation time, whereas the time-independent Green's-function-based approach, \autoref{eq:spectrum_gf}, allows us to target 
particular peaks with a resolution determined numerically through a finite value of $\eta$ (the smaller $\eta$, the higher the resolution but the more difficult it is to compute \autoref{eq:spectrum_gf}). The spectrum is then described as a linear combination of Lorentzians.\cite{dyn-dmrg-2002}

The state ${\hat G(E)}\ket{\Psi(t=0)}$ provides further insights into the spectrum. Expanding it in terms of eigenstates $\{\ket{u_i}\}$ with energies $\{E_i\}$,
\begin{equation}
  \hat G(E)\ket{\Psi(t=0)} =\frac{1}{\pi} \lim_{\eta \to 0} \sum_i  \frac{\braket{u_i}{\Psi(t=0)}}{E+ \ii \eta - E_i} \ket{u_i},
  \label{eq:gf_cv}
\end{equation}
shows that  ${\hat G(E)}\ket{\Psi(t=0)}$ has dominant contributions from eigenstates close to the target energy $E$, and, on resonance with $E\to E_i$, it turns into eigenstate $\ket{u_i}$ after normalization.

Expressing $\ket{\Psi(t=0)}$ as an MPS, \autoref{eq:spectrum_gf} can be solved using DMRG-like procedures that consist of solving linear systems of equations sequentially for each of the matrices 
$\mathbf M^{\sigma_i}$ in \autoref{eq:mps_form}. For more details, we refer to \lits{dyn-dmrg-2002,ddmrg++-2017,block2-2023,comp-eigenst-ttns-2025}.

\subsection{Charge migration \label{sec:cm_analysis}}
Charge migration happens when, after ionization,
a positively charged region or \textit{electron hole} moves through the molecule on an ultrafast timescale of attoseconds to few femtoseconds.
To model charge migration, we use the common\cite{koopman-inner-ion-1970, levine-select-react-1997,%
cm-breidbach-2003, cm-breidbach-num-2007, cm-peptide-2007, loc-dyn-dens-sudden-2012, mr-photoel-so-2015, cm-molmode-2021, rhodyn-2022}
sudden ionization approximation,\cite{fast-photoion-1977,mo-breakdown-advchem-1986} 
and describe the initial state $\ket{\Psi(0)}$
by removing an electron from the ground state $\ket{\Psi_\text{GS}}$.
In second quantization, this is described by 
\begin{equation}
  \ket{\Psi(0)} = \matrixe{\Psi_\text{GS}}{\hat a_i^\dagger \hat a_i}{\Psi_\text{GS}}^{-1/2} \, \hat{a_i} \ket{\Psi_\text{GS}},
  \label{eq:psi_0_gen}
\end{equation}
where 
$\hat a_i$ is the annihilation operator of a localized orbital $\ket{i}$. 
Sudden ionization can be achieved, e.g., by using laser pulses with dominant frequencies that overlap with orbital energies of one particular functional group in the molecule. 
Next to sudden ionization, we assume that the nuclei are frozen during the electron propagation, which is a good assumption for the first few femtoseconds.
While this setup is idealized, such simulations still 
illustrate charge migration,  
often lead to significant insights, and match experimental observables.\cite{cm-cederbaum-1999,cm-breidbach-2003,cm-breidbach-num-2007,cm-iodo-2015}

A central quantity of charge migration is the hole density $h(\mathbf r,t)$, which is defined as the difference between the spin-summed ground-state one-particle density $\rho_\text{GS}(\mathbf r)$ and the density $\rho(\mathbf r,t)$ of the ionized wavepacket at time $t$,
\begin{equation}
  h(\mathbf r,t) = \rho_\text{GS}(\mathbf r) - \rho(\mathbf r,t).\label{eq:h_dens}
\end{equation}
For single ionization, $h(\mathbf r,t)$ describes the motion of one hole particle (it integrates to $1$).
The contribution $q_i(t)$ of orbital $\ket{i}$ to $h(\mathbf r,t)$ is called hole occupancy, and computed as $q_i(t) = \matrixe{i}{\hat h(t)}{i}$,
where $\hat h(t)$ is the corresponding operator of $h(\mathbf r,t)$.
According to \autoref{eq:h_dens}, $h(\mathbf{r},t)<0$ indicates an electron excess relative to the neutral molecule, whereas $h(\mathbf{r},t)>0$ indicates a hole.
The same sign convention applies to the hole occupancies $q_i(t)$: positive and negative values denote hole occupancy and electron excess, respectively, in orbital $\ket{i}$.
Analogous to natural orbitals, the time-dependent orbitals that diagonalize $\hat h(t)$ are called natural charge orbitals and can aid the analysis.\cite{cm-breidbach-2003} However, their time dependence and often complex-valued form make them less convenient than time-independent, real-valued orbitals.

Another useful way to analyze charge migration is to expand the time-dependent cationic state $\ket{\Psi(t)}$ in terms of a CI-based expansion, \cite{cm-breidbach-2003}
\begin{equation}
  \ket{\Psi(t)} = \sum_j c_j(t) \hat a_j \ket{\Phi_\text{GS}} + \sum_{r,k < l} c_{rkl}(t) \hat a_r^\dagger \hat a_k \hat a_l \ket{\Phi_\text{GS}} + \dots
  \label{eq:hole_expansion}
\end{equation}
where $\ket{\Phi_\text{GS}}$ is an approximation to the neutral ground state.
In many cases,  $\ket{\Phi_\text{GS}}$ is a Hartree-Fock-based configurations but
in our case, it will be the DMRG-approximated ground state,
 $\ket{\Psi_\text{GS}}$.
The configurations or states $\hat a_j \ket{\Phi_\text{GS}}$ are called one-hole (1h) configurations and $\hat a_r^\dagger \hat a_k \hat a_l \ket{\Phi_\text{GS}}$ are called  two-hole one-particle (\mbox{2h1p}) configurations.%
\footnote{Note that, in the following, we will use the terms ``2h1p configurations'' and ``2h1p states'' interchangeably. In our case, the operators act on a correlated state, giving a ``2h1p state.`` However, the term ``2h1p configuration'' is more commonly used.}
In contrast to 1h configurations, 2h1p and higher-order configurations extend the mean-field ansatz and are required to recover electron correlation:
In ground state electronic structure, 1h (1h and 2h1p) configurations correspond to CIS (CISD).
Furthermore, 2h1p and higher-order configurations
can involve orbitals of different irreducible representations (irreps), which leads to important consequences in charge migration. 
Since 2h1p configurations are formed through a removal of two electrons from $\ket{\Phi_\text{GS}}$ followed by an \emph{addition} of another electron in a different orbital, the occurrence of 2h1p (and higher-order, 3h2p etc.) configurations can be identified by orbitals with negative hole occupancies.
These orbitals are often antibonding, and their occupancy weakens the corresponding bond.
We will discuss this below in \autoref{sec:symmetry} and
\autoref{sec:init-hole-effect}.

\subsection{Intrinsic bond orbitals and their extensions \label{sec:ibo}}

IBOs are convenient to create and provide orbital shapes that match chemical intuition.\cite{ibo-2013}
The original IBO procedure is for bonding orbitals. Below, we discuss an extension to antibonding orbitals.
IBO constructions start with the minimal-basis \textit{intrinsic atomic orbitals} (IAOs), which are not only  
localized around atoms but also encode the polarization environment of the molecule. 
IAOs are  computed by projecting $n_\text{min}$ 
free-atom minimal basis set orbitals onto polarized Hartree-Fock (HF) occupied and unoccupied molecular orbitals (MOs) $\ket{o}$
that are computed using a larger basis set.
More details about the IAO construction are described in \sisec{2}.

The  IBOs $\{\ket{i}\}$ are then constructed by 
localizing the $n_\text{occ}$ occupied HF MOs
using a variation of Pipek-Mezey orbital localization,\cite{pm_loc-1989} which minimizes the number of atoms an orbital is centered on using IAO-based atomic charges.
That is, IBOs are a rotation of the occupied MOs,
\begin{equation}
\ket{i} = \sum_{o} T_{oi} \ket{o}.
\label{eq:ibo_rot}
\end{equation}
The unitary matrix $\matr T$ is chosen to maximize the electron occupancies $n_A(i)$ of IBO $\ket{i}$ at atom $A$, i.e., 
    $\text{max}_{\matr T}%
    \sum_A^{n_\text{atoms}} \sum_i^{n_\text{occ}}\left[n_A(i)\right]^4$.
Due to \autoref{eq:ibo_rot},
the IBOs span exactly the same space as the occupied MOs and generate the same single-determinantal wavefunction.
\autoref{fig:furfural-sym} displays examples of IBOs.

For excited states and non-equilibrium dynamics  such as those occurring in charge migration, excited configurations with occupied antibonding orbitals 
play a crucial role. However, by construction, IBOs are all bonding. 
To generate antibonding orbitals, we here extend the IBO procedure in a straightforward way using split localization.
As we have $n_\text{occ}$ IBOs, we generate  $n_\text{min} -n_\text{occ}$ localized, antibonding orbitals that we dub here  \textit{complementary IBOs} (cIBOs).
These cIBOs are related to, e.g., the valence-virtual orbitals from \lits{molintrsc_minbas-i-2004, molintrsc_analysis-i-2013},
and have previously been generated by similar procedures.\cite{livvo-core-excite-2017, masked-phenyl-niiv-2019, gen-intrn-orb-quart-2021}

We generate cIBOs by first projecting the 
$n_\text{min}$  IAOs $\{\ket{\nu}\}$ into the space orthogonal to the IBOs $\{\ket{i}\}$, yielding $\ket{\tilde c} = \left( \hat 1 - \sum_{i=1}^{n_\text{occ}} \ketbra{i}{i} \right) \ket{\nu}$.
We then remove linearly dependent vectors from the set $\{ \ket{\tilde c} \}$ using
singular value decomposition,
$\mathbf{C} = \mathbf{U} \, \text{diag}({\matrgreek{\sigma}}) \, \mathbf{V}^\dagger$, with $C_{\kappa c}=\braket{\kappa}{\tilde c}$, 
where $\{\ket{\kappa}\}$ are orthogonalized atomic orbitals. Since $\ket{\tilde c}$ are constructed by projecting the  $n_\text{min}$ IAOs to the space orthogonal to the $n_\text{occ}$ IBOs, 
there are $n_\text{min} -n_\text{occ}$ nonvanishing singular values $\sigma$. Thus, we use the first $n_\text{min} -n_\text{occ}$ columns of $\matr U$ to construct the orthogonal orbitals $\ket{c}=\sum_{\kappa} U_{\kappa c}\ket{\kappa}$.
Finally, we obtain the cIBOs  by localizing $\{\ket{c}\}$ in the same way as the IBOs, as described around \autoref{eq:ibo_rot}.
For the cIBO localization, we use the same IAO charges as for the IBOs, for consistency.
Different charges can  produce nearly identical results.\cite{livvo-core-excite-2017}
By construction, the cIBOs form a set of localized orbitals that are orthogonal to the IBOs, and 
the union of IBOs and cIBOs exactly spans the same space as that of the IAOs. Hence, in the following, when no context is needed, we will collectively refer to IBOs or cIBOs as just IBOs.

IBOs and cIBOs together form a valence-internal space\cite{molintrsc_analysis-i-2013}
and in most cases will suffice for a comprehensive qualitative analysis of our electron dynamics.
The remaining orbitals correspond to what is called hard virtual space\cite{fast_local_vorb-2005}
or valence-external space.\cite{molintrsc_analysis-i-2013}
Here, we dub these remaining orbitals \mbox{oIBOs}, where ``o'' refers to the fact that these orbitals are orthogonal to the IBO and cIBO space.
We construct the oIBOs in the same way we construct the cIBOs.
\subsection{Global and local symmetries \label{sec:symmetry}}

\begin{figure}[t]
     \centering
     \includegraphics[width=0.9\linewidth]{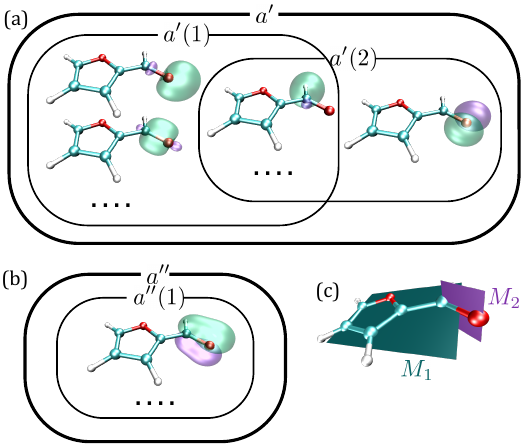}
     \caption{
     Comparison of global and local symmetries.
       (a, b) Venn-diagram illustrations of the relationship between the global and local symmetries for the IBOs localized around the \ch{C=O} bond in furfural. The $a'(2)$ IBOs are oriented parallel to the molecular plane.
       (c) The two mirror planes that define the global ($M_1$) and local ($M_2$) symmetries illustrated in the Venn diagrams. 
       The local symmetries are labeled as $a'(x)$ and $a''(x)$, respectively, where $x$ arbitrarily 
       enumerates them.
     }
     \label{fig:furfural-sym}
\end{figure}

To characterize hole motion in our electron-dynamics analysis, we distinguish global and local symmetries. 
Global symmetry refers to the irrep of the molecular point group. 
Local symmetry is an additional classification
within a spatially restricted region spanning only a subset of atoms. 
The role of global symmetry in charge-migration dynamics has been explored previously,\cite{cm-break-symm-2020, 5-heterocyclic-cm-2023, symmetry-reduction-probe-2023, cm-adc-chordiya-2023}
and mainly leads to the appearance of orbitals 
whose symmetry differs from that of the total wavefunction, which is often due to the 2h1p configurations defined in \autoref{eq:hole_expansion}.

As shown below, restricting symmetry to local molecular regions --- rather than the molecule as a whole --- provides a useful framework for interpreting key features of charge migration. 
Such local symmetry is meaningful only for localized orbitals. 
This is not a restriction,
as the wavefunction is invariant under a 
unitary transformation that localizes the orbitals.
Unlike other uses of local symmetry in different contexts,\cite{local-symm-1d-2013, dyn-local-symm-2017}
we apply local symmetry here as an approximate classification and neglect small IBO distortions that violate the chosen local symmetry.

\autoref{fig:furfural-sym} highlights the interplay of global and local symmetries in selected IBOs of  the $C_s$-symmetric \furf{}. The orbitals in panel (a) transform as $a'$ because they are even under reflection in the molecular plane $M_1$ in panel (c), whereas the orbital in panel (b) transforms as $a''$ due to its odd $M_1$ parity. Focusing on the formyl group, the $a'$ orbitals can be further classified by their parity with respect to a second reflection plane $M_2$ defined in panel (c), which contains the \ch{C=O} bond and is perpendicular to $M_1$. This leads to a local  symmetry with irreps $a'(1)$ and $a'(2)$ that denote even and odd $M_2$ parity, respectively. Orbitals from different global irreps necessarily fall into different local symmetries, but some orbitals lack a particular local symmetry. These can be decomposed into orbitals with different symmetries, as indicated by the overlap of the Venn diagrams for $a'(1)$ and $a'(2)$.
Next to the local symmetries shown in \autoref{fig:furfural-sym}, other local symmetries are possible for \furf.

\section{Results and Discussions \label{sec:results}}

In the following, we present our investigation of charge migration dynamics using IBO hole occupancies as the main tool of analysis. We start by validating and introducing the IBO-based analysis in the simple scenario of \clace{} in \autoref{sec:clace}. We then apply it to study changes of hole symmetries in the charge migration of \pheace{} and \furf{} in \autoref{sec:init-hole-effect}, followed by revealing 
different charge migration efficiencies in conformers 
in \autoref{sec:conformer}. 
Details on the computational setup are given in \compdetails %
and in the \siref.

\subsection{Introduction of the IBO analysis in \clace{}\label{sec:clace}}

\begin{figure*}[!t]
  \centering
  \includegraphics[width=\figfull]{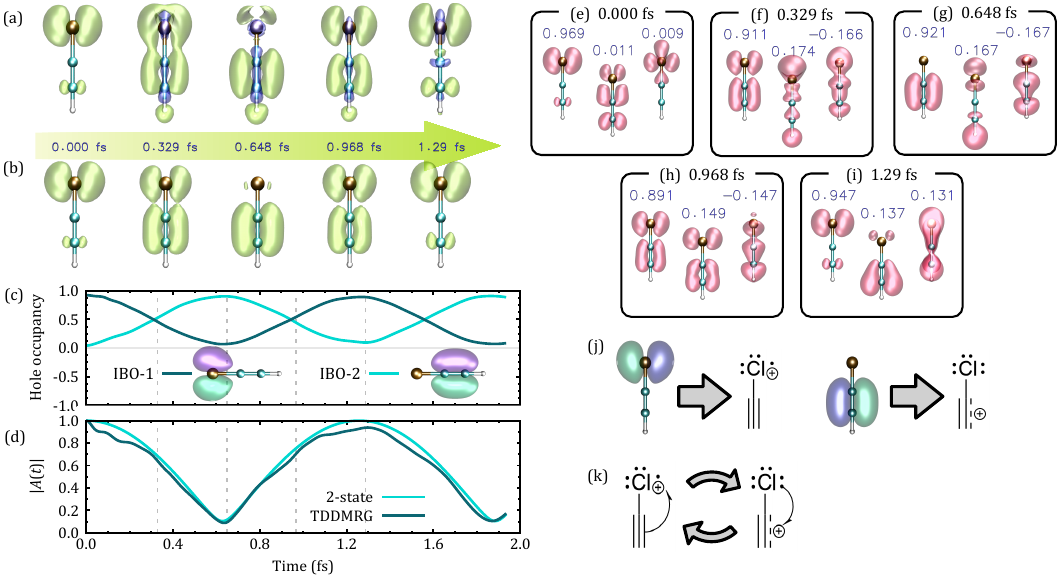}
  \caption{
  Charge migration analysis of \clace.
  (a) Hole density snapshots. %
  Regions with positive (negative) density are shown as green (purple) lobes.
  (b) Hole density snapshots %
  obtained by projecting the density from (a) to the space of the two dominant IBOs. %
  (c) Hole occupancies of the two IBOs that dominate the hole dynamics. The IBOs are shown on the graph.
The times at which the snapshots in (a) are taken are indicated with dashed vertical lines.
  (d) Absolute value of the autocorrelation function (dark green curve) in comparison to a two-state model (cyan curve). 
  (e)--(i) Orbital densities of the three most dominant (largest absolute eigenvalues) natural charge orbitals at the times of the snapshots in (a).
  (j) Mapping of IBOs to cationic Lewis structures. 
  (k) Curly-arrow representation of the periodic hole oscillations. %
  }
  \label{fig:draft-chloro}
\end{figure*}

One of the first experimental observations of charge migration was made in iodoacetylene.\cite{cm-iodo-2015}
Since iodoacetylene is small and linear, its dynamics provide a simple setting, which we use to introduce the IBO analysis workflow before turning to more complex scenarios.
In this molecule, 
a $p$ hole initially localized on iodine migrates to the opposite end and returns to iodine in about \qty{1.9}{fs}.
The initial state is approximately a 50-50 superposition of the ground state $\ket{u_0}$ and first excited state $\ket{u_1}$
of the cation. %
In general, for such superpositions of the cationic state, $\ket{\Psi(t)} = c_0 \ket{u_0} + c_1 \ket{u_1}$, 
the hole density is given by 
\begin{align}
  h(\mathbf r, t) =
  &\,
  \rho_\text{GS}(\mathbf r)
  -
  c_0^2 \rho_0(\mathbf r)
  -
  c_1^2 \rho_1(\mathbf r)
  \nonumber\\
  &\,
  -
  2 c_0 c_1 \rho_{01}(\mathbf r) \cos(\Delta E t),
  \label{eq:clac_2state}
\end{align}
where $\rho_0$ and $\rho_1$ are the electron densities of \ket{u_0} and \ket{u_1}, respectively, $\rho_{01}$ is the transition density between these two eigenstates, and $\Delta E$ is their energy difference.
Such a situation has been dubbed \textit{hole mixing}.\cite{cm-breidbach-2003}

Here, we model the charge migration in iodoacetylene using \clace{}, which has fewer electrons and weaker relativistic effects and hence is easier  to simulate.
Our TDDMRG simulation uses a bond dimension of $D=1,000$ and includes all $41$ non-core orbitals of the def2-SV(P) basis; see \compdetails for more details.
We construct the initial  $p$ hole state by 
removing an electron from the neutral in an orbital that is a superposition of the HF HOMO and HOMO-1. 

The snapshots of the evolving hole density in \clace{} %
are shown in \sfigref{fig:draft-chloro}{a}. 
Given that only the cationic ground state and the first excited state %
should dominate the dynamics, we expect that also only two IBOs will be dominant.\footnote[4]{This assumes that the first excited state is approximately described by an orbital excitation.}
Indeed, %
only two IBOs %
contribute to the hole density. %
To gauge how well the two dominant IBOs reproduce the dynamics, we recompute the hole density using just these two dominant IBOs, ignoring all other orbitals, and
show it in \sfigref{fig:draft-chloro}{b}. The main hole lobes (positive, green surfaces) in \sfigref{fig:draft-chloro}{a} are faithfully reproduced by the two IBOs. The missing features in the IBO-projected hole density are mostly electron lobes (negative, purple surfaces), which are mainly located near the molecular axis. 
These differences are due to additional correlation effects not accounted for when using the two dominant IBOs only. 
Importantly, by using the \emph{full} IBO, cIBO, and oIBO spaces, the dynamics can be exactly reproduced. 

The shapes and the time-dependent hole occupancies of the two dominant IBOs are shown in \sfigref{fig:draft-chloro}{c}.
They are close to the initial $p$ hole state at the Cl atom  (IBO-1) and a $\pi$ hole state at the \ce{C=C} bond (IBO-2).
In agreement with 
\autoref{eq:clac_2state},
the hole occupancies  exhibit a clear periodic oscillation with a period of ${\sim}\qty{1.29}{fs}$. %
To further confirm this, \sfigref{fig:draft-chloro}{d} compares the absolute value of the autocorrelation function, $|A(t)|$ with a reproduction from the 2-state model in \autoref{eq:clac_2state} (see \sisec{3} for more details). %
As the hole density reproduced with two IBOs only,
the two-state autocorrelation function closely resembles the 
TDDMRG reference.

As discussed in \autoref{sec:cm_analysis},
a common way to analyze charge migration is to use time-dependent natural charge orbitals, obtained by diagonalizing the hole density at different time steps.
The densities of the three most dominant natural charge orbitals at different times are shown in \sfigrref{fig:draft-chloro}{e}{i}.
For all times shown, the most dominant natural charge orbital reproduces the hole density almost as good as the  two IBOs.
Note that some small details are better captured by the two IBOs than by the dominant natural charge orbital, e.g., a small contribution to the hole density at the \ce{Cl} atom at \qty{0.648}{fs}.
Like the excluded IBOs, the additional natural charge orbitals lead to a full reproduction of the hole density. 
However, unlike IBOs, the natural charge orbitals do not directly reveal the two-state characteristic of the dynamics. The time dependence of the natural charge orbitals and, in most scenarios, their complex-valued nature further complicate an analysis based on them.

Previously, IBOs have been used to draw curly-arrow mechanisms of chemical reactions by analyzing IBOs along  reaction paths.\cite{electron-flow-2015, cpcet-vs-hat-2018, alkane-activate-2024, reduction-CO2-2025}
For the simple charge migration scenario of \clace, we follow a similar procedure but analyze the IBOs not along a reaction path but along the real simulation time for a fixed nuclear structure.
We first map each of the two IBOs to Lewis structures of the cation, see \sfigref{fig:draft-chloro}{j}. 
Following the hole occupancy analysis in \sfigref{fig:draft-chloro}{c}, we then identify the main Lewis structures at times when only one of them dominates and connect them through curly arrows as shown in   \sfigref{fig:draft-chloro}{k}. 
As the single-headed curly arrows (harpoons) correspond to the ``motion'' of a single electron, their direction is opposite to the motion of the electron hole.\footnote[5]{
For simplicity, 
we avoid here the introduction of additional arrows that would depict the motion of a hole/positive charge. }
For this simple case, the Lewis-structure mechanism fully agrees with \autoref{eq:clac_2state}.

The presented concepts used for the charge migration in \clace{} can be straightforwardly extended to larger linear molecules. We demonstrate this in \sisec{4} for \clbut{}, which contains  one additional conjugated \ce{C#C}  pair and hence requires one more $\pi$ IBO and hole state for the analysis. Therein, we also discuss additional convergence criteria for the IBO analysis.

\subsection{Effects of the initial hole symmetry \label{sec:init-hole-effect}}

The dynamics of \clace{} in the preceding Section are relatively simple, for instance, the dominant IBOs all share the same global symmetry as the initial hole. 
In the remainder, we will analyze charge migration in more complicated scenarios---\pheace{} in \autoref{sec:pheace} and \furf{} in \autoref{sec:furf}, which reveal both local and global symmetry effects. %

\subsubsection{\Pheace{}\label{sec:pheace}}

\begin{figure*}[!h]
     \centering
     \includegraphics[width=\figfull]{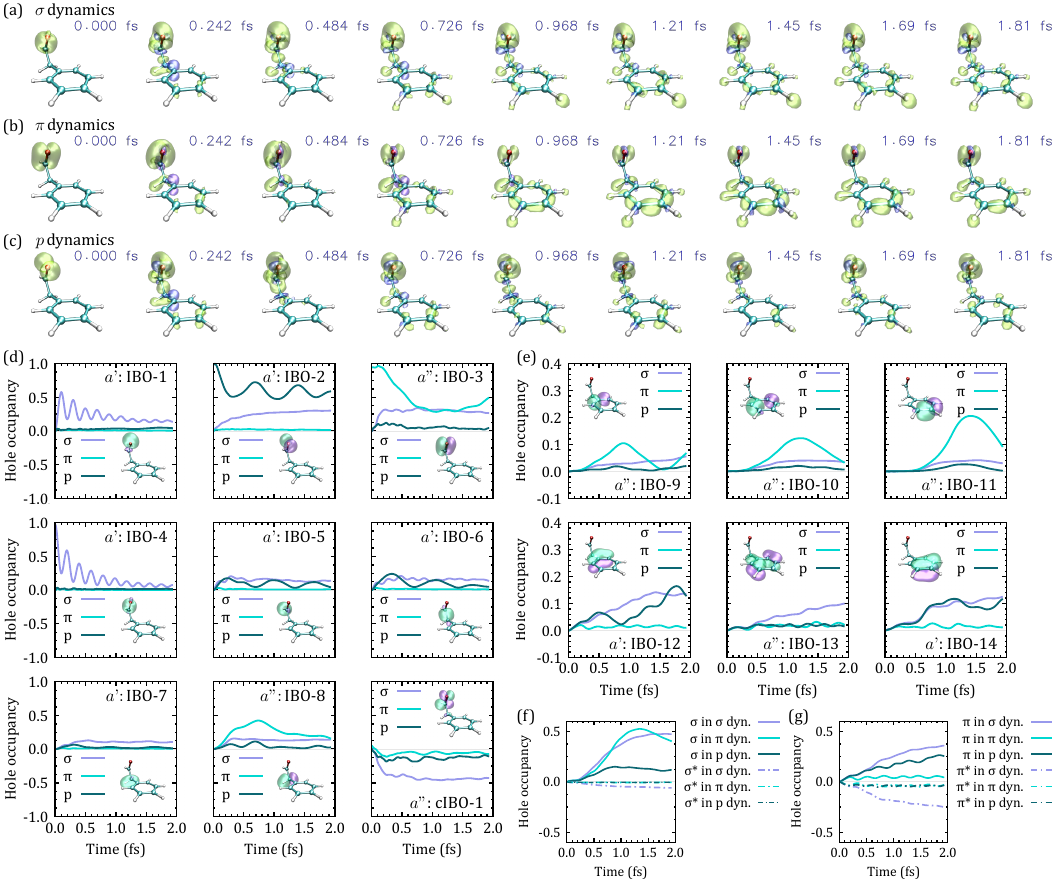}
      \caption{
  Charge migration analysis of \pheace.
     (a)--(c) Hole density snapshots for the $\sigma$, $\pi$, and $p$ dynamics, respectively. The color code is the same as that in \sfigref{fig:draft-chloro}{a}.
     (d) Hole occupancies of IBOs localized in the 2-oxoethyl group during the three dynamics. The IBO associated with each hole occupancy and its irrep label are shown in each panel. 
     (e) Hole occupancies of selected $\sigma$ and $\pi$ IBOs localized at the phenyl group. Note that the ordinate range differs from that in (d).
     (f, g) summed hole occupancies of IBOs within the phenyl ring having $\sigma$ and $\pi$ local symmetries, respectively. This includes IBOs localized at the \ce{H} atoms and those shown in (e).
     }
     \label{fig:draft-phenylacet}
\end{figure*}

Our first more complex charge migration scenario is that of
a $C_s$-symmetric conformer of \pheace
that is
eclipsed between the %
carbonyl group and the phenyl group%
; see \sfigref{fig:draft-phenylacet}{a}.
This raises the question of how an initial hole at the carbonyl group migrates into the phenyl ring and how this depends on the hole's symmetry %
with respect to the phenyl-ring plane.
To answer this, we perform three TDDMRG simulations
with different initial holes, each localized at the carbonyl group: 
1) a $\sigma$ hole, 
2) a $\pi$ hole, 
and
3) a hole in the \ce{O} atom's $p$ lone-pair orbital.
We refer to the resulting dynamics %
as $\sigma$, $\pi$, and $p$ dynamics, respectively. 
See the leftmost panels in ~\sfigrref{fig:draft-phenylacet}{a}{c} for the shapes of the initial holes.

Snapshots of the hole densities during these three dynamics are shown in \sfigrref{fig:draft-phenylacet}{a}{c}. The hole occupancies of dominant IBOs localized at the 2-oxoethyl group are shown in \sfigref{fig:draft-phenylacet}{d},  
while the occupancies of selected IBOs localized at the phenyl ring are shown in \sfigref{fig:draft-phenylacet}{e} (the other phenyl IBO occupancies are shown in \sisec{5.1}).
Furthermore,  we display summed hole occupancies of all phenyl IBOs
 in \sfigref{fig:draft-phenylacet}{e, f} for $\sigma$ and $\pi$ IBOs, respectively.
 Note that, here, $\sigma$ and $\pi$ refer to the local symmetry with respect to the phenyl plane.

The three dynamics have various features that can be explained by different physical mechanisms. In the following, we analyze and discuss three particular mechanisms and the resulting dynamics: 
(a) hyperconjugation-based through-space orbital interactions that 
are responsible for different charge migration efficiencies and that lead to a change of hole shape when migrating from the carbonyl group into the phenyl ring;
(b) excited 2h1p configurations that lead to 
the appearance of orbitals with different global symmetries; 
and (c) the ``breakdown of the molecular orbital picture'' that leads to quasiexponential decays of IBO hole occupancies for the $\sigma$ and $p$ dynamics.
We summarize these effects in \autoref{tab:dyn-summary}.   

\begin{table*}%
   \caption{\label{tab:dyn-summary} 
   Changes of the hole symmetry in \pheace.
   For each dynamics, the hole conversion is shown next to the relevant IBOs at the 2-oxoethyl group and at the entry of the phenyl group (c.f.~\autoref{fig:draft-phenylacet}), and a qualitative explanation.
   For the change of local symmetry,
   the initial hole symmetry corresponds to that of the orbital at the \ce{CO} group\nmydot
   }
   \begin{tblr}{colspec={c c c X[l,m]}, rowhead=1, row{1}={c}, width=\textwidth, rulesep=1pt}
        \toprule
         & Local symmetry\\
         Dynamics/initial hole symmetry &
         Hole symmetry in the Ph ring &
         Relevant IBOs &
         Hyperconjugations around Ph entry \\
         \midrule
         $\sigma$   &
         $\sigma + \pi$  &
         4-8, 12 
          &
         $\sigma-\sigma$ and $\sigma-\pi$ \\%
         $\pi$   &
         $\sigma$  &
         3, 8, 9
          &
         $\sigma-\sigma$\\%
         $p$   &
         $\pi$  &
         2, 5-9, 12
         &
         $\sigma-\pi$\\%
         \midrule
         & Global symmetry\\
         Dynamics/initial hole irrep &
         Final hole irreps &
         Most relevant IBOs &
         Explanation \\
         \midrule
         $\sigma\ (a')$   &
         $\sigma\ (a')+ \pi (a'')$  &
         3, 4,  cIBO-1
         &
         Coupling through 2h1p configurations; 
         ``orbital breakdown''\\
         \bottomrule
      \end{tblr}
\end{table*}

\begin{figure}[t]
     \centering
     \includegraphics[width=\linewidth]{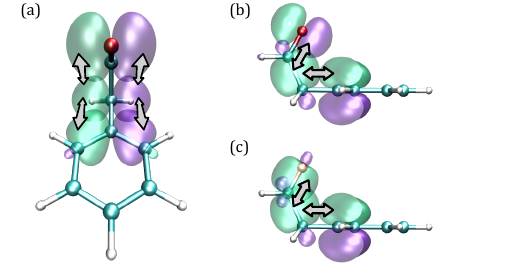}
     \caption{
       Main IBOs taking part in the hyperconjugation interactions
       in \pheace. (a) $\pi$, (b) $p$, and (c) $\sigma$ dynamics. The interactions lead to ring $\sigma$ holes in the first and ring $\pi$ holes in the last two dynamics, respectively. The arrows indicate  the interactions.
     }
     \label{fig:pheace-ibos}
\end{figure}

\paragraph{Hyperconjugation modulates charge migration efficiency and converts the local hole symmetry\mydot}
A striking difference between the three dynamics is the amount of charge migration into the phenyl ring. 
For the time range studied, 
the $p$ ($\pi$) dynamics exhibits the smallest (largest) migration, and the maximum total hole occupancy in the ring is $0.34$ ($0.53$). %
Compared to the $\pi$ dynamics,
the $\sigma$ dynamics features a similar amount of migration into the ring, whose %
origin we will discuss later in the context of 2h1p configurations and global symmetries.

In conjunction with different charge migration efficiencies observed above, for all three dynamics, the initial hole changes its local symmetry when migrating from the carbonyl group into the phenyl ring as follows:  
1) $\sigma \rightarrow \sigma + \pi$, 2) $\pi \rightarrow \sigma$, and 3) $p \rightarrow \pi$. 
This is most noticeable from the phenyl IBO hole occupancies in \sfigref{fig:draft-phenylacet}{e}, and from  the  summed phenyl IBO occupancies in panels (f) and (g).
Changes in local hole symmetry are also evident in the time-dependent natural charge orbitals (\sisec{5.2}), but the time-independent, localized IBOs dramatically speed up the analysis and thus are shown here.

This change of local symmetry is most obvious for the $\pi$ dynamics,
as shown by the hole density at $t = \qty{0.968}{fs}$ in \sfigref{fig:draft-phenylacet}{b} and by the
selected phenyl $\sigma$ IBO occupancies in panel (e). 
What leads to these phenyl ring $\sigma$ holes?
Initially, the \ce{C=O} $\pi$ hole (IBO-3) migrates into the 
two mirror-symmetric \ch{C-H} $\sigma$ bonds of 2-oxoethyl; compare the initial hole density with that at $t=\qty{0.242}{fs}$ in \sfigref{fig:draft-phenylacet}{b} and the occupancies of IBO-8 in panel (d).
Note that, aside from IBOs 3 and 8, no other 2-oxoethyl IBOs have large occupancies. 
After arriving at the \ce{C-H} %
IBO-8,
the hole migrates through the nearby phenyl \ch{C-C} $\sigma$ IBO-9 in the ring, compare the snapshots at $t=\qty{0.484}{fs}$ and $t=\qty{0.726}{fs}$ in panel (b),
and the $\sigma$-ring IBO occupancies in panel (e).\footnote[6]{This mechanism is independent from a negative hole density appearing in the phenyl ring, which appears for all three dynamics and is due to 2h1p configurations and global symmetry changes discussed later.}
Comparing the occupancies of IBOs 9 and 10 shows that migration from one end of the ring to the other requires only \qty{0.5}{fs}.

Qualitatively, we explain
both the different charge migration efficiencies and the change of local symmetries
in terms of 
hyperconjugations/through-space orbital interactions\cite{hyperconjugation-2019} 
that occur
as the hole migrates down the 2-oxoethyl group and arrives at the entry point into the ring.\footnote[7]{Note that quantifying these interactions is beyond the scope of this work.} \sfigref{fig:pheace-ibos}{a} illustrates this mechanism.
These interactions first occur between the initial \ce{C=O} $\pi$ IBO-3 and the \ce{C-H} $\sigma$ IBO-8. %
Once the hole reaches the latter IBO,
additional interactions occur with
the nearby phenyl \ch{C-C} $\sigma$ IBO-9,
leading to the migration into the ring.

In contrast to the $\pi$ dynamics,  the $p$ dynamics
leads to $\pi$ character in the ring.
Note that there is no $\pi$-type conjugation visible in phenyl in \sfigref{fig:draft-phenylacet}{c}, instead, we see $p$-orbital-shaped lobes on the \ce{C} atoms.
However, plots at a different isosurface value reveal conjugation, see \sisec{5.4}. %
We confirm this from the
$\pi$ IBO occupancies in panel (e) and from the
summed IBO hole occupancies in panels (f) and (g), where, for the shown time range,
the summed $\pi$ ring IBO occupancy of the $p$-dynamics reaches a maximum value of $0.22$, %
whereas that of the $\sigma$ ring IBOs is only $0.14$. %

The change of local symmetry in the $p$ dynamics is
due to the 2-oxoethyl \ce{C-C} $\sigma$ IBO 6, which interacts through hyperconjugation with the phenyl $\pi$ IBO-12, as shown in \sfigref{fig:pheace-ibos}{b}.
IBO 6 is not active  in the $\pi$ dynamics, as its symmetry is different to that of the initial hole. A similar analysis holds for IBO 5.
The limited interactions between the initial $p$ hole (IBO-2) and the \ce{C-H} $\sigma$-bonding IBOs 7 and 8
leads to a  reduced migration into the ring through these IBOs 
and consequently almost no $\sigma$ holes in the ring for the $p$ dynamics.

The $\sigma$ dynamics lead to hole densities in the ring of both $\sigma$ and $\pi$ character. Because the initial IBO-4 is mirror-symmetric, like the $p$-type IBO-2,
the same orbitals active in the $p$ dynamics give rise to the $\pi$-ring hole occupancies; see \sfigref{fig:draft-phenylacet}{e} for the occupancies and \sfigref{fig:pheace-ibos}{c} for the mechanism.
Through another mechanism  that we will explain  in the next paragraph, in the $\sigma$ dynamics, the $\pi$ IBO-3 becomes active and hence also leads to active $\sigma$ ring IBOs.

\paragraph{2h1p interactions lead to a change of global hole symmetry\mydot}
In the preceding paragraph, we have analyzed how the hole shape and thereby its local symmetry changes when migrating from the carbonyl group into the phenyl ring.
Now, we shift our analysis to a change of the global symmetry, i.e., the occurrence of dominant hole orbitals with irreps different from that of the initial hole.
This can occur through an excitation mechanism 
that contain orbitals with different irreps.\cite{cm-breidbach-2003} 
Note that this does not affect the symmetry of the total wavefunction, which  is conserved.

This global symmetry effect is very pronounced in the $\sigma$ dynamics, where
within the first \qty{0.8}{fs} the occupancy of the initial hole in
the $\sigma$-type \ce{C=O} IBO-4 ($a'$) rapidly decreases while other IBOs become dominant, 
particularly the
positive occupancy of the $\pi$-type \ce{C=O} IBO-3 ($a''$) and the negative occupancy of the $\pi^\ast$ cIBO-1 ($a''$), c.f.~\sfigref{fig:draft-phenylacet}{d}.
These orbitals of different global symmetries are coupled through excited configurations as shown in \autoref{eq:hole_expansion}: %
the initial state corresponds to a one-hole (\mbox{1h}) configuration,
$\hat a_{\text{IBO-4}} \ket{\Psi_\text{GS}}$,
and it couples with \mbox{2h1p} configurations
of type $\hat a_{\text{cIBO-1}}^\dag \hat a_{\text{IBO-3}} \hat a_{\text{IBO-$X$}} \ket{\Psi_\text{GS}}$,
where $X$ is one of the orbitals of $a'$ symmetry whose hole occupancy is increasing, e.g., $X=5$, while that of cIBO-1 is decreasing.
The 2h1p configurations with electrons in the $\pi^\ast$  \ce{C=O} bond weaken this bond and hence facilitate the charge migration into other regions of the molecule. 
Through this mechanism, the $\pi$-bonding \ce{C=O} IBO-3 becomes active
during the $\sigma$ dynamics
and enables a similar mechanism and charge migration efficiency as that of the $\pi$ dynamics.

When starting from a 1h state, electron-correlation-induced couplings to 2h1p states and higher-order states (3h2p etc.) are easily recognizable by the appearance of negative occupancies in the antibonding cIBOs (see \autoref{sec:cm_analysis}).
For example,
common for all three dynamics is a negative $p$-type lobe
in the hole density of the phenyl ring at \qty{0.242}{fs} in \sfigref{fig:draft-phenylacet}{a}. This leads to very similar $\pi$ and $\pi^\ast$ ring IBO occupancies at the onset of the dynamics, c.f.,~\sfigref{fig:draft-phenylacet}{e} and \sifig{2}, where negative occupancies of $\pi^\ast$ IBOs are shown.
As another, related example,
 at later times, once the ring becomes populated, the  $\sigma$ dynamics  exhibit large negative occupancies in the antibonding $\pi$ orbitals of the phenyl ring; see \sfigref{fig:draft-phenylacet}{g}.
However, the occupancies of the $\pi$ and $\pi^\ast$ IBOs almost cancel each other and the maximum value of the combined $\pi$ and $\pi^\ast$  occupancy is $0.11$, which explains why only a few $\pi$-like features are visible in the hole density in \sfigref{fig:draft-phenylacet}{a}. 
The large negative occupancies %
in the $\sigma$ dynamics that are due to many 2h1p configurations
indicates that there are many more states in the $\sigma$ dynamics that are excited than in the other two dynamics. We will analyze this in more detail in the next paragraph.

\paragraph{The ``breakdown of the molecular orbital picture'' leads to damped oscillations\mydot}

Both the $\sigma$ and $p$ dynamics 
lead to  quasiexponentially decaying
oscillations of IBO occupancies; %
compare the hole occupancies in \sfigref{fig:draft-phenylacet}{d} of IBOs 1, 4, 5, and 6 for the $\sigma$ dynamics  
and those of IBOs 2, 5, and 6 for the $p$ dynamics, respectively.
This quasiexponential decay has previously been  analyzed by Breidbach and Cederbaum,\cite{cm-breidbach-2003, cm-breidbach-num-2007} and was identified as a signature of a highly correlated initial state produced by ionization of a core orbital or, as in our case, an inner valence orbital. 
A signature of such a scenario is contributions to the initial state from a quasicontinuum of states in the ionization spectrum. Consequently, this scenario is dubbed ``breakdown of the molecular orbital (MO) picture.''\cite{mo-breakdown-advchem-1986} 

\begin{figure}[t]
     \centering
     \includegraphics[width=\linewidth]{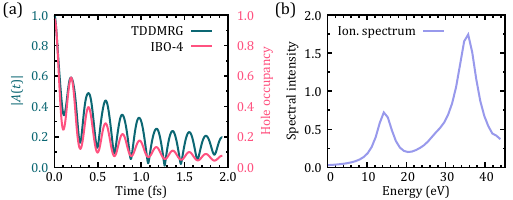}
         \caption{
       Autocorrelation function and ionization spectrum of the $\sigma$ dynamics of \pheace.
       (a) TDDMRG autocorrelation function (dark green) and IBO-4 hole occupancy from \sfigref{fig:draft-phenylacet}{d} (red).
       (b) %
       DMRG ionization spectrum of the $\sigma$ dynamics (light purple). %
     }
     \label{fig:pheace-autocor}
\end{figure}

In the following, we will analyze this behavior for the $\sigma$ dynamics, where it is most dominant. 
Indeed, the initial state in the $\sigma$ dynamics corresponds to inner valence ionization with an ionization potential  of \qty{34}{eV}, which is a typical energy for an orbital breakdown. %
The exponentially damped oscillation is not only visible in 
the hole occupancies, but also in the autocorrelation function in \sfigref{fig:pheace-autocor}{a}, dark green curve. %
Remarkably, the autocorrelation closely corresponds to the hole occupancy of IBO-4 (red) for all shown times.
We used this IBO to create the initial state.

The ionization spectrum provides further insights (see \autoref{sec:gf}), and it is shown  in \sfigref{fig:pheace-autocor}{b}. %
While these spectra are difficult to converge, 
with the used  spectral broadening of $\eta=0.08$, we clearly see two peaks at \qty{14.5}{eV} and \qty{35.5}{eV},
and a shoulder near \qty{27.4}{eV}. The shoulder is due to another peak, which we confirmed through unconverged computations with smaller broadening (not shown).
Since the broadening used does not allow for resolution at the eigenstate level, we assume that multiple eigenstates contribute to the peaks.
The spectrum in \sfigref{fig:pheace-autocor}{b} can be fully described by three Lorentzians, and, the Fourier transform qualitatively  matches the TDDMRG autocorrelation function; see \sisec{5.3}: The energy difference of the two main peaks leads to the oscillations whereas the Lorentzian shape leads to the exponential decay.
These observations  are in full agreement with the breakdown of the MO picture.\cite{cm-breidbach-num-2007} %

We gain additional insights by analyzing 
the response wavefunctions that contribute to the peaks (see \autoref{eq:gf_cv} and \sisec{5.3}). %
The hole densities for these wavefunctions are all localized at the 2-oxoethyl group (shown in \sifig{6}).
The wavefunction at \qty{14.5}{eV}
has large contributions from IBOs 1, 3, 4, and cIBO-1, indicating that this peak is due to both 1h and the 2h1p states discussed in the previous paragraph. 
In contrast, the wavefunction at \qty{35.5}{eV} has large contributions from IBOs 1 and 4 only, hence no 2h1p configurations contribute to this peak.
This discussion 
not only shows that IBOs can be used to analyze ionization spectra, but also that, remarkably, only the few IBOs shown in \autoref{fig:draft-phenylacet} are necessary to analyze a breakdown of the molecular orbital picture.

\subsubsection{\Furf{}\label{sec:furf}}

\begin{figure*}[th]
     \centering
     \includegraphics[width=\textwidth]{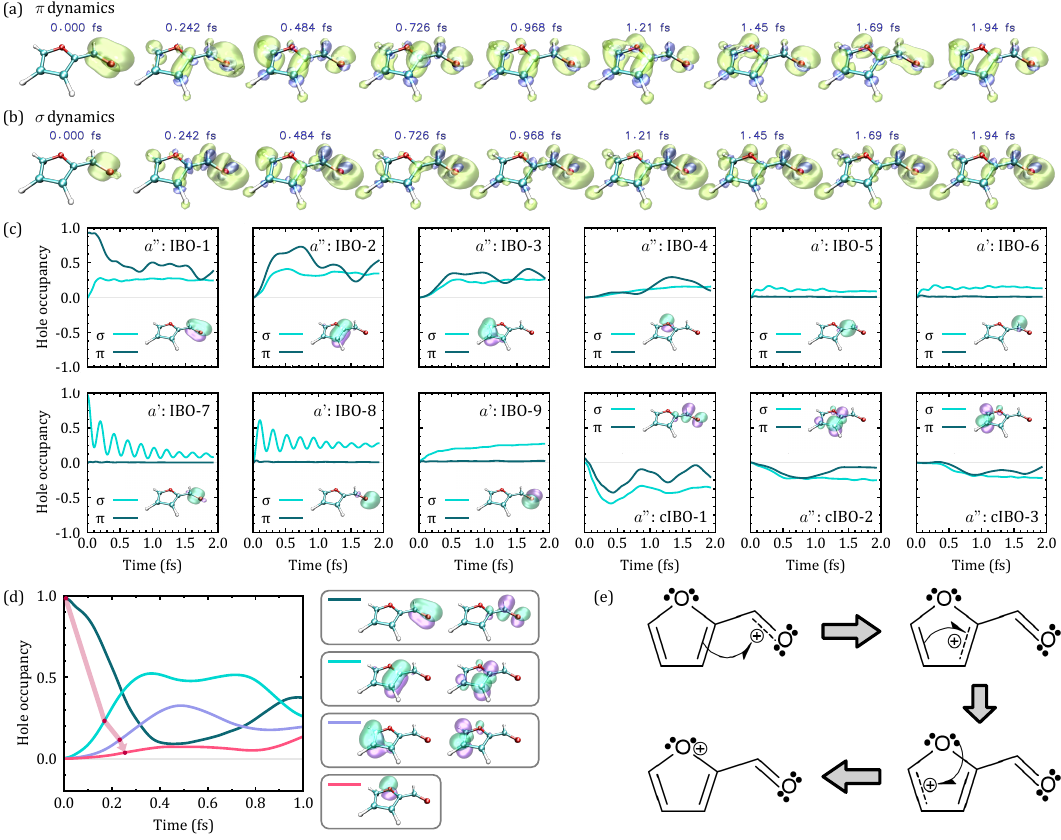}
     \caption{
  Charge migration analysis of \furf.
     (a, b) Hole density snapshots for the $\pi$ and $\sigma$ dynamics, respectively. The color code is the same as that in \sfigref{fig:draft-chloro}{a}.
     (c) Dominant IBO hole occupancies. %
     The IBO associated with each hole occupancy and its irrep label are shown in each panel.
     (d) Summed hole occupancies of the dominant IBOs and the corresponding cIBOs within the first \qty{1}{fs} during the $\pi$ dynamics. The red arrow suggests the movement of the hole.
     (e) Curly arrow mechanism demonstrating the initial hole migration; compare with the  red arrow in panel (d).
   }
     \label{fig:draft-furfural}
\end{figure*}

Following a setting similar to that in \pheace, here we compare charge migration dynamics in \furf, where the initial hole is either  a $\pi$ \ch{C=O} orbital (dubbed $\pi$ dynamics) or a $\sigma$ \ch{C=O} orbital (dubbed $\sigma$ dynamics).
Unlike \pheace, \furf is planar, and the carbonyl group is $\pi$-conjugated with the furyl group.
We have already studied \furf in \lit{wahyutama-tddmrg-2024} in terms of natural orbitals and dipole moments. Here, we use IBOs, which provide a much more straightforward analysis and more insights.

Snapshots of the hole density for the $\pi$ dynamics are shown in \sfigref{fig:draft-furfural}{a}.
As the molecule is fully $\pi$ conjugated, the initial  $\pi$ hole rapidly migrates into the ring, which starts 
within the first \qty{0.2}{fs}. %
This is more clearly visible through the IBO occupancies shown in 
\sfigref{fig:draft-furfural}{c}.
The $\pi$-conjugated IBOs 2-4 in the furyl ring quickly become populated, which is due to conjugation of these orbitals with the initial hole. %
Note that these orbitals have the same $a''$ symmetry and no $\sigma$-type orbitals with $a'$ symmetry become populated during the $\pi$ dynamics. 
A closer inspection of \sfigref{fig:draft-furfural}{c}
shows that %
the hole occupancy first increases rapidly
for the furyl \ce{C=C} $\pi$ IBO-2, followed by a similar but slightly delayed increase for IBO-3, which is the farthest away from the initial hole. %
At later times, the furyl oxygen $p$ orbital also gains hole occupancy. %
To complete this analysis,
next to the IBO occupancies we need to include the negative occupancies of the antibonding cIBOs localized at the same bond as the bonding IBOs, %
which correspond to active 2h1p configurations and weaken a chemical bond.
Adding the IBO and cIBO occupancies leads to an effective bond occupancy.
These occupancies are shown in \sfigref{fig:draft-furfural}{d} and reveal a sequential migration to the different orbitals.
We have seen a similar type of migration in the $\pi$ dynamics of \pheace. %

Like our analysis of \clace in \autoref{sec:clace}, the $\pi$ dynamics can be understood in terms of Lewis structures and curly arrows; see \sfigref{fig:draft-furfural}{e}.
While helpful,
such a curly-arrow mechanism of course only provides a rough qualitative understanding of the dynamics. 
It is difficult to extend curly-arrow mechanisms
to more complex migrations that display symmetry effects, strong entanglement and/or wavepacket bifurcations.
For such situations, methods that explicitly analyze the electron flux %
may be useful in determining the curly arrows.\cite{elflux-benzene-cm-2016,qcontrol-elflux-2017}

Snapshots of the hole density for the $\sigma$ dynamics are shown in \sfigref{fig:draft-furfural}{b}.
Surprisingly,
despite the vastly different initial states
of the $\sigma$ and $\pi$ dynamics, within \qty{0.484}{fs},
parts of the $\sigma$ dynamics resemble the $\pi$ dynamics.
This is more clearly seen in the IBO analysis in panel (c); 
compare the occupancies of the $\pi$ IBOs 2-4, and that of the cIBOs 1-3. All of these have $a''$ symmetry whereas the initial $\sigma$ hole has $a'$ symmetry.
Hence, 
while the hole shape in the $\pi$ dynamics remains unchanged, the hole in the $\sigma$ dynamics changes to that of a $\pi$ type as the hole enters the furyl group.
Note the  absence of active $\sigma$ orbitals within the ring in \sfigref{fig:draft-furfural}{c}.
As the migration into the $\pi$ furyl orbitals is the main mechanism, the $\sigma$ dynamics is less efficient than the $\pi$ dynamics
and the total sum of the furyl IBO occupancies  reaches a maximum value  of only $0.60$ for the $\sigma$ dynamics, compared to $0.89$ for the $\pi$ dynamics (c.f.~\sisec{6}). %

As in \pheace in  \autoref{sec:pheace}, the furyl $\pi$ holes that appear in the $\sigma$ dynamics are accompanied by strongly negative occupancies of antibonding cIBOs and a quasiexponential decay of key IBOs, e.g.~IBO-7 in \sfigref{fig:draft-furfural}{c}.
Thus, 
also the \furf $\sigma$ dynamics exhibits a breakdown of the MO picture and 2h1p configurations 
transfer the dominant hole character to orbitals of different symmetry. 
However,  in contrast to \pheace,
the migration into the ring is mediated through $\pi$ conjugation rather than hyperconjugation.

\subsection{Conformer effects \label{sec:conformer}}

\begin{figure*}[!ht]
     \centering
     \includegraphics[width=\textwidth]{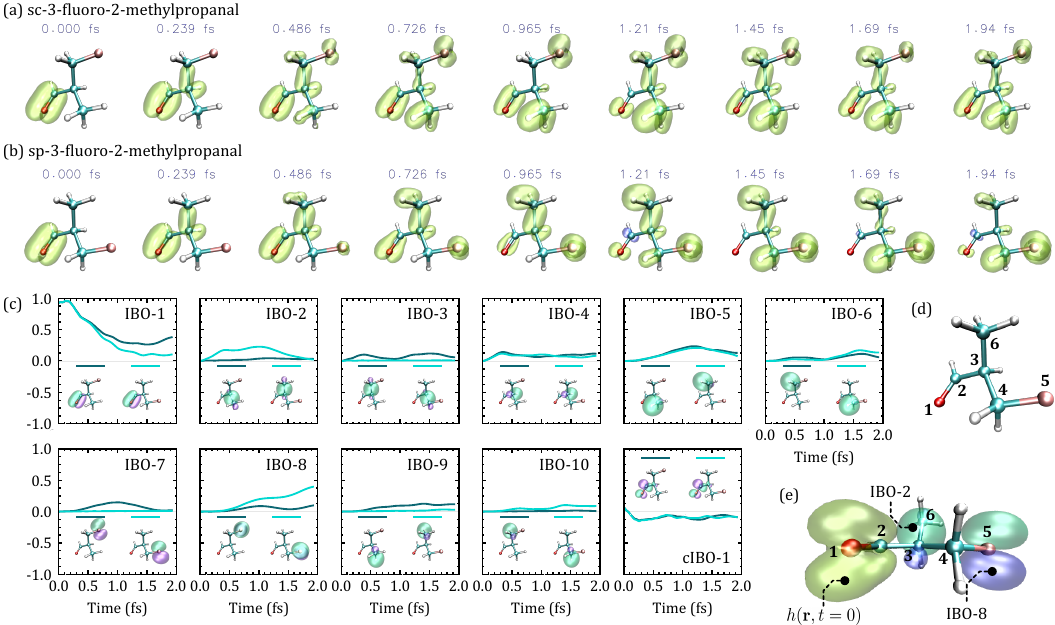}
     \caption{
  Charge migration analysis of \isobutfa.
     (a, b)  Hole density snapshots of \isobutfa{} (\isobutfaB) and \isobutfb{} (\isobutfbB), respectively. Note that only the densities projected into the dominant IBO space  are shown.
     (c) Dominant IBO hole occupancies. %
     The IBOs of the two conformers associated with each hole occupancy are shown in each panel.
     (d) Structure of \isobutfbB{} highlighting the six atoms whose IBOs are in a nearly planar arrangement and therefore facilitate charge migration.
     (e) Same as (d) but viewed at an angle that reveals the relative alignment of the initial hole with IBOs 2 and 8.
     }
     \label{fig:draft-isobutyrald}
\end{figure*}

In \autoref{sec:init-hole-effect}, we have shown that both global and  local symmetries can play a pivotal role in determining how the hole shape changes as it migrates through the molecule. So far, all molecules studied have global symmetry. What if the molecule has local symmetries but no global symmetry? 
To address this for a particular scenario, we study two $C_1$-symmetric conformers 
of \butfa, 
namely the synclinal one (\isobutfa{}, dubbed \isobutfaB),
and the  synperiplanar one (\isobutfb{}, dubbed \isobutfbB);
see \sfigref{fig:draft-isobutyrald}{a, b} for their geometries.
\isobutfbB is the global minimum, and \isobutfaB is about \qty{3.1}{kJ/mol} higher in energy.
The conformers primarily differ in the placement of fluorine, whose lone pairs and high electronegativity can significantly affect charge migration.  We will show below that the placement of fluorine %
can be crucial for efficient charge migration and thus suggest design principles for optimal charge migration.

For both conformers, our simulations start with a hole in the $\pi$ bond of \ch{C=O}.
In \sfigref{fig:draft-isobutyrald}{a, b} 
we show snapshots of the hole density
 for \isobutfaB{} and \isobutfbB{}, respectively.
Unlike in our previous studies, 
we here only show the hole densities projected into the dominant IBO space, which facilitate the analysis.\footnote[8]{The unprojected hole densities are semi-quantitatively identical to the projected ones and are shown in \sisec{7}.} %
As both conformers have the same initial hole, the conformer effect is not large during the initial dynamics but becomes larger at later times.
In particular,
the hole density snapshots
indicate that dominant changes in the charge migration
start to appear at $\sim$\qty{0.7}{fs}.
After $\qty{\sim1}{fs}$, the initial hole in 
\isobutfbB{} becomes almost completely drained out. 
Hence, 
the placement of fluorine in \isobutfbB{} leads to more efficient charge migration than in \isobutfaB{}.
The IBO hole occupancies provide more details and are shown in \sfigref{fig:draft-isobutyrald}{c}.
The charge migration efficiency is clearly seen  from
IBO-1,   whose occupancy starts close to $1$ and for \isobutfbB{} reaches a minimum of $0.090$, whereas for \isobutfaB{} it only reaches $0.26$ for the time range studied. 

The enhanced charge migration in \isobutfbB{} arises from the molecular geometry and orbital alignment highlighted in \sfigref{fig:draft-isobutyrald}{d, e},
in particular a nearly planar arrangement of the shown atoms. 
This leads to a favorable mutual orientation of IBOs 1, 2, and 8, 
which then enables sequential hyperconjugative coupling, first from IBO-1 to IBO-2 and then from IBO-2 to IBO-8.
As shown in panels (b) and (c), the hole therefore migrates from the \ce{C=O} $\pi$ IBO-1 to the
 \ch{C$_3$-C$_6$} $\sigma$ IBO-2 and finally reaches the \ch{F} atom's $p$ lone-pair IBO-8.

The local symmetry of the initial hole further determines which orbitals participate in the dynamics.
For example, the \ch{C2-C3} $\sigma$ IBO (not shown) remains inactive because its local symmetry is incompatible with that of
IBO-1. The same restriction applies to IBOs 3 and 7 (see their SP occupancies in \sfigref{fig:draft-isobutyrald}{c}).

This mechanism is less effective in \isobutfaB{} because its geometry does not yield a similarly favorable alignment involving
the occupied fluorine orbitals. IBO-1 nevertheless couples weakly to \isobutfaB{} IBO-7, which explains the IBO-7 occupancy in \sfigref{fig:draft-isobutyrald}{c};
however, their incomplete alignment limits the strength of this coupling.

While the migration toward the \ce{F} atom differs markedly between the two conformers, we  observe some similarities in other migration paths:
the occupancy curves of cIBO-1, IBOs 4-6,  and that of SC/SP IBOs  9/10 are very similar for both conformers. These IBOs are associated to migrations to other parts of the molecule, e.g.~the hydrogen atoms. The similarity between the two conformers is due to  the same type of initial dynamics, which is not affected by the placement of the $F$ atom at the opposite end of the molecule.

Our results on \isobutfaB{} and \isobutfbB{} have demonstrated an interplay between initial hole symmetry, location of a functional group with lone pairs, and geometry in enhancing charge migration, and thus will be useful in the search for molecules that exhibit a clear charge migration signal. 
As the dynamics involve both $\sigma$, $\pi$, and $p$ orbitals,
they also present an example of the underlying similarity between conjugation and hyperconjugation\cite{hyperconjugation-2019} in the context of electron dynamics.

\section{Conclusions\label{sec:conclusions}}%

In this work, we introduced a new way of analyzing charge migration dynamics that is based on mapping the hole density and related quantities to an extension of intrinsic bond orbitals (IBOs). 
This extension includes antibonding orbitals and leads to a very compact representation of the essentials of the real-time many-body quantum dynamics, even in challenging cases such as correlated inner-valence ionized wavefunctions. The real-valued and time-independent nature of IBOs makes them particularly convenient to use for time-dependent simulations.
Our simulations were based on an efficient setup of the time-dependent density matrix renormalization group (TDDMRG), %
which allowed us to simulate charge migration in molecules as large as \pheace with 45 fully correlated electrons in 50 orbitals.

IBOs bridge quantum mechanics with chemical concepts, and, in our case, enabled us to not only connect IBOs to established quantities in charge migration  such as hole configurations, but also to understand complex charge migration in terms of through-bond/-space interactions such as hyperconjugation and Lewis structures.
We demonstrated this capability for various molecules, with a focus on how ionization from orbitals of various shapes affect the charge migration.
For example, we showed that in \pheace a sudden ionization at the carbonyl group leads to a migration into the $\sigma$ bonds of the phenyl ring regardless of whether the initial hole has a $\pi$ or a $\sigma$ shape.
We found that through-bond  and through-space hyperconjugation interactions between different IBOs 
of similar local symmetry
are responsible for this surprising effect. We also observed a rapid change of character from a $\sigma$ hole to a $\pi$ hole at the carbonyl bond, which we explained by connecting antibonding extensions of IBOs to excited two-hole one-particle (2h1p) configurations that  couple orbitals with $\pi$ and $\sigma$ symmetries.
Surprisingly, a small set of IBOs can provide a semiquantitative description of the correlated dynamics
even in the so-called ``breakdown of the molecular orbital picture'' regime.

In contrast to \pheace, in \furf we found that charge  at the carbonyl group migrates into the $\pi$ bonds of the formyl group, regardless of whether the ionization created a $\pi$ or $\sigma$ hole at the carbonyl group.
The $\pi$-hole-initiated migration is driven by a full conjugation of the $\pi$ orbitals throughout the molecule, while the $\sigma$-hole-initiated migration is mediated by excited 2h1p configurations. %
Taking advantage of the straightforward mapping of IBOs to Lewis structures, we also demonstrated how an IBO analysis leads to a qualitative mechanism in terms of curly arrows
for charge migrations without strong wavefunction bifurcations and interference effects.

Lastly, we used our extension of IBOs to explain
different charge migration efficiencies of two conformers of \butfa. The localized character of the IBOs dramatically sped up this analysis.
We found that, for one conformer, the charge migration is facilitated by through-space interactions of three key IBOs that are favorably aligned. This leads to an almost complete drain of the initial hole, in contrast to the other conformer, where the IBOs are aligned less favorably.
This illustrates a potential way to find molecules and functional groups with efficient, long-lasting, coherent charge migration, which is an ongoing quest and crucial for clear experiments.\cite{mol-long-cm-2023, cm-localion-2025, cm-increase-size-flexlty-2025}
As such, we believe that the present work %
might help in realizing 
future experiments, and by extension, ultrafast control of chemical reactions.

While we focused on charge migration,
the proposed IBO  analysis can be extended to understand not only other types of nonequilibrium real-time molecular dynamics but also time-independent quantities such as response functions in ionization spectra, which we briefly demonstrated. Importantly, IBOs are not restricted to DMRG wavefunctions %
but can be combined with other quantum chemistry methods, including density functional theory.

\section*{Associated Content}
  Supporting Information available. Details on the used geometries, 
  fitting procedure of the $n$-state model, and conformers;
  natural charge orbital analysis of the $\sigma$, $\pi$, and $p$ dynamics, additional hole density snapshots of the $p$ dynamics, %
  and Green's function analysis of the $\sigma$ dynamics in \pheace;
  total hole occupancies in the phenyl and furyl rings;
  snapshots of the unprojected hole densities of the \butfa conformers; %
  convergence studies for all dynamics. %

\if\USEACHEMSO1
\begin{acknowledgement}
\else
\acknowledgements
\fi

This work was supported by the U.S.~Department of Energy, Office of Science, Office of Basic Energy Sciences under Award Number DE-SC0026361 (Green's function computations, simulation analysis, Lewis structures),
by the donors of American Chemical Society Petroleum Research Fund via
grant no.~67511-DNI6 (IBO procedure, TDDMRG simulations), 
and 
through computational time on the Pinnacles/CENVAL-ARC cluster at the Cyberinfrastructure and Research Technologies (CIRT) at University of California, Merced,
supported by NSF Grants No.~2346744 and 2019144.

\if\USEACHEMSO1
\end{acknowledgement}
\fi

\appendix
\section{computational details}
\label{app:comp_details}
Our TDDMRG implementation is based on the \textsc{Block2}\cite{block2-2023, block2-gh-2024} and \textsc{PySCF}\cite{pyscf-wires-2018, pyscf-jcp-2020, pyscf-gh-2024, preprint-pyscf-2026} codes. %
We follow our recently established guidelines for efficient TDDMRG charge migration simulations.\cite{wahyutama-tddmrg-2024} 
To enforce the spin SU(2) symmetry of open-shell wavefunctions resulting from ionization, we use a spin-adapted MPS. 
This leads to a slightly different basis than that given in  \autoref{eq:mps_form}, which only supports particle-number U(1) symmetry.
Based on the findings in our previous work,\cite{wahyutama-tddmrg-2024} 
we use singlet-embedding for enforcing the nonzero total spin within the spin-adapted MPS, that is,
the nonzero spin state is embedded in a state  that has a total spin of zero.\cite{tatsuaki-singlet-embed-2000, dmrg-spin-symm-2012} 
To propagate the complex-valued wavefunction, we use a fully complex MPS representation, i.e., all matrices 
$\mathbf M^{\sigma_i}(t)$  in \autoref{eq:mps_form} are complex-valued.
The individual MPS matrices are 
propagated using the short iterative Lanczos algorithm\cite{sil-light-1986}
with a relative convergence tolerance of $5\times10^{-6}$.  
The main TDDMRG propagation parameters are outlined in \autoref{tab:tddmrg_params}.
See \sisec{8} for convergence studies.

\begin{table}
  \centering
  \small
\caption{TDDMRG propagation parameters. $D_{\text{prelim}}$ and $D$ are the bond dimensions for the preliminary and main TDDMRG propagations, respectively. The former is used for selecting the CAS orbitals\nmydot}%
\label{tab:tddmrg_params}
\begin{tabular*}{0.48\textwidth}{lllll}
\toprule
System / dynamics & $D_{\text{prelim}}$ & $D$ & Active space & $\Delta t/\unit{as}$ \\
\midrule
\clace            & {N/A} & $1000$ & CAS($15e,41o$) & $0.968$ \\
\pheace           & {100} & $500$  & CAS($45e,50o$) & $0.484$ \\
\furf\ ($\pi$)    & {200} & $500$  & CAS($35e,40o$) & $0.968$ \\
\furf\ ($\sigma$) & {200} & $500$  & CAS($35e,40o$) & $0.484$ \\
\butfa            & {150} & $500$  & CAS($35e,40o$) & $0.726$ \\
\bottomrule
\end{tabular*}
\end{table}

We use 6-31G basis set\cite{6-31g_1-1971, 6-31g_2-1972} for all molecules except for \clace{}, %
for which we use the  def2-SV(P) basis set.\cite{def2-sv_p_-2005}
All simulations use the frozen core approximation.
The geometries were optimized using density functional theory (see \sisec{1}).
For \pheace, 
the particular conformer is not the global minimum (other conformers that are \qty{2.7}{kJ/mol} lower in energy exist), %
but it is accessible at room temperature.

To select the active orbitals for the CAS, 
we use our procedures developed in \lit{wahyutama-tddmrg-2024}.
In short,
we select the active orbitals for the dynamics from the dominant orbitals of both the reduced density matrix (RDM) of the neutral molecule's ground state and, in addition, from the time-averaged hole density matrix of a preliminary TDDMRG simulation using all orbitals and a small bond dimension $D_{\text{prelim}}$ shown in \autoref{tab:tddmrg_params}. 
To improve the DMRG efficiency, the selected orbitals are then split-localized using Pipek-Mezey localization.\cite{pm_loc-1989}
We employ these hole-DM-selected orbitals for all molecules except for \clace{}, for which we use split-localized Hartree-Fock orbitals and the full orbital space.
Note that these orbitals used for the TDDMRG simulations are not IBOs. The IBO analysis is performed through a change of basis.

For all molecules except \clace{}, we use the annihilation operator $\hat a_k$ of a particular IBO $\ket{k}$ to create a localized initial hole. For \clace{}, we create the initial hole from an equal-weight superposition of Hartree-Fock HOMO and HOMO-1, which leads to a $p$ lone-pair orbital on \ch{Cl}.
We compute restricted-Hartee-Fock-based IBOs  using the implementation in \textsc{PySCF}, %
which differs slightly from the original IBO procedure outlined in \lit{ibo-2013}.
We compute the IBOs by localizing the MOs without directly enforcing symmetry, save for \pheace, where we  localize the MOs within each irrep of the $C_s$ point group.
This makes the IBOs less localized but facilitates the charge migration analysis.

For the hole density and natural orbital density plots, we use isosurface values of  $\pm 0.006$, while for the IBOs, this value is $\pm 0.06$ for all molecules except for \furf, which is $\pm 0.08$. 
The ionization potentials of the states in the 
$\sigma$, $\pi$, and $p$ dynamics of  \pheace in \autoref{sec:pheace}
are \qty{34.0}{eV}, \qty{15.6}{eV}, and \qty{14.9}{eV}, respectively.
For \furf, the ionization potentials are 
\qty{15.6}{eV}  and  \qty{33.6}{eV} for
the $\pi$ dynamics and $\sigma$ dynamics, respectively.
We note that
the energy of the $\sigma$ dynamics for both molecules is higher than the second ionization potential,
which means that some influence from Auger-Meitner decay to this dynamics should be expected at times longer than our simulation time.%
\cite{core-ionize-cm-2016}

The Green's function computations for \pheace  in \autoref{sec:pheace}
are based on a bond dimension of $D=550$,
use a final relative convergence tolerance of $10^{-8}$, a maximal sweep number of $8$, and a linear solver tolerance of $10^{-6}$. To steer the optimization problem out of local minima, we use noise on the order of $10^{-5}$ for the first 4 sweeps.

\providecommand{\noopsort}[1]{}\providecommand{\singleletter}[1]{#1}%

\if\USEACHEMSO1
\clearpage
\fi

\clearpage


\begin{thebibliography}{156}%
\makeatletter
\providecommand \@ifxundefined [1]{%
 \@ifx{#1\undefined}
}%
\providecommand \@ifnum [1]{%
 \ifnum #1\expandafter \@firstoftwo
 \else \expandafter \@secondoftwo
 \fi
}%
\providecommand \@ifx [1]{%
 \ifx #1\expandafter \@firstoftwo
 \else \expandafter \@secondoftwo
 \fi
}%
\providecommand \natexlab [1]{#1}%
\providecommand \enquote  [1]{``#1''}%
\providecommand \bibnamefont  [1]{#1}%
\providecommand \bibfnamefont [1]{#1}%
\providecommand \citenamefont [1]{#1}%
\providecommand \href@noop [0]{\@secondoftwo}%
\providecommand \href [0]{\begingroup \@sanitize@url \@href}%
\providecommand \@href[1]{\@@startlink{#1}\@@href}%
\providecommand \@@href[1]{\endgroup#1\@@endlink}%
\providecommand \@sanitize@url [0]{\catcode `\\12\catcode `\$12\catcode
  `\&12\catcode `\#12\catcode `\^12\catcode `\_12\catcode `\%12\relax}%
\providecommand \@@startlink[1]{}%
\providecommand \@@endlink[0]{}%
\providecommand \url  [0]{\begingroup\@sanitize@url \@url }%
\providecommand \@url [1]{\endgroup\@href {#1}{\urlprefix }}%
\providecommand \urlprefix  [0]{URL }%
\providecommand \Eprint [0]{\href }%
\providecommand \doibase [0]{https://doi.org/}%
\providecommand \selectlanguage [0]{\@gobble}%
\providecommand \bibinfo  [0]{\@secondoftwo}%
\providecommand \bibfield  [0]{\@secondoftwo}%
\providecommand \translation [1]{[#1]}%
\providecommand \BibitemOpen [0]{}%
\providecommand \bibitemStop [0]{}%
\providecommand \bibitemNoStop [0]{.\EOS\space}%
\providecommand \EOS [0]{\spacefactor3000\relax}%
\providecommand \BibitemShut  [1]{\csname bibitem#1\endcsname}%
\let\auto@bib@innerbib\@empty
%
\bibitem [{\citenamefont {Nisoli}\ \emph {et~al.}(2017)\citenamefont {Nisoli},
  \citenamefont {Decleva}, \citenamefont {Calegari}, \citenamefont {Palacios},\
  and\ \citenamefont {Martín}}]{atto-el-dyn-mol-2017}%
  \BibitemOpen
  \bibfield  {author} {\bibinfo {author} {\bibfnamefont {M.}~\bibnamefont
  {Nisoli}}, \bibinfo {author} {\bibfnamefont {P.}~\bibnamefont {Decleva}},
  \bibinfo {author} {\bibfnamefont {F.}~\bibnamefont {Calegari}}, \bibinfo
  {author} {\bibfnamefont {A.}~\bibnamefont {Palacios}},\ and\ \bibinfo
  {author} {\bibfnamefont {F.}~\bibnamefont {Martín}},\ }\bibfield  {title}
  {\enquote {\bibinfo {title} {Attosecond electron dynamics in molecules},}\
  }\href {https://doi.org/10.1021/acs.chemrev.6b00453} {\bibfield  {journal}
  {\bibinfo  {journal} {Chem. Rev.}\ }\textbf {\bibinfo {volume} {117}},\
  \bibinfo {pages} {10760--10825} (\bibinfo {year} {2017})}\BibitemShut
  {NoStop}%
\bibitem [{\citenamefont {Vrakking}\ and\ \citenamefont
  {Lepine}(2018)}]{atto-mol-dyn-2018}%
  \BibitemOpen
  \bibfield  {author} {\bibinfo {author} {\bibfnamefont {M.~J.~J.}\
  \bibnamefont {Vrakking}}\ and\ \bibinfo {author} {\bibfnamefont
  {F.}~\bibnamefont {Lepine}},\ }\href {https://doi.org/10.1039/9781788012669}
  {\emph {\bibinfo {title} {Attosecond Molecular Dynamics}}}\ (\bibinfo
  {publisher} {The Royal Society of Chemistry},\ \bibinfo {year}
  {2018})\BibitemShut {NoStop}%
\bibitem [{\citenamefont {Palacios}\ and\ \citenamefont
  {Martín}(2020)}]{quantum-chem-atto-2020}%
  \BibitemOpen
  \bibfield  {author} {\bibinfo {author} {\bibfnamefont {A.}~\bibnamefont
  {Palacios}}\ and\ \bibinfo {author} {\bibfnamefont {F.}~\bibnamefont
  {Martín}},\ }\bibfield  {title} {\enquote {\bibinfo {title} {The quantum
  chemistry of attosecond molecular science},}\ }\href
  {https://doi.org/https://doi.org/10.1002/wcms.1430} {\bibfield  {journal}
  {\bibinfo  {journal} {Wiley Interdiscip. Rev.: Comput. Mol. Sci.}\ }\textbf
  {\bibinfo {volume} {10}},\ \bibinfo {pages} {e1430} (\bibinfo {year}
  {2020})}\BibitemShut {NoStop}%
\bibitem [{\citenamefont {Merritt}, \citenamefont {Jacquemin},\ and\
  \citenamefont {Vacher}(2021)}]{attochemistry-future-2021}%
  \BibitemOpen
  \bibfield  {author} {\bibinfo {author} {\bibfnamefont {I.~C.~D.}\
  \bibnamefont {Merritt}}, \bibinfo {author} {\bibfnamefont {D.}~\bibnamefont
  {Jacquemin}},\ and\ \bibinfo {author} {\bibfnamefont {M.}~\bibnamefont
  {Vacher}},\ }\bibfield  {title} {\enquote {\bibinfo {title} {Attochemistry:
  Is controlling electrons the future of photochemistry?}}\ }\href
  {https://doi.org/10.1021/acs.jpclett.1c02016} {\bibfield  {journal} {\bibinfo
   {journal} {J. Phys. Chem. Lett.}\ }\textbf {\bibinfo {volume} {12}},\
  \bibinfo {pages} {8404--8415} (\bibinfo {year} {2021})}\BibitemShut {NoStop}%
\bibitem [{\citenamefont {Månsson}\ \emph {et~al.}(2017)\citenamefont
  {Månsson}, \citenamefont {De~Camillis}, \citenamefont {Castrovilli},
  \citenamefont {Galli}, \citenamefont {Nisoli}, \citenamefont {Calegari},\
  and\ \citenamefont {Greenwood}}]{dna-ionize-2017}%
  \BibitemOpen
  \bibfield  {author} {\bibinfo {author} {\bibfnamefont {E.~P.}\ \bibnamefont
  {Månsson}}, \bibinfo {author} {\bibfnamefont {S.}~\bibnamefont
  {De~Camillis}}, \bibinfo {author} {\bibfnamefont {M.~C.}\ \bibnamefont
  {Castrovilli}}, \bibinfo {author} {\bibfnamefont {M.}~\bibnamefont {Galli}},
  \bibinfo {author} {\bibfnamefont {M.}~\bibnamefont {Nisoli}}, \bibinfo
  {author} {\bibfnamefont {F.}~\bibnamefont {Calegari}},\ and\ \bibinfo
  {author} {\bibfnamefont {J.~B.}\ \bibnamefont {Greenwood}},\ }\bibfield
  {title} {\enquote {\bibinfo {title} {{Ultrafast dynamics in the DNA building
  blocks thymidine and thymine initiated by ionizing radiation}},}\ }\href
  {https://doi.org/10.1039/C7CP02803B} {\bibfield  {journal} {\bibinfo
  {journal} {Phys. Chem. Chem. Phys.}\ }\textbf {\bibinfo {volume} {19}},\
  \bibinfo {pages} {19815--19821} (\bibinfo {year} {2017})}\BibitemShut
  {NoStop}%
\bibitem [{\citenamefont {Swiderek}(2006)}]{dna-radiation-dmg-2006}%
  \BibitemOpen
  \bibfield  {author} {\bibinfo {author} {\bibfnamefont {P.}~\bibnamefont
  {Swiderek}},\ }\bibfield  {title} {\enquote {\bibinfo {title} {Fundamental
  {{Processes}} in {{Radiation Damage}} of {{DNA}}},}\ }\href
  {https://doi.org/10.1002/anie.200600614} {\bibfield  {journal} {\bibinfo
  {journal} {Angew. Chem. Int. Ed.}\ }\textbf {\bibinfo {volume} {45}},\
  \bibinfo {pages} {4056--4059} (\bibinfo {year} {2006})}\BibitemShut {NoStop}%
\bibitem [{\citenamefont {Weinkauf}\ \emph {et~al.}(1996)\citenamefont
  {Weinkauf}, \citenamefont {Schanen}, \citenamefont {Metsala}, \citenamefont
  {Schlag}, \citenamefont {Bürgle},\ and\ \citenamefont
  {Kessler}}]{charge-transfer-weinkauf-1996}%
  \BibitemOpen
  \bibfield  {author} {\bibinfo {author} {\bibfnamefont {R.}~\bibnamefont
  {Weinkauf}}, \bibinfo {author} {\bibfnamefont {P.}~\bibnamefont {Schanen}},
  \bibinfo {author} {\bibfnamefont {A.}~\bibnamefont {Metsala}}, \bibinfo
  {author} {\bibfnamefont {E.~W.}\ \bibnamefont {Schlag}}, \bibinfo {author}
  {\bibfnamefont {M.}~\bibnamefont {Bürgle}},\ and\ \bibinfo {author}
  {\bibfnamefont {H.}~\bibnamefont {Kessler}},\ }\bibfield  {title} {\enquote
  {\bibinfo {title} {Highly efficient charge transfer in peptide cations in the
  gas phase: Threshold effects and mechanism},}\ }\href
  {https://doi.org/10.1021/jp960926m} {\bibfield  {journal} {\bibinfo
  {journal} {J. Phys. Chem.}\ }\textbf {\bibinfo {volume} {100}},\ \bibinfo
  {pages} {18567--18585} (\bibinfo {year} {1996})}\BibitemShut {NoStop}%
\bibitem [{\citenamefont {Weinkauf}\ \emph {et~al.}(1997)\citenamefont
  {Weinkauf}, \citenamefont {Schlag}, \citenamefont {Martinez},\ and\
  \citenamefont {Levine}}]{levine-select-react-1997}%
  \BibitemOpen
  \bibfield  {author} {\bibinfo {author} {\bibfnamefont {R.}~\bibnamefont
  {Weinkauf}}, \bibinfo {author} {\bibfnamefont {E.~W.}\ \bibnamefont
  {Schlag}}, \bibinfo {author} {\bibfnamefont {T.~J.}\ \bibnamefont
  {Martinez}},\ and\ \bibinfo {author} {\bibfnamefont {R.~D.}\ \bibnamefont
  {Levine}},\ }\bibfield  {title} {\enquote {\bibinfo {title} {Nonstationary
  electronic states and site-selective reactivity},}\ }\href
  {https://doi.org/10.1021/jp9715742} {\bibfield  {journal} {\bibinfo
  {journal} {J. Phys. Chem. A}\ }\textbf {\bibinfo {volume} {101}},\ \bibinfo
  {pages} {7702--7710} (\bibinfo {year} {1997})}\BibitemShut {NoStop}%
\bibitem [{\citenamefont {Remacle}, \citenamefont {Levine},\ and\ \citenamefont
  {Ratner}(1998)}]{charge-direct-react-1998}%
  \BibitemOpen
  \bibfield  {author} {\bibinfo {author} {\bibfnamefont {F.}~\bibnamefont
  {Remacle}}, \bibinfo {author} {\bibfnamefont {R.}~\bibnamefont {Levine}},\
  and\ \bibinfo {author} {\bibfnamefont {M.}~\bibnamefont {Ratner}},\
  }\bibfield  {title} {\enquote {\bibinfo {title} {Charge directed reactivity:
  a simple electronic model, exhibiting site selectivity, for the dissociation
  of ions},}\ }\href
  {https://doi.org/https://doi.org/10.1016/S0009-2614(97)01314-6} {\bibfield
  {journal} {\bibinfo  {journal} {Chem. Phys. Lett.}\ }\textbf {\bibinfo
  {volume} {285}},\ \bibinfo {pages} {25--33} (\bibinfo {year}
  {1998})}\BibitemShut {NoStop}%
\bibitem [{\citenamefont {Remacle}\ \emph {et~al.}(1999)\citenamefont
  {Remacle}, \citenamefont {Levine}, \citenamefont {Schlag},\ and\
  \citenamefont {Weinkauf}}]{control-select-reaction-1999}%
  \BibitemOpen
  \bibfield  {author} {\bibinfo {author} {\bibfnamefont {F.}~\bibnamefont
  {Remacle}}, \bibinfo {author} {\bibfnamefont {R.~D.}\ \bibnamefont {Levine}},
  \bibinfo {author} {\bibfnamefont {E.~W.}\ \bibnamefont {Schlag}},\ and\
  \bibinfo {author} {\bibfnamefont {R.}~\bibnamefont {Weinkauf}},\ }\bibfield
  {title} {\enquote {\bibinfo {title} {Electronic control of site selective
  reactivity: A model combining charge migration and dissociation},}\ }\href
  {https://doi.org/10.1021/jp991853k} {\bibfield  {journal} {\bibinfo
  {journal} {J. Phys. Chem. A}\ }\textbf {\bibinfo {volume} {103}},\ \bibinfo
  {pages} {10149--10158} (\bibinfo {year} {1999})}\BibitemShut {NoStop}%
\bibitem [{\citenamefont {Cui}, \citenamefont {Thompson},\ and\ \citenamefont
  {Reilly}(2005)}]{peptide-fragment-pathways-2005}%
  \BibitemOpen
  \bibfield  {author} {\bibinfo {author} {\bibfnamefont {W.}~\bibnamefont
  {Cui}}, \bibinfo {author} {\bibfnamefont {M.~S.}\ \bibnamefont {Thompson}},\
  and\ \bibinfo {author} {\bibfnamefont {J.~P.}\ \bibnamefont {Reilly}},\
  }\bibfield  {title} {\enquote {\bibinfo {title} {Pathways of peptide ion
  fragmentation induced by vacuum ultraviolet light},}\ }\href
  {https://doi.org/https://doi.org/10.1016/j.jasms.2005.03.050} {\bibfield
  {journal} {\bibinfo  {journal} {J. Am. Soc. Mass Spectrom.}\ }\textbf
  {\bibinfo {volume} {16}},\ \bibinfo {pages} {1384--1398} (\bibinfo {year}
  {2005})}\BibitemShut {NoStop}%
\bibitem [{\citenamefont {Remacle}\ and\ \citenamefont
  {Levine}(2007)}]{cm-peptide-2007}%
  \BibitemOpen
  \bibfield  {author} {\bibinfo {author} {\bibfnamefont {F.}~\bibnamefont
  {Remacle}}\ and\ \bibinfo {author} {\bibfnamefont {R.~D.}\ \bibnamefont
  {Levine}},\ }\bibfield  {title} {\enquote {\bibinfo {title} {Probing
  ultrafast purely electronic charge migration in small peptides},}\ }\href
  {https://doi.org/doi:10.1524/zpch.2007.221.5.647} {\bibfield  {journal}
  {\bibinfo  {journal} {Z. Phys. Chem.}\ }\textbf {\bibinfo {volume} {221}},\
  \bibinfo {pages} {647--661} (\bibinfo {year} {2007})}\BibitemShut {NoStop}%
\bibitem [{\citenamefont {Reilly}(2009)}]{uv-photogragment-2009}%
  \BibitemOpen
  \bibfield  {author} {\bibinfo {author} {\bibfnamefont {J.~P.}\ \bibnamefont
  {Reilly}},\ }\bibfield  {title} {\enquote {\bibinfo {title} {Ultraviolet
  photofragmentation of biomolecular ions},}\ }\href
  {https://doi.org/https://doi.org/10.1002/mas.20214} {\bibfield  {journal}
  {\bibinfo  {journal} {Mass Spectrom. Rev.}\ }\textbf {\bibinfo {volume}
  {28}},\ \bibinfo {pages} {425--447} (\bibinfo {year} {2009})}\BibitemShut
  {NoStop}%
\bibitem [{\citenamefont {Remacle}\ and\ \citenamefont
  {Levine}(1999)}]{cm-select-reaction-covion-1999}%
  \BibitemOpen
  \bibfield  {author} {\bibinfo {author} {\bibfnamefont {F.}~\bibnamefont
  {Remacle}}\ and\ \bibinfo {author} {\bibfnamefont {R.~D.}\ \bibnamefont
  {Levine}},\ }\bibfield  {title} {\enquote {\bibinfo {title} {{Charge
  migration and control of site selective reactivity: The role of covalent and
  ionic states}},}\ }\href {https://doi.org/10.1063/1.478406} {\bibfield
  {journal} {\bibinfo  {journal} {J. Chem. Phys.}\ }\textbf {\bibinfo {volume}
  {110}},\ \bibinfo {pages} {5089--5099} (\bibinfo {year} {1999})}\BibitemShut
  {NoStop}%
\bibitem [{\citenamefont {Kling}\ \emph {et~al.}(2013)\citenamefont {Kling},
  \citenamefont {von~den Hoff}, \citenamefont {Znakovskaya},\ and\
  \citenamefont {de~Vivie-Riedle}}]{reaction-control-kling-2013}%
  \BibitemOpen
  \bibfield  {author} {\bibinfo {author} {\bibfnamefont {M.~F.}\ \bibnamefont
  {Kling}}, \bibinfo {author} {\bibfnamefont {P.}~\bibnamefont {von~den Hoff}},
  \bibinfo {author} {\bibfnamefont {I.}~\bibnamefont {Znakovskaya}},\ and\
  \bibinfo {author} {\bibfnamefont {R.}~\bibnamefont {de~Vivie-Riedle}},\
  }\bibfield  {title} {\enquote {\bibinfo {title} {{(Sub-)femtosecond control
  of molecular reactions via tailoring the electric field of light}},}\ }\href
  {https://doi.org/10.1039/C3CP50591J} {\bibfield  {journal} {\bibinfo
  {journal} {Phys. Chem. Chem. Phys.}\ }\textbf {\bibinfo {volume} {15}},\
  \bibinfo {pages} {9448--9467} (\bibinfo {year} {2013})}\BibitemShut {NoStop}%
\bibitem [{\citenamefont {Leone}\ \emph {et~al.}(2014)\citenamefont {Leone},
  \citenamefont {McCurdy}, \citenamefont {Burgd{\"o}rfer}, \citenamefont
  {Cederbaum}, \citenamefont {Chang}, \citenamefont {Dudovich}, \citenamefont
  {Feist}, \citenamefont {Greene}, \citenamefont {Ivanov}, \citenamefont
  {Kienberger} \emph {et~al.}}]{what-observe-real-time-2014}%
  \BibitemOpen
  \bibfield  {author} {\bibinfo {author} {\bibfnamefont {S.~R.}\ \bibnamefont
  {Leone}}, \bibinfo {author} {\bibfnamefont {C.~W.}\ \bibnamefont {McCurdy}},
  \bibinfo {author} {\bibfnamefont {J.}~\bibnamefont {Burgd{\"o}rfer}},
  \bibinfo {author} {\bibfnamefont {L.~S.}\ \bibnamefont {Cederbaum}}, \bibinfo
  {author} {\bibfnamefont {Z.}~\bibnamefont {Chang}}, \bibinfo {author}
  {\bibfnamefont {N.}~\bibnamefont {Dudovich}}, \bibinfo {author}
  {\bibfnamefont {J.}~\bibnamefont {Feist}}, \bibinfo {author} {\bibfnamefont
  {C.~H.}\ \bibnamefont {Greene}}, \bibinfo {author} {\bibfnamefont
  {M.}~\bibnamefont {Ivanov}}, \bibinfo {author} {\bibfnamefont
  {R.}~\bibnamefont {Kienberger}}, \emph {et~al.},\ }\bibfield  {title}
  {\enquote {\bibinfo {title} {What will it take to observe processes in'real
  time'?}}\ }\href {https://doi.org/https://doi.org/10.1038/nphoton.2014.48}
  {\bibfield  {journal} {\bibinfo  {journal} {Nat. Photonics}\ }\textbf
  {\bibinfo {volume} {8}},\ \bibinfo {pages} {162--166} (\bibinfo {year}
  {2014})}\BibitemShut {NoStop}%
\bibitem [{\citenamefont {Golubev}\ and\ \citenamefont
  {Kuleff}(2015)}]{cm-control-laser-2015}%
  \BibitemOpen
  \bibfield  {author} {\bibinfo {author} {\bibfnamefont {N.~V.}\ \bibnamefont
  {Golubev}}\ and\ \bibinfo {author} {\bibfnamefont {A.~I.}\ \bibnamefont
  {Kuleff}},\ }\bibfield  {title} {\enquote {\bibinfo {title} {Control of
  charge migration in molecules by ultrashort laser pulses},}\ }\href
  {https://doi.org/10.1103/PhysRevA.91.051401} {\bibfield  {journal} {\bibinfo
  {journal} {Phys. Rev. A}\ }\textbf {\bibinfo {volume} {91}},\ \bibinfo
  {pages} {051401} (\bibinfo {year} {2015})}\BibitemShut {NoStop}%
\bibitem [{\citenamefont {Larsson}\ \emph {et~al.}(2016)\citenamefont
  {Larsson}, \citenamefont {Bauch}, \citenamefont {S{\o}rensen},\ and\
  \citenamefont {Bonitz}}]{Correlation2016larsson-2016}%
  \BibitemOpen
  \bibfield  {author} {\bibinfo {author} {\bibfnamefont {H.~R.}\ \bibnamefont
  {Larsson}}, \bibinfo {author} {\bibfnamefont {S.}~\bibnamefont {Bauch}},
  \bibinfo {author} {\bibfnamefont {L.~K.}\ \bibnamefont {S{\o}rensen}},\ and\
  \bibinfo {author} {\bibfnamefont {M.}~\bibnamefont {Bonitz}},\ }\bibfield
  {title} {\enquote {\bibinfo {title} {Correlation effects in strong-field
  ionization of heteronuclear diatomic molecules},}\ }\href
  {https://doi.org/10.1103/PhysRevA.93.013426} {\bibfield  {journal} {\bibinfo
  {journal} {Phys. Rev. A}\ }\textbf {\bibinfo {volume} {93}},\ \bibinfo
  {pages} {013426} (\bibinfo {year} {2016})}\BibitemShut {NoStop}%
\bibitem [{\citenamefont {Nikolay V.~Golubev}\ and\ \citenamefont
  {Kuleff}(2017)}]{control-smooth-pulse-2017}%
  \BibitemOpen
  \bibfield  {author} {\bibinfo {author} {\bibfnamefont {V.~D.}\ \bibnamefont
  {Nikolay V.~Golubev}}\ and\ \bibinfo {author} {\bibfnamefont {A.~I.}\
  \bibnamefont {Kuleff}},\ }\bibfield  {title} {\enquote {\bibinfo {title}
  {Quantum control with smoothly varying pulses: general theory and application
  to charge migration},}\ }\href
  {https://doi.org/10.1080/09500340.2016.1275855} {\bibfield  {journal}
  {\bibinfo  {journal} {J. Mod. Opt.}\ }\textbf {\bibinfo {volume} {64}},\
  \bibinfo {pages} {1031--1041} (\bibinfo {year} {2017})}\BibitemShut {NoStop}%
\bibitem [{\citenamefont {Azoury}\ \emph {et~al.}(2017)\citenamefont {Azoury},
  \citenamefont {Kr{\"u}ger}, \citenamefont {Orenstein}, \citenamefont
  {Larsson}, \citenamefont {Bauch}, \citenamefont {Bruner},\ and\ \citenamefont
  {Dudovich}}]{Selfprobing2017azoury-2017}%
  \BibitemOpen
  \bibfield  {author} {\bibinfo {author} {\bibfnamefont {D.}~\bibnamefont
  {Azoury}}, \bibinfo {author} {\bibfnamefont {M.}~\bibnamefont {Kr{\"u}ger}},
  \bibinfo {author} {\bibfnamefont {G.}~\bibnamefont {Orenstein}}, \bibinfo
  {author} {\bibfnamefont {H.~R.}\ \bibnamefont {Larsson}}, \bibinfo {author}
  {\bibfnamefont {S.}~\bibnamefont {Bauch}}, \bibinfo {author} {\bibfnamefont
  {B.~D.}\ \bibnamefont {Bruner}},\ and\ \bibinfo {author} {\bibfnamefont
  {N.}~\bibnamefont {Dudovich}},\ }\bibfield  {title} {\enquote {\bibinfo
  {title} {Self-probing spectroscopy of {{XUV}} photo-ionization dynamics in
  atoms subjected to a strong-field environment},}\ }\href
  {https://doi.org/10.1038/s41467-017-01723-w} {\bibfield  {journal} {\bibinfo
  {journal} {Nat. Commun.}\ }\textbf {\bibinfo {volume} {8}},\ \bibinfo {pages}
  {1453} (\bibinfo {year} {2017})}\BibitemShut {NoStop}%
\bibitem [{\citenamefont {Cederbaum}\ and\ \citenamefont
  {Zobeley}(1999)}]{cm-cederbaum-1999}%
  \BibitemOpen
  \bibfield  {author} {\bibinfo {author} {\bibfnamefont {L.}~\bibnamefont
  {Cederbaum}}\ and\ \bibinfo {author} {\bibfnamefont {J.}~\bibnamefont
  {Zobeley}},\ }\bibfield  {title} {\enquote {\bibinfo {title} {Ultrafast
  charge migration by electron correlation},}\ }\href
  {https://doi.org/https://doi.org/10.1016/S0009-2614(99)00508-4} {\bibfield
  {journal} {\bibinfo  {journal} {Chem. Phys. Lett.}\ }\textbf {\bibinfo
  {volume} {307}},\ \bibinfo {pages} {205--210} (\bibinfo {year}
  {1999})}\BibitemShut {NoStop}%
\bibitem [{\citenamefont {Breidbach}\ and\ \citenamefont
  {Cederbaum}(2005)}]{cederbaum-atto-response-2005}%
  \BibitemOpen
  \bibfield  {author} {\bibinfo {author} {\bibfnamefont {J.}~\bibnamefont
  {Breidbach}}\ and\ \bibinfo {author} {\bibfnamefont {L.~S.}\ \bibnamefont
  {Cederbaum}},\ }\bibfield  {title} {\enquote {\bibinfo {title} {Universal
  attosecond response to the removal of an electron},}\ }\href
  {https://doi.org/10.1103/PhysRevLett.94.033901} {\bibfield  {journal}
  {\bibinfo  {journal} {Phys. Rev. Lett.}\ }\textbf {\bibinfo {volume} {94}},\
  \bibinfo {pages} {033901} (\bibinfo {year} {2005})}\BibitemShut {NoStop}%
\bibitem [{\citenamefont {Remacle}\ and\ \citenamefont
  {Levine}(2006)}]{timescale-chem-2006}%
  \BibitemOpen
  \bibfield  {author} {\bibinfo {author} {\bibfnamefont {F.}~\bibnamefont
  {Remacle}}\ and\ \bibinfo {author} {\bibfnamefont {R.~D.}\ \bibnamefont
  {Levine}},\ }\bibfield  {title} {\enquote {\bibinfo {title} {An electronic
  time scale in chemistry},}\ }\href {https://doi.org/10.1073/pnas.0601855103}
  {\bibfield  {journal} {\bibinfo  {journal} {Proc. Natl. Acad. Sci. USA}\
  }\textbf {\bibinfo {volume} {103}},\ \bibinfo {pages} {6793--6798} (\bibinfo
  {year} {2006})}\BibitemShut {NoStop}%
\bibitem [{\citenamefont {Belshaw}\ \emph {et~al.}(2012)\citenamefont
  {Belshaw}, \citenamefont {Calegari}, \citenamefont {Duffy}, \citenamefont
  {Trabattoni}, \citenamefont {Poletto}, \citenamefont {Nisoli},\ and\
  \citenamefont {Greenwood}}]{cm-amino-acid-2012}%
  \BibitemOpen
  \bibfield  {author} {\bibinfo {author} {\bibfnamefont {L.}~\bibnamefont
  {Belshaw}}, \bibinfo {author} {\bibfnamefont {F.}~\bibnamefont {Calegari}},
  \bibinfo {author} {\bibfnamefont {M.~J.}\ \bibnamefont {Duffy}}, \bibinfo
  {author} {\bibfnamefont {A.}~\bibnamefont {Trabattoni}}, \bibinfo {author}
  {\bibfnamefont {L.}~\bibnamefont {Poletto}}, \bibinfo {author} {\bibfnamefont
  {M.}~\bibnamefont {Nisoli}},\ and\ \bibinfo {author} {\bibfnamefont {J.~B.}\
  \bibnamefont {Greenwood}},\ }\bibfield  {title} {\enquote {\bibinfo {title}
  {Observation of ultrafast charge migration in an amino acid},}\ }\href
  {https://doi.org/10.1021/jz3016028} {\bibfield  {journal} {\bibinfo
  {journal} {J. Phys. Chem. Lett.}\ }\textbf {\bibinfo {volume} {3}},\ \bibinfo
  {pages} {3751--3754} (\bibinfo {year} {2012})}\BibitemShut {NoStop}%
\bibitem [{\citenamefont {Calegari}\ \emph {et~al.}(2014)\citenamefont
  {Calegari}, \citenamefont {Ayuso}, \citenamefont {Trabattoni}, \citenamefont
  {Belshaw}, \citenamefont {Camillis}, \citenamefont {Anumula}, \citenamefont
  {Frassetto}, \citenamefont {Poletto}, \citenamefont {Palacios}, \citenamefont
  {Decleva}, \citenamefont {Greenwood}, \citenamefont {Martín},\ and\
  \citenamefont {Nisoli}}]{cm-phenylalanine-2014}%
  \BibitemOpen
  \bibfield  {author} {\bibinfo {author} {\bibfnamefont {F.}~\bibnamefont
  {Calegari}}, \bibinfo {author} {\bibfnamefont {D.}~\bibnamefont {Ayuso}},
  \bibinfo {author} {\bibfnamefont {A.}~\bibnamefont {Trabattoni}}, \bibinfo
  {author} {\bibfnamefont {L.}~\bibnamefont {Belshaw}}, \bibinfo {author}
  {\bibfnamefont {S.~D.}\ \bibnamefont {Camillis}}, \bibinfo {author}
  {\bibfnamefont {S.}~\bibnamefont {Anumula}}, \bibinfo {author} {\bibfnamefont
  {F.}~\bibnamefont {Frassetto}}, \bibinfo {author} {\bibfnamefont
  {L.}~\bibnamefont {Poletto}}, \bibinfo {author} {\bibfnamefont
  {A.}~\bibnamefont {Palacios}}, \bibinfo {author} {\bibfnamefont
  {P.}~\bibnamefont {Decleva}}, \bibinfo {author} {\bibfnamefont {J.~B.}\
  \bibnamefont {Greenwood}}, \bibinfo {author} {\bibfnamefont {F.}~\bibnamefont
  {Martín}},\ and\ \bibinfo {author} {\bibfnamefont {M.}~\bibnamefont
  {Nisoli}},\ }\bibfield  {title} {\enquote {\bibinfo {title} {Ultrafast
  electron dynamics in phenylalanine initiated by attosecond pulses},}\ }\href
  {https://doi.org/10.1126/science.1254061} {\bibfield  {journal} {\bibinfo
  {journal} {Science}\ }\textbf {\bibinfo {volume} {346}},\ \bibinfo {pages}
  {336--339} (\bibinfo {year} {2014})}\BibitemShut {NoStop}%
\bibitem [{\citenamefont {Calegari}\ \emph {et~al.}(2015)\citenamefont
  {Calegari}, \citenamefont {Ayuso}, \citenamefont {Trabattoni}, \citenamefont
  {Belshaw}, \citenamefont {De~Camillis}, \citenamefont {Frassetto},
  \citenamefont {Poletto}, \citenamefont {Palacios}, \citenamefont {Decleva},
  \citenamefont {Greenwood}, \citenamefont {Martín},\ and\ \citenamefont
  {Nisoli}}]{cm-amino-acid-2015}%
  \BibitemOpen
  \bibfield  {author} {\bibinfo {author} {\bibfnamefont {F.}~\bibnamefont
  {Calegari}}, \bibinfo {author} {\bibfnamefont {D.}~\bibnamefont {Ayuso}},
  \bibinfo {author} {\bibfnamefont {A.}~\bibnamefont {Trabattoni}}, \bibinfo
  {author} {\bibfnamefont {L.}~\bibnamefont {Belshaw}}, \bibinfo {author}
  {\bibfnamefont {S.}~\bibnamefont {De~Camillis}}, \bibinfo {author}
  {\bibfnamefont {F.}~\bibnamefont {Frassetto}}, \bibinfo {author}
  {\bibfnamefont {L.}~\bibnamefont {Poletto}}, \bibinfo {author} {\bibfnamefont
  {A.}~\bibnamefont {Palacios}}, \bibinfo {author} {\bibfnamefont
  {P.}~\bibnamefont {Decleva}}, \bibinfo {author} {\bibfnamefont {J.~B.}\
  \bibnamefont {Greenwood}}, \bibinfo {author} {\bibfnamefont {F.}~\bibnamefont
  {Martín}},\ and\ \bibinfo {author} {\bibfnamefont {M.}~\bibnamefont
  {Nisoli}},\ }\bibfield  {title} {\enquote {\bibinfo {title} {Ultrafast charge
  dynamics in an amino acid induced by attosecond pulses},}\ }\href
  {https://doi.org/10.1109/JSTQE.2015.2419218} {\bibfield  {journal} {\bibinfo
  {journal} {IEEE J. Sel. Top. Quantum Electron.}\ }\textbf {\bibinfo {volume}
  {21}},\ \bibinfo {pages} {1--12} (\bibinfo {year} {2015})}\BibitemShut
  {NoStop}%
\bibitem [{\citenamefont {Kraus}\ \emph {et~al.}(2015)\citenamefont {Kraus},
  \citenamefont {Mignolet}, \citenamefont {Baykusheva}, \citenamefont
  {Rupenyan}, \citenamefont {Horný}, \citenamefont {Penka}, \citenamefont
  {Grassi}, \citenamefont {Tolstikhin}, \citenamefont {Schneider},
  \citenamefont {Jensen}, \citenamefont {Madsen}, \citenamefont {Bandrauk},
  \citenamefont {Remacle},\ and\ \citenamefont {Wörner}}]{cm-iodo-2015}%
  \BibitemOpen
  \bibfield  {author} {\bibinfo {author} {\bibfnamefont {P.~M.}\ \bibnamefont
  {Kraus}}, \bibinfo {author} {\bibfnamefont {B.}~\bibnamefont {Mignolet}},
  \bibinfo {author} {\bibfnamefont {D.}~\bibnamefont {Baykusheva}}, \bibinfo
  {author} {\bibfnamefont {A.}~\bibnamefont {Rupenyan}}, \bibinfo {author}
  {\bibfnamefont {L.}~\bibnamefont {Horný}}, \bibinfo {author} {\bibfnamefont
  {E.~F.}\ \bibnamefont {Penka}}, \bibinfo {author} {\bibfnamefont
  {G.}~\bibnamefont {Grassi}}, \bibinfo {author} {\bibfnamefont {O.~I.}\
  \bibnamefont {Tolstikhin}}, \bibinfo {author} {\bibfnamefont
  {J.}~\bibnamefont {Schneider}}, \bibinfo {author} {\bibfnamefont
  {F.}~\bibnamefont {Jensen}}, \bibinfo {author} {\bibfnamefont {L.~B.}\
  \bibnamefont {Madsen}}, \bibinfo {author} {\bibfnamefont {A.~D.}\
  \bibnamefont {Bandrauk}}, \bibinfo {author} {\bibfnamefont {F.}~\bibnamefont
  {Remacle}},\ and\ \bibinfo {author} {\bibfnamefont {H.~J.}\ \bibnamefont
  {Wörner}},\ }\bibfield  {title} {\enquote {\bibinfo {title} {Measurement and
  laser control of attosecond charge migration in ionized iodoacetylene},}\
  }\href {https://doi.org/10.1126/science.aab2160} {\bibfield  {journal}
  {\bibinfo  {journal} {Science}\ }\textbf {\bibinfo {volume} {350}},\ \bibinfo
  {pages} {790--795} (\bibinfo {year} {2015})}\BibitemShut {NoStop}%
\bibitem [{\citenamefont {Despr{\'e}}\ \emph
  {et~al.}(2015{\natexlab{a}})\citenamefont {Despr{\'e}}, \citenamefont
  {Marciniak}, \citenamefont {Loriot}, \citenamefont {Galbraith}, \citenamefont
  {Rouz{\'e}e}, \citenamefont {Vrakking}, \citenamefont {L{\'e}pine},\ and\
  \citenamefont {Kuleff}}]{hole-survive-nucl-2015}%
  \BibitemOpen
  \bibfield  {author} {\bibinfo {author} {\bibfnamefont {V.}~\bibnamefont
  {Despr{\'e}}}, \bibinfo {author} {\bibfnamefont {A.}~\bibnamefont
  {Marciniak}}, \bibinfo {author} {\bibfnamefont {V.}~\bibnamefont {Loriot}},
  \bibinfo {author} {\bibfnamefont {M.~C.~E.}\ \bibnamefont {Galbraith}},
  \bibinfo {author} {\bibfnamefont {A.}~\bibnamefont {Rouz{\'e}e}}, \bibinfo
  {author} {\bibfnamefont {M.~J.~J.}\ \bibnamefont {Vrakking}}, \bibinfo
  {author} {\bibfnamefont {F.}~\bibnamefont {L{\'e}pine}},\ and\ \bibinfo
  {author} {\bibfnamefont {A.~I.}\ \bibnamefont {Kuleff}},\ }\bibfield  {title}
  {\enquote {\bibinfo {title} {Attosecond hole migration in benzene molecules
  surviving nuclear motion},}\ }\href {https://doi.org/10.1021/jz502493j}
  {\bibfield  {journal} {\bibinfo  {journal} {J. Phys. Chem. Lett.}\ }\textbf
  {\bibinfo {volume} {6}},\ \bibinfo {pages} {426--431} (\bibinfo {year}
  {2015}{\natexlab{a}})}\BibitemShut {NoStop}%
\bibitem [{\citenamefont {Vacher}\ \emph {et~al.}(2015)\citenamefont {Vacher},
  \citenamefont {Steinberg}, \citenamefont {Jenkins}, \citenamefont
  {Bearpark},\ and\ \citenamefont {Robb}}]{cm-quantum-nucl-2015}%
  \BibitemOpen
  \bibfield  {author} {\bibinfo {author} {\bibfnamefont {M.}~\bibnamefont
  {Vacher}}, \bibinfo {author} {\bibfnamefont {L.}~\bibnamefont {Steinberg}},
  \bibinfo {author} {\bibfnamefont {A.~J.}\ \bibnamefont {Jenkins}}, \bibinfo
  {author} {\bibfnamefont {M.~J.}\ \bibnamefont {Bearpark}},\ and\ \bibinfo
  {author} {\bibfnamefont {M.~A.}\ \bibnamefont {Robb}},\ }\bibfield  {title}
  {\enquote {\bibinfo {title} {Electron dynamics following photoionization:
  Decoherence due to the nuclear-wave-packet width},}\ }\href
  {https://doi.org/10.1103/PhysRevA.92.040502} {\bibfield  {journal} {\bibinfo
  {journal} {Phys. Rev. A}\ }\textbf {\bibinfo {volume} {92}},\ \bibinfo
  {pages} {040502} (\bibinfo {year} {2015})}\BibitemShut {NoStop}%
\bibitem [{\citenamefont {Despr{\'e}}\ \emph
  {et~al.}(2015{\natexlab{b}})\citenamefont {Despr{\'e}}, \citenamefont
  {Marciniak}, \citenamefont {Loriot}, \citenamefont {Galbraith}, \citenamefont
  {Rouz{\'e}e}, \citenamefont {Vrakking}, \citenamefont {L{\'e}pine},\ and\
  \citenamefont {Kuleff}}]{atto-migrate-benzene-2015}%
  \BibitemOpen
  \bibfield  {author} {\bibinfo {author} {\bibfnamefont {V.}~\bibnamefont
  {Despr{\'e}}}, \bibinfo {author} {\bibfnamefont {A.}~\bibnamefont
  {Marciniak}}, \bibinfo {author} {\bibfnamefont {V.}~\bibnamefont {Loriot}},
  \bibinfo {author} {\bibfnamefont {M.~C.~E.}\ \bibnamefont {Galbraith}},
  \bibinfo {author} {\bibfnamefont {A.}~\bibnamefont {Rouz{\'e}e}}, \bibinfo
  {author} {\bibfnamefont {M.~J.~J.}\ \bibnamefont {Vrakking}}, \bibinfo
  {author} {\bibfnamefont {F.}~\bibnamefont {L{\'e}pine}},\ and\ \bibinfo
  {author} {\bibfnamefont {A.~I.}\ \bibnamefont {Kuleff}},\ }\bibfield  {title}
  {\enquote {\bibinfo {title} {Attosecond hole migration in benzene molecules
  surviving nuclear motion},}\ }\href {https://doi.org/10.1021/jz502493j}
  {\bibfield  {journal} {\bibinfo  {journal} {J. Phys. Chem. Lett.}\ }\textbf
  {\bibinfo {volume} {6}},\ \bibinfo {pages} {426--431} (\bibinfo {year}
  {2015}{\natexlab{b}})}\BibitemShut {NoStop}%
\bibitem [{\citenamefont {Arnold}, \citenamefont {Vendrell},\ and\
  \citenamefont {Santra}(2017)}]{decohere-full-quantum-2017}%
  \BibitemOpen
  \bibfield  {author} {\bibinfo {author} {\bibfnamefont {C.}~\bibnamefont
  {Arnold}}, \bibinfo {author} {\bibfnamefont {O.}~\bibnamefont {Vendrell}},\
  and\ \bibinfo {author} {\bibfnamefont {R.}~\bibnamefont {Santra}},\
  }\bibfield  {title} {\enquote {\bibinfo {title} {Electronic decoherence
  following photoionization: Full quantum-dynamical treatment of the influence
  of nuclear motion},}\ }\href {https://doi.org/10.1103/PhysRevA.95.033425}
  {\bibfield  {journal} {\bibinfo  {journal} {Phys. Rev. A}\ }\textbf {\bibinfo
  {volume} {95}},\ \bibinfo {pages} {033425} (\bibinfo {year}
  {2017})}\BibitemShut {NoStop}%
\bibitem [{\citenamefont {Vacher}\ \emph {et~al.}(2017)\citenamefont {Vacher},
  \citenamefont {Bearpark}, \citenamefont {Robb},\ and\ \citenamefont
  {Malhado}}]{cm-quantum-nucl-2017}%
  \BibitemOpen
  \bibfield  {author} {\bibinfo {author} {\bibfnamefont {M.}~\bibnamefont
  {Vacher}}, \bibinfo {author} {\bibfnamefont {M.~J.}\ \bibnamefont
  {Bearpark}}, \bibinfo {author} {\bibfnamefont {M.~A.}\ \bibnamefont {Robb}},\
  and\ \bibinfo {author} {\bibfnamefont {J.~a.~P.}\ \bibnamefont {Malhado}},\
  }\bibfield  {title} {\enquote {\bibinfo {title} {Electron dynamics upon
  ionization of polyatomic molecules: Coupling to quantum nuclear motion and
  decoherence},}\ }\href {https://doi.org/10.1103/PhysRevLett.118.083001}
  {\bibfield  {journal} {\bibinfo  {journal} {Phys. Rev. Lett.}\ }\textbf
  {\bibinfo {volume} {118}},\ \bibinfo {pages} {083001} (\bibinfo {year}
  {2017})}\BibitemShut {NoStop}%
\bibitem [{\citenamefont {Despr\'e}, \citenamefont {Golubev},\ and\
  \citenamefont {Kuleff}(2018)}]{cm-propiol-full-quantum-2018}%
  \BibitemOpen
  \bibfield  {author} {\bibinfo {author} {\bibfnamefont {V.}~\bibnamefont
  {Despr\'e}}, \bibinfo {author} {\bibfnamefont {N.~V.}\ \bibnamefont
  {Golubev}},\ and\ \bibinfo {author} {\bibfnamefont {A.~I.}\ \bibnamefont
  {Kuleff}},\ }\bibfield  {title} {\enquote {\bibinfo {title} {Charge migration
  in propiolic acid: A full quantum dynamical study},}\ }\href
  {https://doi.org/10.1103/PhysRevLett.121.203002} {\bibfield  {journal}
  {\bibinfo  {journal} {Phys. Rev. Lett.}\ }\textbf {\bibinfo {volume} {121}},\
  \bibinfo {pages} {203002} (\bibinfo {year} {2018})}\BibitemShut {NoStop}%
\bibitem [{\citenamefont {Ma}\ \emph {et~al.}(2023)\citenamefont {Ma},
  \citenamefont {Manz}, \citenamefont {Wang}, \citenamefont {Yan},\ and\
  \citenamefont {Yang}}]{de-re-pyrene-2023}%
  \BibitemOpen
  \bibfield  {author} {\bibinfo {author} {\bibfnamefont {H.}~\bibnamefont
  {Ma}}, \bibinfo {author} {\bibfnamefont {J.}~\bibnamefont {Manz}}, \bibinfo
  {author} {\bibfnamefont {H.}~\bibnamefont {Wang}}, \bibinfo {author}
  {\bibfnamefont {Y.}~\bibnamefont {Yan}},\ and\ \bibinfo {author}
  {\bibfnamefont {Y.}~\bibnamefont {Yang}},\ }\bibfield  {title} {\enquote
  {\bibinfo {title} {Ultrafast laser induced charge migration with de- and
  re-coherences in polyatomic molecules: A general method with application to
  pyrene},}\ }\href {https://doi.org/10.1063/5.0141631} {\bibfield  {journal}
  {\bibinfo  {journal} {J. Chem. Phys.}\ }\textbf {\bibinfo {volume} {158}},\
  \bibinfo {pages} {124306} (\bibinfo {year} {2023})}\BibitemShut {NoStop}%
\bibitem [{\citenamefont {Scheidegger}, \citenamefont {Golubev},\ and\
  \citenamefont {Vaníček}(2023)}]{mol-long-cm-2023}%
  \BibitemOpen
  \bibfield  {author} {\bibinfo {author} {\bibfnamefont {A.}~\bibnamefont
  {Scheidegger}}, \bibinfo {author} {\bibfnamefont {N.~V.}\ \bibnamefont
  {Golubev}},\ and\ \bibinfo {author} {\bibfnamefont {J.}~\bibnamefont
  {Vaníček}},\ }\bibfield  {title} {\enquote {\bibinfo {title} {How to find
  molecules with long-lasting charge migration?}}\ }\href
  {https://doi.org/10.2533/chimia.2023.201} {\bibfield  {journal} {\bibinfo
  {journal} {Chimia}\ }\textbf {\bibinfo {volume} {77}},\ \bibinfo {pages}
  {201–205} (\bibinfo {year} {2023})}\BibitemShut {NoStop}%
\bibitem [{\citenamefont {Chordiya}\ \emph {et~al.}(2023)\citenamefont
  {Chordiya}, \citenamefont {Despré}, \citenamefont {Nagyillés},
  \citenamefont {Zeller}, \citenamefont {Diveki}, \citenamefont {Kuleff},\ and\
  \citenamefont {Kahaly}}]{cm-adc-chordiya-2023}%
  \BibitemOpen
  \bibfield  {author} {\bibinfo {author} {\bibfnamefont {K.}~\bibnamefont
  {Chordiya}}, \bibinfo {author} {\bibfnamefont {V.}~\bibnamefont {Despré}},
  \bibinfo {author} {\bibfnamefont {B.}~\bibnamefont {Nagyillés}}, \bibinfo
  {author} {\bibfnamefont {F.}~\bibnamefont {Zeller}}, \bibinfo {author}
  {\bibfnamefont {Z.}~\bibnamefont {Diveki}}, \bibinfo {author} {\bibfnamefont
  {A.~I.}\ \bibnamefont {Kuleff}},\ and\ \bibinfo {author} {\bibfnamefont
  {M.~U.}\ \bibnamefont {Kahaly}},\ }\bibfield  {title} {\enquote {\bibinfo
  {title} {Photo-ionization initiated differential ultrafast charge migration:
  impacts of molecular symmetries and tautomeric forms},}\ }\href
  {https://doi.org/10.1039/D2CP02681C} {\bibfield  {journal} {\bibinfo
  {journal} {Phys. Chem. Chem. Phys.}\ }\textbf {\bibinfo {volume} {25}},\
  \bibinfo {pages} {4472--4480} (\bibinfo {year} {2023})}\BibitemShut {NoStop}%
\bibitem [{\citenamefont {Haase}, \citenamefont {Manz},\ and\ \citenamefont
  {Tremblay}(2020)}]{cm-break-symm-2020}%
  \BibitemOpen
  \bibfield  {author} {\bibinfo {author} {\bibfnamefont {D.}~\bibnamefont
  {Haase}}, \bibinfo {author} {\bibfnamefont {J.}~\bibnamefont {Manz}},\ and\
  \bibinfo {author} {\bibfnamefont {J.~C.}\ \bibnamefont {Tremblay}},\
  }\bibfield  {title} {\enquote {\bibinfo {title} {Attosecond charge migration
  can break electron symmetry while conserving nuclear symmetry},}\ }\href
  {https://doi.org/10.1021/acs.jpca.0c00404} {\bibfield  {journal} {\bibinfo
  {journal} {J. Phys. Chem. A}\ }\textbf {\bibinfo {volume} {124}},\ \bibinfo
  {pages} {3329--3334} (\bibinfo {year} {2020})}\BibitemShut {NoStop}%
\bibitem [{\citenamefont {Giri}, \citenamefont {Dixit},\ and\ \citenamefont
  {Tremblay}(2023)}]{5-heterocyclic-cm-2023}%
  \BibitemOpen
  \bibfield  {author} {\bibinfo {author} {\bibfnamefont {S.}~\bibnamefont
  {Giri}}, \bibinfo {author} {\bibfnamefont {G.}~\bibnamefont {Dixit}},\ and\
  \bibinfo {author} {\bibfnamefont {J.~C.}\ \bibnamefont {Tremblay}},\
  }\bibfield  {title} {\enquote {\bibinfo {title} {Attosecond charge migration
  in heterocyclic five-membered rings},}\ }\href
  {https://doi.org/10.1140/epjs/s11734-023-00942-1} {\bibfield  {journal}
  {\bibinfo  {journal} {Eur. Phys. J.: Spec. Top.}\ }\textbf {\bibinfo {volume}
  {232}},\ \bibinfo {pages} {1935--1943} (\bibinfo {year} {2023})}\BibitemShut
  {NoStop}%
\bibitem [{\citenamefont {Tremblay}\ \emph {et~al.}(2023)\citenamefont
  {Tremblay}, \citenamefont {Blanc}, \citenamefont {Krause}, \citenamefont
  {Giri},\ and\ \citenamefont {Dixit}}]{symmetry-reduction-probe-2023}%
  \BibitemOpen
  \bibfield  {author} {\bibinfo {author} {\bibfnamefont {J.~C.}\ \bibnamefont
  {Tremblay}}, \bibinfo {author} {\bibfnamefont {A.}~\bibnamefont {Blanc}},
  \bibinfo {author} {\bibfnamefont {P.}~\bibnamefont {Krause}}, \bibinfo
  {author} {\bibfnamefont {S.}~\bibnamefont {Giri}},\ and\ \bibinfo {author}
  {\bibfnamefont {G.}~\bibnamefont {Dixit}},\ }\bibfield  {title} {\enquote
  {\bibinfo {title} {{Probing Electronic Symmetry Reduction during Charge
  Migration via Time-Resolved X-Ray Diffraction}},}\ }\href
  {https://doi.org/https://doi.org/10.1002/cphc.202200463} {\bibfield
  {journal} {\bibinfo  {journal} {Chem. Phys. Chem.}\ }\textbf {\bibinfo
  {volume} {24}},\ \bibinfo {pages} {e202200463} (\bibinfo {year}
  {2023})}\BibitemShut {NoStop}%
\bibitem [{\citenamefont {Lünnemann}, \citenamefont {Kuleff},\ and\
  \citenamefont {Cederbaum}(2008)}]{cm-phenylethyl-2008}%
  \BibitemOpen
  \bibfield  {author} {\bibinfo {author} {\bibfnamefont {S.}~\bibnamefont
  {Lünnemann}}, \bibinfo {author} {\bibfnamefont {A.~I.}\ \bibnamefont
  {Kuleff}},\ and\ \bibinfo {author} {\bibfnamefont {L.~S.}\ \bibnamefont
  {Cederbaum}},\ }\bibfield  {title} {\enquote {\bibinfo {title} {Ultrafast
  charge migration in 2-phenylethyl-n,n-dimethylamine},}\ }\href
  {https://doi.org/https://doi.org/10.1016/j.cplett.2007.11.031} {\bibfield
  {journal} {\bibinfo  {journal} {Chem. Phys. Lett.}\ }\textbf {\bibinfo
  {volume} {450}},\ \bibinfo {pages} {232--235} (\bibinfo {year}
  {2008})}\BibitemShut {NoStop}%
\bibitem [{\citenamefont {Kuleff}, \citenamefont {Lünnemann},\ and\
  \citenamefont {Cederbaum}(2013)}]{cm-oligopeptides-2013}%
  \BibitemOpen
  \bibfield  {author} {\bibinfo {author} {\bibfnamefont {A.~I.}\ \bibnamefont
  {Kuleff}}, \bibinfo {author} {\bibfnamefont {S.}~\bibnamefont {Lünnemann}},\
  and\ \bibinfo {author} {\bibfnamefont {L.~S.}\ \bibnamefont {Cederbaum}},\
  }\bibfield  {title} {\enquote {\bibinfo {title} {Electron-correlation-driven
  charge migration in oligopeptides},}\ }\href
  {https://doi.org/https://doi.org/10.1016/j.chemphys.2012.02.019} {\bibfield
  {journal} {\bibinfo  {journal} {Chem. Phys.}\ }\textbf {\bibinfo {volume}
  {414}},\ \bibinfo {pages} {100--105} (\bibinfo {year} {2013})}\BibitemShut
  {NoStop}%
\bibitem [{\citenamefont {Bruner}\ \emph {et~al.}(2017)\citenamefont {Bruner},
  \citenamefont {Hernandez}, \citenamefont {Mauger}, \citenamefont {Abanador},
  \citenamefont {LaMaster}, \citenamefont {Gaarde}, \citenamefont {Schafer},\
  and\ \citenamefont {Lopata}}]{cm-tddft3-2017}%
  \BibitemOpen
  \bibfield  {author} {\bibinfo {author} {\bibfnamefont {A.}~\bibnamefont
  {Bruner}}, \bibinfo {author} {\bibfnamefont {S.}~\bibnamefont {Hernandez}},
  \bibinfo {author} {\bibfnamefont {F.}~\bibnamefont {Mauger}}, \bibinfo
  {author} {\bibfnamefont {P.~M.}\ \bibnamefont {Abanador}}, \bibinfo {author}
  {\bibfnamefont {D.~J.}\ \bibnamefont {LaMaster}}, \bibinfo {author}
  {\bibfnamefont {M.~B.}\ \bibnamefont {Gaarde}}, \bibinfo {author}
  {\bibfnamefont {K.~J.}\ \bibnamefont {Schafer}},\ and\ \bibinfo {author}
  {\bibfnamefont {K.}~\bibnamefont {Lopata}},\ }\bibfield  {title} {\enquote
  {\bibinfo {title} {{Attosecond Charge Migration with TDDFT: Accurate Dynamics
  from a Well-Defined Initial State}},}\ }\href
  {https://doi.org/10.1021/acs.jpclett.7b01652} {\bibfield  {journal} {\bibinfo
   {journal} {J. Phys. Chem. Lett.}\ }\textbf {\bibinfo {volume} {8}},\
  \bibinfo {pages} {3991--3996} (\bibinfo {year} {2017})}\BibitemShut {NoStop}%
\bibitem [{\citenamefont {Jia}\ \emph {et~al.}(2017)\citenamefont {Jia},
  \citenamefont {Manz}, \citenamefont {Paulus}, \citenamefont {Pohl},
  \citenamefont {Tremblay},\ and\ \citenamefont {Yang}}]{qcontrol-elflux-2017}%
  \BibitemOpen
  \bibfield  {author} {\bibinfo {author} {\bibfnamefont {D.}~\bibnamefont
  {Jia}}, \bibinfo {author} {\bibfnamefont {J.}~\bibnamefont {Manz}}, \bibinfo
  {author} {\bibfnamefont {B.}~\bibnamefont {Paulus}}, \bibinfo {author}
  {\bibfnamefont {V.}~\bibnamefont {Pohl}}, \bibinfo {author} {\bibfnamefont
  {J.~C.}\ \bibnamefont {Tremblay}},\ and\ \bibinfo {author} {\bibfnamefont
  {Y.}~\bibnamefont {Yang}},\ }\bibfield  {title} {\enquote {\bibinfo {title}
  {Quantum control of electronic fluxes during adiabatic attosecond charge
  migration in degenerate superposition states of benzene},}\ }\href
  {https://doi.org/https://doi.org/10.1016/j.chemphys.2016.09.021} {\bibfield
  {journal} {\bibinfo  {journal} {Chem. Phys.}\ }\textbf {\bibinfo {volume}
  {482}},\ \bibinfo {pages} {146--159} (\bibinfo {year} {2017})}\BibitemShut
  {NoStop}%
\bibitem [{\citenamefont {Lara-Astiaso}\ \emph {et~al.}(2017)\citenamefont
  {Lara-Astiaso}, \citenamefont {Palacios}, \citenamefont {Decleva},
  \citenamefont {Tavernelli},\ and\ \citenamefont
  {Martín}}]{cm-nucl-glycine-2017}%
  \BibitemOpen
  \bibfield  {author} {\bibinfo {author} {\bibfnamefont {M.}~\bibnamefont
  {Lara-Astiaso}}, \bibinfo {author} {\bibfnamefont {A.}~\bibnamefont
  {Palacios}}, \bibinfo {author} {\bibfnamefont {P.}~\bibnamefont {Decleva}},
  \bibinfo {author} {\bibfnamefont {I.}~\bibnamefont {Tavernelli}},\ and\
  \bibinfo {author} {\bibfnamefont {F.}~\bibnamefont {Martín}},\ }\bibfield
  {title} {\enquote {\bibinfo {title} {Role of electron-nuclear coupled
  dynamics on charge migration induced by attosecond pulses in glycine},}\
  }\href {https://doi.org/https://doi.org/10.1016/j.cplett.2017.05.008}
  {\bibfield  {journal} {\bibinfo  {journal} {Chem. Phys. Lett.}\ }\textbf
  {\bibinfo {volume} {683}},\ \bibinfo {pages} {357--364} (\bibinfo {year}
  {2017})}\BibitemShut {NoStop}%
\bibitem [{\citenamefont {Hamer}\ \emph {et~al.}(2022)\citenamefont {Hamer},
  \citenamefont {Mauger}, \citenamefont {Folorunso}, \citenamefont {Lopata},
  \citenamefont {Jones}, \citenamefont {DiMauro}, \citenamefont {Schafer},\
  and\ \citenamefont {Gaarde}}]{cm-particle-2022}%
  \BibitemOpen
  \bibfield  {author} {\bibinfo {author} {\bibfnamefont {K.~A.}\ \bibnamefont
  {Hamer}}, \bibinfo {author} {\bibfnamefont {F.}~\bibnamefont {Mauger}},
  \bibinfo {author} {\bibfnamefont {A.~S.}\ \bibnamefont {Folorunso}}, \bibinfo
  {author} {\bibfnamefont {K.}~\bibnamefont {Lopata}}, \bibinfo {author}
  {\bibfnamefont {R.~R.}\ \bibnamefont {Jones}}, \bibinfo {author}
  {\bibfnamefont {L.~F.}\ \bibnamefont {DiMauro}}, \bibinfo {author}
  {\bibfnamefont {K.~J.}\ \bibnamefont {Schafer}},\ and\ \bibinfo {author}
  {\bibfnamefont {M.~B.}\ \bibnamefont {Gaarde}},\ }\bibfield  {title}
  {\enquote {\bibinfo {title} {Characterizing particle-like charge-migration
  dynamics with high-order harmonic sideband spectroscopy},}\ }\href
  {https://doi.org/10.1103/PhysRevA.106.013103} {\bibfield  {journal} {\bibinfo
   {journal} {Phys. Rev. A}\ }\textbf {\bibinfo {volume} {106}},\ \bibinfo
  {pages} {013103} (\bibinfo {year} {2022})}\BibitemShut {NoStop}%
\bibitem [{\citenamefont {Hamer}\ \emph {et~al.}(2024)\citenamefont {Hamer},
  \citenamefont {Folorunso}, \citenamefont {Lopata}, \citenamefont {Schafer},
  \citenamefont {Gaarde},\ and\ \citenamefont {Mauger}}]{cm-fmatch-2024}%
  \BibitemOpen
  \bibfield  {author} {\bibinfo {author} {\bibfnamefont {K.~A.}\ \bibnamefont
  {Hamer}}, \bibinfo {author} {\bibfnamefont {A.~S.}\ \bibnamefont
  {Folorunso}}, \bibinfo {author} {\bibfnamefont {K.}~\bibnamefont {Lopata}},
  \bibinfo {author} {\bibfnamefont {K.~J.}\ \bibnamefont {Schafer}}, \bibinfo
  {author} {\bibfnamefont {M.~B.}\ \bibnamefont {Gaarde}},\ and\ \bibinfo
  {author} {\bibfnamefont {F.}~\bibnamefont {Mauger}},\ }\bibfield  {title}
  {\enquote {\bibinfo {title} {Tracking charge migration with frequency-matched
  strobo-spectroscopy},}\ }\href {https://doi.org/10.1021/acs.jpca.3c04234}
  {\bibfield  {journal} {\bibinfo  {journal} {J. Phys. Chem. A}\ }\textbf
  {\bibinfo {volume} {128}},\ \bibinfo {pages} {20--27} (\bibinfo {year}
  {2024})}\BibitemShut {NoStop}%
\bibitem [{\citenamefont {Guiot~du Doignon}\ \emph {et~al.}(2025)\citenamefont
  {Guiot~du Doignon}, \citenamefont {Sinha-Roy}, \citenamefont {Rabilloud},\
  and\ \citenamefont {Despr\'e}}]{corr-cm-mpi-2025}%
  \BibitemOpen
  \bibfield  {author} {\bibinfo {author} {\bibfnamefont {C.}~\bibnamefont
  {Guiot~du Doignon}}, \bibinfo {author} {\bibfnamefont {R.}~\bibnamefont
  {Sinha-Roy}}, \bibinfo {author} {\bibfnamefont {F.}~\bibnamefont
  {Rabilloud}},\ and\ \bibinfo {author} {\bibfnamefont {V.}~\bibnamefont
  {Despr\'e}},\ }\bibfield  {title} {\enquote {\bibinfo {title}
  {Correlation-driven charge migration triggered by infrared multi-photon
  ionization},}\ }\href {https://doi.org/10.1039/D5SC02374B} {\bibfield
  {journal} {\bibinfo  {journal} {Chem. Sci.}\ }\textbf {\bibinfo {volume}
  {16}},\ \bibinfo {pages} {16729--16736} (\bibinfo {year} {2025})}\BibitemShut
  {NoStop}%
\bibitem [{\citenamefont {Breidbach}\ and\ \citenamefont
  {Cederbaum}(2003)}]{cm-breidbach-2003}%
  \BibitemOpen
  \bibfield  {author} {\bibinfo {author} {\bibfnamefont {J.}~\bibnamefont
  {Breidbach}}\ and\ \bibinfo {author} {\bibfnamefont {L.~S.}\ \bibnamefont
  {Cederbaum}},\ }\bibfield  {title} {\enquote {\bibinfo {title} {{Migration of
  holes: Formalism, mechanisms, and illustrative applications}},}\ }\href
  {https://doi.org/10.1063/1.1540618} {\bibfield  {journal} {\bibinfo
  {journal} {J. Chem. Phys.}\ }\textbf {\bibinfo {volume} {118}},\ \bibinfo
  {pages} {3983--3996} (\bibinfo {year} {2003})}\BibitemShut {NoStop}%
\bibitem [{\citenamefont {Breidbach}\ and\ \citenamefont
  {Cederbaum}(2007)}]{cm-breidbach-num-2007}%
  \BibitemOpen
  \bibfield  {author} {\bibinfo {author} {\bibfnamefont {J.}~\bibnamefont
  {Breidbach}}\ and\ \bibinfo {author} {\bibfnamefont {L.~S.}\ \bibnamefont
  {Cederbaum}},\ }\bibfield  {title} {\enquote {\bibinfo {title} {{Migration of
  holes: Numerical algorithms and implementation}},}\ }\href
  {https://doi.org/10.1063/1.2428292} {\bibfield  {journal} {\bibinfo
  {journal} {J. Chem. Phys.}\ }\textbf {\bibinfo {volume} {126}},\ \bibinfo
  {pages} {034101} (\bibinfo {year} {2007})}\BibitemShut {NoStop}%
\bibitem [{\citenamefont {Folorunso}\ \emph {et~al.}(2021)\citenamefont
  {Folorunso}, \citenamefont {Bruner}, \citenamefont {Mauger}, \citenamefont
  {Hamer}, \citenamefont {Hernandez}, \citenamefont {Jones}, \citenamefont
  {DiMauro}, \citenamefont {Gaarde}, \citenamefont {Schafer},\ and\
  \citenamefont {Lopata}}]{cm-molmode-2021}%
  \BibitemOpen
  \bibfield  {author} {\bibinfo {author} {\bibfnamefont {A.~S.}\ \bibnamefont
  {Folorunso}}, \bibinfo {author} {\bibfnamefont {A.}~\bibnamefont {Bruner}},
  \bibinfo {author} {\bibfnamefont {F.}~\bibnamefont {Mauger}}, \bibinfo
  {author} {\bibfnamefont {K.~A.}\ \bibnamefont {Hamer}}, \bibinfo {author}
  {\bibfnamefont {S.}~\bibnamefont {Hernandez}}, \bibinfo {author}
  {\bibfnamefont {R.~R.}\ \bibnamefont {Jones}}, \bibinfo {author}
  {\bibfnamefont {L.~F.}\ \bibnamefont {DiMauro}}, \bibinfo {author}
  {\bibfnamefont {M.~B.}\ \bibnamefont {Gaarde}}, \bibinfo {author}
  {\bibfnamefont {K.~J.}\ \bibnamefont {Schafer}},\ and\ \bibinfo {author}
  {\bibfnamefont {K.}~\bibnamefont {Lopata}},\ }\bibfield  {title} {\enquote
  {\bibinfo {title} {Molecular modes of attosecond charge migration},}\ }\href
  {https://doi.org/10.1103/PhysRevLett.126.133002} {\bibfield  {journal}
  {\bibinfo  {journal} {Phys. Rev. Lett.}\ }\textbf {\bibinfo {volume} {126}},\
  \bibinfo {pages} {133002} (\bibinfo {year} {2021})}\BibitemShut {NoStop}%
\bibitem [{\citenamefont {Folorunso}\ \emph {et~al.}(2023)\citenamefont
  {Folorunso}, \citenamefont {Mauger}, \citenamefont {Hamer}, \citenamefont
  {Jayasinghe}, \citenamefont {Wahyutama}, \citenamefont {Ragains},
  \citenamefont {Jones}, \citenamefont {DiMauro}, \citenamefont {Gaarde},
  \citenamefont {Schafer},\ and\ \citenamefont {Lopata}}]{cm-attochem-2023}%
  \BibitemOpen
  \bibfield  {author} {\bibinfo {author} {\bibfnamefont {A.~S.}\ \bibnamefont
  {Folorunso}}, \bibinfo {author} {\bibfnamefont {F.}~\bibnamefont {Mauger}},
  \bibinfo {author} {\bibfnamefont {K.~A.}\ \bibnamefont {Hamer}}, \bibinfo
  {author} {\bibfnamefont {D.~D.}\ \bibnamefont {Jayasinghe}}, \bibinfo
  {author} {\bibfnamefont {I.~S.}\ \bibnamefont {Wahyutama}}, \bibinfo {author}
  {\bibfnamefont {J.~R.}\ \bibnamefont {Ragains}}, \bibinfo {author}
  {\bibfnamefont {R.~R.}\ \bibnamefont {Jones}}, \bibinfo {author}
  {\bibfnamefont {L.~F.}\ \bibnamefont {DiMauro}}, \bibinfo {author}
  {\bibfnamefont {M.~B.}\ \bibnamefont {Gaarde}}, \bibinfo {author}
  {\bibfnamefont {K.~J.}\ \bibnamefont {Schafer}},\ and\ \bibinfo {author}
  {\bibfnamefont {K.}~\bibnamefont {Lopata}},\ }\bibfield  {title} {\enquote
  {\bibinfo {title} {Attochemistry regulation of charge migration},}\ }\href
  {https://doi.org/10.1021/acs.jpca.3c00568} {\bibfield  {journal} {\bibinfo
  {journal} {J. Phys. Chem. A}\ }\textbf {\bibinfo {volume} {127}},\ \bibinfo
  {pages} {1894--1900} (\bibinfo {year} {2023})}\BibitemShut {NoStop}%
\bibitem [{\citenamefont {Hua}, \citenamefont {Kurkowski},\ and\ \citenamefont
  {Lopata}(2025)}]{cm-core-hole-shape-2025}%
  \BibitemOpen
  \bibfield  {author} {\bibinfo {author} {\bibfnamefont {T.}~\bibnamefont
  {Hua}}, \bibinfo {author} {\bibfnamefont {L.}~\bibnamefont {Kurkowski}},\
  and\ \bibinfo {author} {\bibfnamefont {K.}~\bibnamefont {Lopata}},\
  }\bibfield  {title} {\enquote {\bibinfo {title} {The effect of core-hole
  shape on attosecond valence electron dynamics},}\ }\href
  {https://doi.org/10.1021/acs.jpca.5c01706} {\bibfield  {journal} {\bibinfo
  {journal} {J. Phys. Chem. A}\ }\textbf {\bibinfo {volume} {129}},\ \bibinfo
  {pages} {7742--7750} (\bibinfo {year} {2025})}\BibitemShut {NoStop}%
\bibitem [{\citenamefont {Hamer}\ \emph {et~al.}(2025)\citenamefont {Hamer},
  \citenamefont {Mauger}, \citenamefont {Lopata}, \citenamefont {Schafer},\
  and\ \citenamefont {Gaarde}}]{cm-localion-2025}%
  \BibitemOpen
  \bibfield  {author} {\bibinfo {author} {\bibfnamefont {K.~A.}\ \bibnamefont
  {Hamer}}, \bibinfo {author} {\bibfnamefont {F.}~\bibnamefont {Mauger}},
  \bibinfo {author} {\bibfnamefont {K.}~\bibnamefont {Lopata}}, \bibinfo
  {author} {\bibfnamefont {K.~J.}\ \bibnamefont {Schafer}},\ and\ \bibinfo
  {author} {\bibfnamefont {M.~B.}\ \bibnamefont {Gaarde}},\ }\bibfield  {title}
  {\enquote {\bibinfo {title} {Strong-field ionization with few-cycle,
  midinfrared laser pulses inducing a localized ionization followed by
  long-lasting charge migration in halogenated organic molecules},}\ }\href
  {https://doi.org/10.1103/PhysRevA.111.L011101} {\bibfield  {journal}
  {\bibinfo  {journal} {Phys. Rev. A}\ }\textbf {\bibinfo {volume} {111}},\
  \bibinfo {pages} {L011101} (\bibinfo {year} {2025})}\BibitemShut {NoStop}%
\bibitem [{\citenamefont {Mendive-Tapia}\ \emph {et~al.}(2013)\citenamefont
  {Mendive-Tapia}, \citenamefont {Vacher}, \citenamefont {Bearpark},\ and\
  \citenamefont {Robb}}]{elnuc-cm-ct-cone-2013}%
  \BibitemOpen
  \bibfield  {author} {\bibinfo {author} {\bibfnamefont {D.}~\bibnamefont
  {Mendive-Tapia}}, \bibinfo {author} {\bibfnamefont {M.}~\bibnamefont
  {Vacher}}, \bibinfo {author} {\bibfnamefont {M.~J.}\ \bibnamefont
  {Bearpark}},\ and\ \bibinfo {author} {\bibfnamefont {M.~A.}\ \bibnamefont
  {Robb}},\ }\bibfield  {title} {\enquote {\bibinfo {title} {Coupled
  electron-nuclear dynamics: Charge migration and charge transfer initiated
  near a conical intersection},}\ }\href {https://doi.org/10.1063/1.4815914}
  {\bibfield  {journal} {\bibinfo  {journal} {J. Chem. Phys.}\ }\textbf
  {\bibinfo {volume} {139}},\ \bibinfo {pages} {044110} (\bibinfo {year}
  {2013})}\BibitemShut {NoStop}%
\bibitem [{\citenamefont {Khalili}\ \emph {et~al.}(2021)\citenamefont
  {Khalili}, \citenamefont {Vafaee}, \citenamefont {Cho},\ and\ \citenamefont
  {Shokri}}]{cm-caffeine-2021}%
  \BibitemOpen
  \bibfield  {author} {\bibinfo {author} {\bibfnamefont {F.}~\bibnamefont
  {Khalili}}, \bibinfo {author} {\bibfnamefont {M.}~\bibnamefont {Vafaee}},
  \bibinfo {author} {\bibfnamefont {D.}~\bibnamefont {Cho}},\ and\ \bibinfo
  {author} {\bibfnamefont {B.}~\bibnamefont {Shokri}},\ }\bibfield  {title}
  {\enquote {\bibinfo {title} {Charge migration in caffeine: A real-time
  time-dependent density functional theory study},}\ }\href
  {https://doi.org/https://doi.org/10.1002/qua.26754} {\bibfield  {journal}
  {\bibinfo  {journal} {Int. J. Quantum Chem.}\ }\textbf {\bibinfo {volume}
  {121}},\ \bibinfo {pages} {e26754} (\bibinfo {year} {2021})}\BibitemShut
  {NoStop}%
\bibitem [{\citenamefont {Wahyutama}\ and\ \citenamefont
  {Larsson}(2024)}]{wahyutama-tddmrg-2024}%
  \BibitemOpen
  \bibfield  {author} {\bibinfo {author} {\bibfnamefont {I.~S.}\ \bibnamefont
  {Wahyutama}}\ and\ \bibinfo {author} {\bibfnamefont {H.~R.}\ \bibnamefont
  {Larsson}},\ }\bibfield  {title} {\enquote {\bibinfo {title} {Simulating
  real-time molecular electron dynamics efficiently using the time-dependent
  density matrix renormalization group},}\ }\href
  {https://doi.org/10.1021/acs.jctc.4c01185} {\bibfield  {journal} {\bibinfo
  {journal} {J. Chem. Theory Comput.}\ }\textbf {\bibinfo {volume} {20}},\
  \bibinfo {pages} {9814--9831} (\bibinfo {year} {2024})}\BibitemShut {NoStop}%
\bibitem [{\citenamefont {Pohl}, \citenamefont {Hermann},\ and\ \citenamefont
  {Tremblay}(2017)}]{analyze-n-dyn-2017}%
  \BibitemOpen
  \bibfield  {author} {\bibinfo {author} {\bibfnamefont {V.}~\bibnamefont
  {Pohl}}, \bibinfo {author} {\bibfnamefont {G.}~\bibnamefont {Hermann}},\ and\
  \bibinfo {author} {\bibfnamefont {J.~C.}\ \bibnamefont {Tremblay}},\
  }\bibfield  {title} {\enquote {\bibinfo {title} {{An open-source framework
  for analyzing N-electron dynamics. I. Multideterminantal wave functions}},}\
  }\href {https://doi.org/https://doi.org/10.1002/jcc.24792} {\bibfield
  {journal} {\bibinfo  {journal} {J. Comput. Chem.}\ }\textbf {\bibinfo
  {volume} {38}},\ \bibinfo {pages} {1515--1527} (\bibinfo {year}
  {2017})}\BibitemShut {NoStop}%
\bibitem [{\citenamefont {Hermann}\ \emph {et~al.}(2020)\citenamefont
  {Hermann}, \citenamefont {Pohl}, \citenamefont {Dixit},\ and\ \citenamefont
  {Tremblay}}]{probe-flux-trxrs-2020}%
  \BibitemOpen
  \bibfield  {author} {\bibinfo {author} {\bibfnamefont {G.}~\bibnamefont
  {Hermann}}, \bibinfo {author} {\bibfnamefont {V.}~\bibnamefont {Pohl}},
  \bibinfo {author} {\bibfnamefont {G.}~\bibnamefont {Dixit}},\ and\ \bibinfo
  {author} {\bibfnamefont {J.~C.}\ \bibnamefont {Tremblay}},\ }\bibfield
  {title} {\enquote {\bibinfo {title} {{Probing Electronic Fluxes via
  Time-Resolved X-Ray Scattering}},}\ }\href
  {https://doi.org/10.1103/PhysRevLett.124.013002} {\bibfield  {journal}
  {\bibinfo  {journal} {Phys. Rev. Lett.}\ }\textbf {\bibinfo {volume} {124}},\
  \bibinfo {pages} {013002} (\bibinfo {year} {2020})}\BibitemShut {NoStop}%
\bibitem [{\citenamefont {Meng}\ \emph {et~al.}(2023)\citenamefont {Meng},
  \citenamefont {Wang}, \citenamefont {Zhang},\ and\ \citenamefont
  {Yang}}]{ultrafast-cm-iccnh-2023}%
  \BibitemOpen
  \bibfield  {author} {\bibinfo {author} {\bibfnamefont {Y.}~\bibnamefont
  {Meng}}, \bibinfo {author} {\bibfnamefont {H.}~\bibnamefont {Wang}}, \bibinfo
  {author} {\bibfnamefont {Y.}~\bibnamefont {Zhang}},\ and\ \bibinfo {author}
  {\bibfnamefont {Y.}~\bibnamefont {Yang}},\ }\bibfield  {title} {\enquote
  {\bibinfo {title} {{Ultrafast charge migration in ionized iodo-alkyne chain
  I(CC)$_n$H$^+$}},}\ }\href {https://doi.org/10.1063/5.0142214} {\bibfield
  {journal} {\bibinfo  {journal} {AIP Adv.}\ }\textbf {\bibinfo {volume}
  {13}},\ \bibinfo {pages} {045301} (\bibinfo {year} {2023})}\BibitemShut
  {NoStop}%
\bibitem [{\citenamefont {Wang}\ \emph
  {et~al.}(2025{\natexlab{a}})\citenamefont {Wang}, \citenamefont {Ren},
  \citenamefont {Ren}, \citenamefont {Xiao}, \citenamefont {Jia},\ and\
  \citenamefont {Yang}}]{cm-coherence-increase-2025}%
  \BibitemOpen
  \bibfield  {author} {\bibinfo {author} {\bibfnamefont {H.}~\bibnamefont
  {Wang}}, \bibinfo {author} {\bibfnamefont {X.}~\bibnamefont {Ren}}, \bibinfo
  {author} {\bibfnamefont {X.}~\bibnamefont {Ren}}, \bibinfo {author}
  {\bibfnamefont {L.}~\bibnamefont {Xiao}}, \bibinfo {author} {\bibfnamefont
  {S.}~\bibnamefont {Jia}},\ and\ \bibinfo {author} {\bibfnamefont
  {Y.}~\bibnamefont {Yang}},\ }\bibfield  {title} {\enquote {\bibinfo {title}
  {{Coherences of charge migration increasing with molecular size: A case study
  for chloro-alkyne ions H(CC)$_n$Cl$^+$}},}\ }\href
  {https://doi.org/10.1103/PhysRevResearch.7.L022003} {\bibfield  {journal}
  {\bibinfo  {journal} {Phys. Rev. Res.}\ }\textbf {\bibinfo {volume} {7}},\
  \bibinfo {pages} {L022003} (\bibinfo {year}
  {2025}{\natexlab{a}})}\BibitemShut {NoStop}%
\bibitem [{\citenamefont {Wörner}\ \emph {et~al.}(2017)\citenamefont
  {Wörner}, \citenamefont {Arrell}, \citenamefont {Banerji}, \citenamefont
  {Cannizzo}, \citenamefont {Chergui}, \citenamefont {Das}, \citenamefont
  {Hamm}, \citenamefont {Keller}, \citenamefont {Kraus}, \citenamefont
  {Liberatore}, \citenamefont {Lopez-Tarifa}, \citenamefont {Lucchini},
  \citenamefont {Meuwly}, \citenamefont {Milne}, \citenamefont {Moser},
  \citenamefont {Rothlisberger}, \citenamefont {Smolentsev}, \citenamefont
  {Teuscher}, \citenamefont {van Bokhoven},\ and\ \citenamefont
  {Wenger}}]{cm-ct-review-2017}%
  \BibitemOpen
  \bibfield  {author} {\bibinfo {author} {\bibfnamefont {H.~J.}\ \bibnamefont
  {Wörner}}, \bibinfo {author} {\bibfnamefont {C.~A.}\ \bibnamefont {Arrell}},
  \bibinfo {author} {\bibfnamefont {N.}~\bibnamefont {Banerji}}, \bibinfo
  {author} {\bibfnamefont {A.}~\bibnamefont {Cannizzo}}, \bibinfo {author}
  {\bibfnamefont {M.}~\bibnamefont {Chergui}}, \bibinfo {author} {\bibfnamefont
  {A.~K.}\ \bibnamefont {Das}}, \bibinfo {author} {\bibfnamefont
  {P.}~\bibnamefont {Hamm}}, \bibinfo {author} {\bibfnamefont {U.}~\bibnamefont
  {Keller}}, \bibinfo {author} {\bibfnamefont {P.~M.}\ \bibnamefont {Kraus}},
  \bibinfo {author} {\bibfnamefont {E.}~\bibnamefont {Liberatore}}, \bibinfo
  {author} {\bibfnamefont {P.}~\bibnamefont {Lopez-Tarifa}}, \bibinfo {author}
  {\bibfnamefont {M.}~\bibnamefont {Lucchini}}, \bibinfo {author}
  {\bibfnamefont {M.}~\bibnamefont {Meuwly}}, \bibinfo {author} {\bibfnamefont
  {C.}~\bibnamefont {Milne}}, \bibinfo {author} {\bibfnamefont {J.-E.}\
  \bibnamefont {Moser}}, \bibinfo {author} {\bibfnamefont {U.}~\bibnamefont
  {Rothlisberger}}, \bibinfo {author} {\bibfnamefont {G.}~\bibnamefont
  {Smolentsev}}, \bibinfo {author} {\bibfnamefont {J.}~\bibnamefont
  {Teuscher}}, \bibinfo {author} {\bibfnamefont {J.~A.}\ \bibnamefont {van
  Bokhoven}},\ and\ \bibinfo {author} {\bibfnamefont {O.}~\bibnamefont
  {Wenger}},\ }\bibfield  {title} {\enquote {\bibinfo {title} {Charge migration
  and charge transfer in molecular systems},}\ }\href
  {https://doi.org/10.1063/1.4996505} {\bibfield  {journal} {\bibinfo
  {journal} {Struct. Dyn.}\ }\textbf {\bibinfo {volume} {4}},\ \bibinfo {pages}
  {061508} (\bibinfo {year} {2017})}\BibitemShut {NoStop}%
\bibitem [{\citenamefont {Despr\'e}\ and\ \citenamefont
  {Kuleff}(2022)}]{cm-init-corr-bands-2022}%
  \BibitemOpen
  \bibfield  {author} {\bibinfo {author} {\bibfnamefont {V.}~\bibnamefont
  {Despr\'e}}\ and\ \bibinfo {author} {\bibfnamefont {A.~I.}\ \bibnamefont
  {Kuleff}},\ }\bibfield  {title} {\enquote {\bibinfo {title}
  {Correlation-driven charge migration as an initial step of the dynamics in
  correlation bands},}\ }\href {https://doi.org/10.1103/PhysRevA.106.L021501}
  {\bibfield  {journal} {\bibinfo  {journal} {Phys. Rev. A}\ }\textbf {\bibinfo
  {volume} {106}},\ \bibinfo {pages} {L021501} (\bibinfo {year}
  {2022})}\BibitemShut {NoStop}%
\bibitem [{\citenamefont {Matselyukh}\ \emph {et~al.}(2022)\citenamefont
  {Matselyukh}, \citenamefont {Despr{\'e}}, \citenamefont {Golubev},
  \citenamefont {Kuleff},\ and\ \citenamefont
  {W{\"o}rner}}]{decohere-revival-attocm-2022}%
  \BibitemOpen
  \bibfield  {author} {\bibinfo {author} {\bibfnamefont {D.~T.}\ \bibnamefont
  {Matselyukh}}, \bibinfo {author} {\bibfnamefont {V.}~\bibnamefont
  {Despr{\'e}}}, \bibinfo {author} {\bibfnamefont {N.~V.}\ \bibnamefont
  {Golubev}}, \bibinfo {author} {\bibfnamefont {A.~I.}\ \bibnamefont
  {Kuleff}},\ and\ \bibinfo {author} {\bibfnamefont {H.~J.}\ \bibnamefont
  {W{\"o}rner}},\ }\bibfield  {title} {\enquote {\bibinfo {title} {Decoherence
  and revival in attosecond charge migration driven by non-adiabatic
  dynamics},}\ }\href {https://doi.org/10.1038/s41567-022-01690-0} {\bibfield
  {journal} {\bibinfo  {journal} {Nat. Phys.}\ }\textbf {\bibinfo {volume}
  {18}},\ \bibinfo {pages} {1206--1213} (\bibinfo {year} {2022})}\BibitemShut
  {NoStop}%
\bibitem [{\citenamefont {Schwickert}\ \emph {et~al.}(2022)\citenamefont
  {Schwickert}, \citenamefont {Ruberti}, \citenamefont {Kolorenč},
  \citenamefont {Usenko}, \citenamefont {Przystawik}, \citenamefont {Baev},
  \citenamefont {Baev}, \citenamefont {Braune}, \citenamefont {Bocklage},
  \citenamefont {Czwalinna}, \citenamefont {Deinert}, \citenamefont
  {Düsterer}, \citenamefont {Hans}, \citenamefont {Hartmann}, \citenamefont
  {Haunhorst}, \citenamefont {Kuhlmann}, \citenamefont {Palutke}, \citenamefont
  {Röhlsberger}, \citenamefont {Rönsch-Schulenburg}, \citenamefont {Schmidt},
  \citenamefont {Toleikis}, \citenamefont {Viefhaus}, \citenamefont {Martins},
  \citenamefont {Knie}, \citenamefont {Kip}, \citenamefont {Averbukh},
  \citenamefont {Marangos},\ and\ \citenamefont
  {Laarmann}}]{quantum-chr-glycine-2022}%
  \BibitemOpen
  \bibfield  {author} {\bibinfo {author} {\bibfnamefont {D.}~\bibnamefont
  {Schwickert}}, \bibinfo {author} {\bibfnamefont {M.}~\bibnamefont {Ruberti}},
  \bibinfo {author} {\bibfnamefont {P.}~\bibnamefont {Kolorenč}}, \bibinfo
  {author} {\bibfnamefont {S.}~\bibnamefont {Usenko}}, \bibinfo {author}
  {\bibfnamefont {A.}~\bibnamefont {Przystawik}}, \bibinfo {author}
  {\bibfnamefont {K.}~\bibnamefont {Baev}}, \bibinfo {author} {\bibfnamefont
  {I.}~\bibnamefont {Baev}}, \bibinfo {author} {\bibfnamefont {M.}~\bibnamefont
  {Braune}}, \bibinfo {author} {\bibfnamefont {L.}~\bibnamefont {Bocklage}},
  \bibinfo {author} {\bibfnamefont {M.~K.}\ \bibnamefont {Czwalinna}}, \bibinfo
  {author} {\bibfnamefont {S.}~\bibnamefont {Deinert}}, \bibinfo {author}
  {\bibfnamefont {S.}~\bibnamefont {Düsterer}}, \bibinfo {author}
  {\bibfnamefont {A.}~\bibnamefont {Hans}}, \bibinfo {author} {\bibfnamefont
  {G.}~\bibnamefont {Hartmann}}, \bibinfo {author} {\bibfnamefont
  {C.}~\bibnamefont {Haunhorst}}, \bibinfo {author} {\bibfnamefont
  {M.}~\bibnamefont {Kuhlmann}}, \bibinfo {author} {\bibfnamefont
  {S.}~\bibnamefont {Palutke}}, \bibinfo {author} {\bibfnamefont
  {R.}~\bibnamefont {Röhlsberger}}, \bibinfo {author} {\bibfnamefont
  {J.}~\bibnamefont {Rönsch-Schulenburg}}, \bibinfo {author} {\bibfnamefont
  {P.}~\bibnamefont {Schmidt}}, \bibinfo {author} {\bibfnamefont
  {S.}~\bibnamefont {Toleikis}}, \bibinfo {author} {\bibfnamefont
  {J.}~\bibnamefont {Viefhaus}}, \bibinfo {author} {\bibfnamefont
  {M.}~\bibnamefont {Martins}}, \bibinfo {author} {\bibfnamefont
  {A.}~\bibnamefont {Knie}}, \bibinfo {author} {\bibfnamefont {D.}~\bibnamefont
  {Kip}}, \bibinfo {author} {\bibfnamefont {V.}~\bibnamefont {Averbukh}},
  \bibinfo {author} {\bibfnamefont {J.~P.}\ \bibnamefont {Marangos}},\ and\
  \bibinfo {author} {\bibfnamefont {T.}~\bibnamefont {Laarmann}},\ }\bibfield
  {title} {\enquote {\bibinfo {title} {{Electronic quantum coherence in glycine
  molecules probed with ultrashort X-ray pulses in real time}},}\ }\href
  {https://doi.org/10.1126/sciadv.abn6848} {\bibfield  {journal} {\bibinfo
  {journal} {Sci. Adv.}\ }\textbf {\bibinfo {volume} {8}},\ \bibinfo {pages}
  {eabn6848} (\bibinfo {year} {2022})}\BibitemShut {NoStop}%
\bibitem [{\citenamefont {Calegari}\ and\ \citenamefont
  {Martin}(2023)}]{attochem-questions-2023}%
  \BibitemOpen
  \bibfield  {author} {\bibinfo {author} {\bibfnamefont {F.}~\bibnamefont
  {Calegari}}\ and\ \bibinfo {author} {\bibfnamefont {F.}~\bibnamefont
  {Martin}},\ }\bibfield  {title} {\enquote {\bibinfo {title} {Open questions
  in attochemistry},}\ }\href {https://doi.org/10.1038/s42004-023-00989-0}
  {\bibfield  {journal} {\bibinfo  {journal} {Commun. Chem.}\ }\textbf
  {\bibinfo {volume} {6}},\ \bibinfo {pages} {184} (\bibinfo {year}
  {2023})}\BibitemShut {NoStop}%
\bibitem [{\citenamefont {Wanie}\ \emph {et~al.}(2024)\citenamefont {Wanie},
  \citenamefont {Bloch}, \citenamefont {M{\aa}nsson}, \citenamefont {Colaizzi},
  \citenamefont {Ryabchuk}, \citenamefont {Saraswathula}, \citenamefont
  {Ordonez}, \citenamefont {Ayuso}, \citenamefont {Smirnova}, \citenamefont
  {Trabattoni} \emph {et~al.}}]{capture-chiral-dyn-2024}%
  \BibitemOpen
  \bibfield  {author} {\bibinfo {author} {\bibfnamefont {V.}~\bibnamefont
  {Wanie}}, \bibinfo {author} {\bibfnamefont {E.}~\bibnamefont {Bloch}},
  \bibinfo {author} {\bibfnamefont {E.~P.}\ \bibnamefont {M{\aa}nsson}},
  \bibinfo {author} {\bibfnamefont {L.}~\bibnamefont {Colaizzi}}, \bibinfo
  {author} {\bibfnamefont {S.}~\bibnamefont {Ryabchuk}}, \bibinfo {author}
  {\bibfnamefont {K.}~\bibnamefont {Saraswathula}}, \bibinfo {author}
  {\bibfnamefont {A.~F.}\ \bibnamefont {Ordonez}}, \bibinfo {author}
  {\bibfnamefont {D.}~\bibnamefont {Ayuso}}, \bibinfo {author} {\bibfnamefont
  {O.}~\bibnamefont {Smirnova}}, \bibinfo {author} {\bibfnamefont
  {A.}~\bibnamefont {Trabattoni}}, \emph {et~al.},\ }\bibfield  {title}
  {\enquote {\bibinfo {title} {{Capturing electron-driven chiral dynamics in
  UV-excited molecules}},}\ }\href {https://doi.org/10.1038/s41586-024-07415-y}
  {\bibfield  {journal} {\bibinfo  {journal} {Nature}\ }\textbf {\bibinfo
  {volume} {630}},\ \bibinfo {pages} {109--115} (\bibinfo {year}
  {2024})}\BibitemShut {NoStop}%
\bibitem [{\citenamefont {Haase}\ \emph {et~al.}(2026)\citenamefont {Haase},
  \citenamefont {Manz}, \citenamefont {Paulus}, \citenamefont {Scherlitzki},\
  and\ \citenamefont {Tremblay}}]{atto-chiral-triatom-2026}%
  \BibitemOpen
  \bibfield  {author} {\bibinfo {author} {\bibfnamefont {D.}~\bibnamefont
  {Haase}}, \bibinfo {author} {\bibfnamefont {J.}~\bibnamefont {Manz}},
  \bibinfo {author} {\bibfnamefont {B.}~\bibnamefont {Paulus}}, \bibinfo
  {author} {\bibfnamefont {J.}~\bibnamefont {Scherlitzki}},\ and\ \bibinfo
  {author} {\bibfnamefont {J.~C.}\ \bibnamefont {Tremblay}},\ }\bibfield
  {title} {\enquote {\bibinfo {title} {A simple approach to attosecond
  electronic chirality flips using triatomic molecules},}\ }\href
  {https://doi.org/10.1039/D5CP04637H} {\bibfield  {journal} {\bibinfo
  {journal} {Phys. Chem. Chem. Phys.}\ }\textbf {\bibinfo {volume} {28}},\
  \bibinfo {pages} {5175--5180} (\bibinfo {year} {2026})}\BibitemShut {NoStop}%
\bibitem [{\citenamefont {Kermack}\ and\ \citenamefont
  {Robinson}(1922)}]{partial-valence-interp-1922}%
  \BibitemOpen
  \bibfield  {author} {\bibinfo {author} {\bibfnamefont {W.~O.}\ \bibnamefont
  {Kermack}}\ and\ \bibinfo {author} {\bibfnamefont {R.}~\bibnamefont
  {Robinson}},\ }\bibfield  {title} {\enquote {\bibinfo {title} {{LI.—An
  explanation of the property of induced polarity of atoms and an
  interpretation of the theory of partial valencies on an electronic basis}},}\
  }\href {https://doi.org/10.1039/CT9222100427} {\bibfield  {journal} {\bibinfo
   {journal} {J. Chem. Soc. Trans.}\ }\textbf {\bibinfo {volume} {121}},\
  \bibinfo {pages} {427--440} (\bibinfo {year} {1922})}\BibitemShut {NoStop}%
\bibitem [{\citenamefont {Bader}(1990)}]{bader1990atoms-1990}%
  \BibitemOpen
  \bibfield  {author} {\bibinfo {author} {\bibfnamefont {R.~F.~W.}\
  \bibnamefont {Bader}},\ }\href
  {https://books.google.ca/books?id=up1pQgAACAAJ} {\emph {\bibinfo {title}
  {Atoms in Molecules: A Quantum Theory}}},\ International series of monographs
  on chemistry\ (\bibinfo  {publisher} {Oxford Univesity Press},\ \bibinfo
  {year} {1990})\BibitemShut {NoStop}%
\bibitem [{\citenamefont {Rodríguez}(2013)}]{efficient-qtaim-2013}%
  \BibitemOpen
  \bibfield  {author} {\bibinfo {author} {\bibfnamefont {J.~I.}\ \bibnamefont
  {Rodríguez}},\ }\bibfield  {title} {\enquote {\bibinfo {title} {An efficient
  method for computing the qtaim topology of a scalar field: The electron
  density case},}\ }\href {https://doi.org/https://doi.org/10.1002/jcc.23180}
  {\bibfield  {journal} {\bibinfo  {journal} {J. Comput. Chem.}\ }\textbf
  {\bibinfo {volume} {34}},\ \bibinfo {pages} {681--686} (\bibinfo {year}
  {2013})}\BibitemShut {NoStop}%
\bibitem [{\citenamefont {Kitaura}\ and\ \citenamefont
  {Morokuma}(1976)}]{kitaura-morokuma-eda-1976}%
  \BibitemOpen
  \bibfield  {author} {\bibinfo {author} {\bibfnamefont {K.}~\bibnamefont
  {Kitaura}}\ and\ \bibinfo {author} {\bibfnamefont {K.}~\bibnamefont
  {Morokuma}},\ }\bibfield  {title} {\enquote {\bibinfo {title} {{A new energy
  decomposition scheme for molecular interactions within the Hartree-Fock
  approximation}},}\ }\href {https://doi.org/10.1002/qua.560100211} {\bibfield
  {journal} {\bibinfo  {journal} {Int. J. Quantum Chem.}\ }\textbf {\bibinfo
  {volume} {10}},\ \bibinfo {pages} {325--340} (\bibinfo {year}
  {1976})}\BibitemShut {NoStop}%
\bibitem [{\citenamefont {Glendening}(1996)}]{neda-1996}%
  \BibitemOpen
  \bibfield  {author} {\bibinfo {author} {\bibfnamefont {E.~D.}\ \bibnamefont
  {Glendening}},\ }\bibfield  {title} {\enquote {\bibinfo {title} {Natural
  energy decomposition analysis: Explicit evaluation of electrostatic and
  polarization effects with application to aqueous clusters of alkali metal
  cations and neutrals},}\ }\href {https://doi.org/10.1021/ja951834y}
  {\bibfield  {journal} {\bibinfo  {journal} {J. Am. Chem. Soc.}\ }\textbf
  {\bibinfo {volume} {118}},\ \bibinfo {pages} {2473--2482} (\bibinfo {year}
  {1996})}\BibitemShut {NoStop}%
\bibitem [{\citenamefont {Su}\ and\ \citenamefont {Li}(2009)}]{lmoeda-2009}%
  \BibitemOpen
  \bibfield  {author} {\bibinfo {author} {\bibfnamefont {P.}~\bibnamefont
  {Su}}\ and\ \bibinfo {author} {\bibfnamefont {H.}~\bibnamefont {Li}},\
  }\bibfield  {title} {\enquote {\bibinfo {title} {Energy decomposition
  analysis of covalent bonds and intermolecular interactions},}\ }\href
  {https://doi.org/10.1063/1.3159673} {\bibfield  {journal} {\bibinfo
  {journal} {J. Chem. Phys.}\ }\textbf {\bibinfo {volume} {131}},\ \bibinfo
  {pages} {014102} (\bibinfo {year} {2009})}\BibitemShut {NoStop}%
\bibitem [{\citenamefont {Su}\ \emph {et~al.}(2014)\citenamefont {Su},
  \citenamefont {Jiang}, \citenamefont {Chen},\ and\ \citenamefont
  {Wu}}]{gkseda-2014}%
  \BibitemOpen
  \bibfield  {author} {\bibinfo {author} {\bibfnamefont {P.}~\bibnamefont
  {Su}}, \bibinfo {author} {\bibfnamefont {Z.}~\bibnamefont {Jiang}}, \bibinfo
  {author} {\bibfnamefont {Z.}~\bibnamefont {Chen}},\ and\ \bibinfo {author}
  {\bibfnamefont {W.}~\bibnamefont {Wu}},\ }\bibfield  {title} {\enquote
  {\bibinfo {title} {{Energy Decomposition Scheme Based on the Generalized
  Kohn–Sham Scheme}},}\ }\href {https://doi.org/10.1021/jp500405s} {\bibfield
   {journal} {\bibinfo  {journal} {J. Phys. Chem. A}\ }\textbf {\bibinfo
  {volume} {118}},\ \bibinfo {pages} {2531--2542} (\bibinfo {year}
  {2014})}\BibitemShut {NoStop}%
\bibitem [{\citenamefont {Horn}, \citenamefont {Mao},\ and\ \citenamefont
  {Head-Gordon}(2016)}]{almoeda-2016}%
  \BibitemOpen
  \bibfield  {author} {\bibinfo {author} {\bibfnamefont {P.~R.}\ \bibnamefont
  {Horn}}, \bibinfo {author} {\bibfnamefont {Y.}~\bibnamefont {Mao}},\ and\
  \bibinfo {author} {\bibfnamefont {M.}~\bibnamefont {Head-Gordon}},\
  }\bibfield  {title} {\enquote {\bibinfo {title} {Probing non-covalent
  interactions with a second generation energy decomposition analysis using
  absolutely localized molecular orbitals},}\ }\href
  {https://doi.org/10.1039/C6CP03784D} {\bibfield  {journal} {\bibinfo
  {journal} {Phys. Chem. Chem. Phys.}\ }\textbf {\bibinfo {volume} {18}},\
  \bibinfo {pages} {23067--23079} (\bibinfo {year} {2016})}\BibitemShut
  {NoStop}%
\bibitem [{\citenamefont {Lu}\ and\ \citenamefont {Chen}(2023)}]{sobeda-2023}%
  \BibitemOpen
  \bibfield  {author} {\bibinfo {author} {\bibfnamefont {T.}~\bibnamefont
  {Lu}}\ and\ \bibinfo {author} {\bibfnamefont {Q.}~\bibnamefont {Chen}},\
  }\bibfield  {title} {\enquote {\bibinfo {title} {Simple, efficient, and
  universal energy decomposition analysis method based on dispersion-corrected
  density functional theory},}\ }\href
  {https://doi.org/10.1021/acs.jpca.3c04374} {\bibfield  {journal} {\bibinfo
  {journal} {J. Phys. Chem. A}\ }\textbf {\bibinfo {volume} {127}},\ \bibinfo
  {pages} {7023--7035} (\bibinfo {year} {2023})}\BibitemShut {NoStop}%
\bibitem [{\citenamefont {Liu}\ \emph {et~al.}(2020)\citenamefont {Liu},
  \citenamefont {Kilby}, \citenamefont {Frankcombe},\ and\ \citenamefont
  {Schmidt}}]{elstru-benzene-tiling-2020}%
  \BibitemOpen
  \bibfield  {author} {\bibinfo {author} {\bibfnamefont {Y.}~\bibnamefont
  {Liu}}, \bibinfo {author} {\bibfnamefont {P.}~\bibnamefont {Kilby}}, \bibinfo
  {author} {\bibfnamefont {T.~J.}\ \bibnamefont {Frankcombe}},\ and\ \bibinfo
  {author} {\bibfnamefont {T.~W.}\ \bibnamefont {Schmidt}},\ }\bibfield
  {title} {\enquote {\bibinfo {title} {The electronic structure of benzene from
  a tiling of the correlated 126-dimensional wavefunction},}\ }\href
  {https://doi.org/10.1038/s41467-020-15039-9} {\bibfield  {journal} {\bibinfo
  {journal} {Nat. Commun.}\ }\textbf {\bibinfo {volume} {11}},\ \bibinfo
  {pages} {1210} (\bibinfo {year} {2020})}\BibitemShut {NoStop}%
\bibitem [{\citenamefont {Liu}, \citenamefont {Frankcombe},\ and\ \citenamefont
  {Schmidt}(2022)}]{hitchhiker-wave-function-2022}%
  \BibitemOpen
  \bibfield  {author} {\bibinfo {author} {\bibfnamefont {Y.}~\bibnamefont
  {Liu}}, \bibinfo {author} {\bibfnamefont {T.~J.}\ \bibnamefont
  {Frankcombe}},\ and\ \bibinfo {author} {\bibfnamefont {T.~W.}\ \bibnamefont
  {Schmidt}},\ }\bibfield  {title} {\enquote {\bibinfo {title} {The
  hitchhiker’s guide to the wave function},}\ }\href
  {https://doi.org/10.1021/acs.jpca.1c07869} {\bibfield  {journal} {\bibinfo
  {journal} {J. Phys. Chem. A}\ }\textbf {\bibinfo {volume} {126}},\ \bibinfo
  {pages} {979--991} (\bibinfo {year} {2022})}\BibitemShut {NoStop}%
\bibitem [{\citenamefont {Boguslawski}\ \emph {et~al.}(2012)\citenamefont
  {Boguslawski}, \citenamefont {Tecmer}, \citenamefont {Legeza},\ and\
  \citenamefont {Reiher}}]{entangle-single-multiref-2012}%
  \BibitemOpen
  \bibfield  {author} {\bibinfo {author} {\bibfnamefont {K.}~\bibnamefont
  {Boguslawski}}, \bibinfo {author} {\bibfnamefont {P.}~\bibnamefont {Tecmer}},
  \bibinfo {author} {\bibfnamefont {{\"O}.}~\bibnamefont {Legeza}},\ and\
  \bibinfo {author} {\bibfnamefont {M.}~\bibnamefont {Reiher}},\ }\bibfield
  {title} {\enquote {\bibinfo {title} {Entanglement measures for single- and
  multireference correlation effects},}\ }\href
  {https://doi.org/10.1021/jz301319v} {\bibfield  {journal} {\bibinfo
  {journal} {J. Phys. Chem. Lett.}\ }\textbf {\bibinfo {volume} {3}},\ \bibinfo
  {pages} {3129--3135} (\bibinfo {year} {2012})}\BibitemShut {NoStop}%
\bibitem [{\citenamefont {Szalay}\ \emph {et~al.}(2017)\citenamefont {Szalay},
  \citenamefont {Barcza}, \citenamefont {Szilv{\'a}si}, \citenamefont {Veis},\
  and\ \citenamefont {Legeza}}]{bond-corr-theory-2017}%
  \BibitemOpen
  \bibfield  {author} {\bibinfo {author} {\bibfnamefont {S.}~\bibnamefont
  {Szalay}}, \bibinfo {author} {\bibfnamefont {G.}~\bibnamefont {Barcza}},
  \bibinfo {author} {\bibfnamefont {T.}~\bibnamefont {Szilv{\'a}si}}, \bibinfo
  {author} {\bibfnamefont {L.}~\bibnamefont {Veis}},\ and\ \bibinfo {author}
  {\bibfnamefont {{\"O}.}~\bibnamefont {Legeza}},\ }\bibfield  {title}
  {\enquote {\bibinfo {title} {The correlation theory of the chemical bond},}\
  }\href {https://doi.org/10.1038/s41598-017-02447-z} {\bibfield  {journal}
  {\bibinfo  {journal} {Sci. Rep.}\ }\textbf {\bibinfo {volume} {7}},\ \bibinfo
  {pages} {2237} (\bibinfo {year} {2017})}\BibitemShut {NoStop}%
\bibitem [{\citenamefont {Ding}, \citenamefont {Dünnweber},\ and\
  \citenamefont {Schilling}(2023)}]{entangle-loc-orb-2023}%
  \BibitemOpen
  \bibfield  {author} {\bibinfo {author} {\bibfnamefont {L.}~\bibnamefont
  {Ding}}, \bibinfo {author} {\bibfnamefont {G.}~\bibnamefont {Dünnweber}},\
  and\ \bibinfo {author} {\bibfnamefont {C.}~\bibnamefont {Schilling}},\
  }\bibfield  {title} {\enquote {\bibinfo {title} {Physical entanglement
  between localized orbitals},}\ }\href
  {https://doi.org/10.1088/2058-9565/ad00d9} {\bibfield  {journal} {\bibinfo
  {journal} {Quantum Sci. Technol.}\ }\textbf {\bibinfo {volume} {9}},\
  \bibinfo {pages} {015005} (\bibinfo {year} {2023})}\BibitemShut {NoStop}%
\bibitem [{\citenamefont {Ding}, \citenamefont {Matito},\ and\ \citenamefont
  {Schilling}(2026)}]{meaos-2026}%
  \BibitemOpen
  \bibfield  {author} {\bibinfo {author} {\bibfnamefont {L.}~\bibnamefont
  {Ding}}, \bibinfo {author} {\bibfnamefont {E.}~\bibnamefont {Matito}},\ and\
  \bibinfo {author} {\bibfnamefont {C.}~\bibnamefont {Schilling}},\ }\bibfield
  {title} {\enquote {\bibinfo {title} {Chemical bonding concepts emerge
  naturally from maximally entangled atomic orbitals},}\ }\href
  {https://doi.org/10.1038/s41467-026-73527-w} {\bibfield  {journal} {\bibinfo
  {journal} {Nat. Commun.}\ }\textbf {\bibinfo {volume} {17}},\ \bibinfo
  {pages} {4732} (\bibinfo {year} {2026})}\BibitemShut {NoStop}%
\bibitem [{\citenamefont {Boys}(1960)}]{boys_loc-1960}%
  \BibitemOpen
  \bibfield  {author} {\bibinfo {author} {\bibfnamefont {S.~F.}\ \bibnamefont
  {Boys}},\ }\bibfield  {title} {\enquote {\bibinfo {title} {Construction of
  some molecular orbitals to be approximately invariant for changes from one
  molecule to another},}\ }\href {https://doi.org/10.1103/RevModPhys.32.296}
  {\bibfield  {journal} {\bibinfo  {journal} {Rev. Mod. Phys.}\ }\textbf
  {\bibinfo {volume} {32}},\ \bibinfo {pages} {296--299} (\bibinfo {year}
  {1960})}\BibitemShut {NoStop}%
\bibitem [{\citenamefont {Edmiston}\ and\ \citenamefont
  {Ruedenberg}(1963)}]{er_loc-1963}%
  \BibitemOpen
  \bibfield  {author} {\bibinfo {author} {\bibfnamefont {C.}~\bibnamefont
  {Edmiston}}\ and\ \bibinfo {author} {\bibfnamefont {K.}~\bibnamefont
  {Ruedenberg}},\ }\bibfield  {title} {\enquote {\bibinfo {title} {Localized
  atomic and molecular orbitals},}\ }\href
  {https://doi.org/10.1103/RevModPhys.35.457} {\bibfield  {journal} {\bibinfo
  {journal} {Rev. Mod. Phys.}\ }\textbf {\bibinfo {volume} {35}},\ \bibinfo
  {pages} {457--464} (\bibinfo {year} {1963})}\BibitemShut {NoStop}%
\bibitem [{\citenamefont {Pipek}\ and\ \citenamefont
  {Mezey}(1989)}]{pm_loc-1989}%
  \BibitemOpen
  \bibfield  {author} {\bibinfo {author} {\bibfnamefont {J.}~\bibnamefont
  {Pipek}}\ and\ \bibinfo {author} {\bibfnamefont {P.~G.}\ \bibnamefont
  {Mezey}},\ }\bibfield  {title} {\enquote {\bibinfo {title} {{A fast intrinsic
  localization procedure applicable for \textit{ab initio} and semiempirical
  linear combination of atomic orbital wave functions}},}\ }\href
  {https://doi.org/10.1063/1.456588} {\bibfield  {journal} {\bibinfo  {journal}
  {J. Chem. Phys.}\ }\textbf {\bibinfo {volume} {90}},\ \bibinfo {pages}
  {4916--4926} (\bibinfo {year} {1989})}\BibitemShut {NoStop}%
\bibitem [{\citenamefont {Lee}\ and\ \citenamefont
  {Head-Gordon}(1997)}]{polar-atom-orb-scf-1997}%
  \BibitemOpen
  \bibfield  {author} {\bibinfo {author} {\bibfnamefont {M.~S.}\ \bibnamefont
  {Lee}}\ and\ \bibinfo {author} {\bibfnamefont {M.}~\bibnamefont
  {Head-Gordon}},\ }\bibfield  {title} {\enquote {\bibinfo {title} {Polarized
  atomic orbitals for self-consistent field electronic structure
  calculations},}\ }\href {https://doi.org/10.1063/1.475199} {\bibfield
  {journal} {\bibinfo  {journal} {J. Chem. Phys.}\ }\textbf {\bibinfo {volume}
  {107}},\ \bibinfo {pages} {9085--9095} (\bibinfo {year} {1997})}\BibitemShut
  {NoStop}%
\bibitem [{\citenamefont {Lee}\ and\ \citenamefont
  {Head-Gordon}(2000)}]{polar-atom-orb-2000}%
  \BibitemOpen
  \bibfield  {author} {\bibinfo {author} {\bibfnamefont {M.~S.}\ \bibnamefont
  {Lee}}\ and\ \bibinfo {author} {\bibfnamefont {M.}~\bibnamefont
  {Head-Gordon}},\ }\bibfield  {title} {\enquote {\bibinfo {title} {Extracting
  polarized atomic orbitals from molecular orbital calculations},}\ }\href
  {https://doi.org/https://doi.org/10.1002/(SICI)1097-461X(2000)76:2<169::AID-QUA7>3.0.CO;2-G}
  {\bibfield  {journal} {\bibinfo  {journal} {Int. J. Quantum Chem.}\ }\textbf
  {\bibinfo {volume} {76}},\ \bibinfo {pages} {169--184} (\bibinfo {year}
  {2000})}\BibitemShut {NoStop}%
\bibitem [{\citenamefont {Subotnik}, \citenamefont {Dutoi},\ and\ \citenamefont
  {Head-Gordon}(2005)}]{fast_local_vorb-2005}%
  \BibitemOpen
  \bibfield  {author} {\bibinfo {author} {\bibfnamefont {J.~E.}\ \bibnamefont
  {Subotnik}}, \bibinfo {author} {\bibfnamefont {A.~D.}\ \bibnamefont
  {Dutoi}},\ and\ \bibinfo {author} {\bibfnamefont {M.}~\bibnamefont
  {Head-Gordon}},\ }\bibfield  {title} {\enquote {\bibinfo {title} {Fast
  localized orthonormal virtual orbitals which depend smoothly on nuclear
  coordinates},}\ }\href {https://doi.org/10.1063/1.2033687} {\bibfield
  {journal} {\bibinfo  {journal} {J. Chem. Phys.}\ }\textbf {\bibinfo {volume}
  {123}},\ \bibinfo {pages} {114108} (\bibinfo {year} {2005})}\BibitemShut
  {NoStop}%
\bibitem [{\citenamefont {Laikov}(2011)}]{intrnc-minbas-wvn-2011}%
  \BibitemOpen
  \bibfield  {author} {\bibinfo {author} {\bibfnamefont {D.~N.}\ \bibnamefont
  {Laikov}},\ }\bibfield  {title} {\enquote {\bibinfo {title} {Intrinsic
  minimal atomic basis representation of molecular electronic wavefunctions},}\
  }\href {https://doi.org/https://doi.org/10.1002/qua.22767} {\bibfield
  {journal} {\bibinfo  {journal} {Int. J. Quantum Chem.}\ }\textbf {\bibinfo
  {volume} {111}},\ \bibinfo {pages} {2851--2867} (\bibinfo {year}
  {2011})}\BibitemShut {NoStop}%
\bibitem [{\citenamefont {Knizia}(2013)}]{ibo-2013}%
  \BibitemOpen
  \bibfield  {author} {\bibinfo {author} {\bibfnamefont {G.}~\bibnamefont
  {Knizia}},\ }\bibfield  {title} {\enquote {\bibinfo {title} {Intrinsic atomic
  orbitals: An unbiased bridge between quantum theory and chemical concepts},}\
  }\href {https://doi.org/10.1021/ct400687b} {\bibfield  {journal} {\bibinfo
  {journal} {J. Chem. Theory Comput.}\ }\textbf {\bibinfo {volume} {9}},\
  \bibinfo {pages} {4834--4843} (\bibinfo {year} {2013})}\BibitemShut {NoStop}%
\bibitem [{\citenamefont {West}\ \emph {et~al.}(2013)\citenamefont {West},
  \citenamefont {Schmidt}, \citenamefont {Gordon},\ and\ \citenamefont
  {Ruedenberg}}]{molintrsc_analysis-i-2013}%
  \BibitemOpen
  \bibfield  {author} {\bibinfo {author} {\bibfnamefont {A.~C.}\ \bibnamefont
  {West}}, \bibinfo {author} {\bibfnamefont {M.~W.}\ \bibnamefont {Schmidt}},
  \bibinfo {author} {\bibfnamefont {M.~S.}\ \bibnamefont {Gordon}},\ and\
  \bibinfo {author} {\bibfnamefont {K.}~\bibnamefont {Ruedenberg}},\ }\bibfield
   {title} {\enquote {\bibinfo {title} {{A comprehensive analysis of
  molecule-intrinsic quasi-atomic, bonding, and correlating orbitals. I.
  Hartree-Fock wave functions}},}\ }\href {https://doi.org/10.1063/1.4840776}
  {\bibfield  {journal} {\bibinfo  {journal} {J. Chem. Phys.}\ }\textbf
  {\bibinfo {volume} {139}},\ \bibinfo {pages} {234107} (\bibinfo {year}
  {2013})}\BibitemShut {NoStop}%
\bibitem [{\citenamefont {Knizia}\ and\ \citenamefont
  {Klein}(2015)}]{electron-flow-2015}%
  \BibitemOpen
  \bibfield  {author} {\bibinfo {author} {\bibfnamefont {G.}~\bibnamefont
  {Knizia}}\ and\ \bibinfo {author} {\bibfnamefont {J.~E. M.~N.}\ \bibnamefont
  {Klein}},\ }\bibfield  {title} {\enquote {\bibinfo {title} {Electron flow in
  reaction mechanisms—revealed from first principles},}\ }\href
  {https://doi.org/https://doi.org/10.1002/anie.201410637} {\bibfield
  {journal} {\bibinfo  {journal} {Angew. Chem., Int. Ed.}\ }\textbf {\bibinfo
  {volume} {54}},\ \bibinfo {pages} {5518--5522} (\bibinfo {year}
  {2015})}\BibitemShut {NoStop}%
\bibitem [{\citenamefont {Nunes~dos Santos~Comprido}\ \emph
  {et~al.}(2015)\citenamefont {Nunes~dos Santos~Comprido}, \citenamefont
  {Klein}, \citenamefont {Knizia}, \citenamefont {Kästner},\ and\
  \citenamefont {Hashmi}}]{stable-in-gold-2015}%
  \BibitemOpen
  \bibfield  {author} {\bibinfo {author} {\bibfnamefont {L.}~\bibnamefont
  {Nunes~dos Santos~Comprido}}, \bibinfo {author} {\bibfnamefont {J.~E. M.~N.}\
  \bibnamefont {Klein}}, \bibinfo {author} {\bibfnamefont {G.}~\bibnamefont
  {Knizia}}, \bibinfo {author} {\bibfnamefont {J.}~\bibnamefont {Kästner}},\
  and\ \bibinfo {author} {\bibfnamefont {A.~S.~K.}\ \bibnamefont {Hashmi}},\
  }\bibfield  {title} {\enquote {\bibinfo {title} {The stabilizing effects in
  gold carbene complexes},}\ }\href
  {https://doi.org/https://doi.org/10.1002/anie.201412401} {\bibfield
  {journal} {\bibinfo  {journal} {Angew. Chem., Int. Ed.}\ }\textbf {\bibinfo
  {volume} {54}},\ \bibinfo {pages} {10336--10340} (\bibinfo {year}
  {2015})}\BibitemShut {NoStop}%
\bibitem [{\citenamefont {Cheng}\ \emph {et~al.}(2024)\citenamefont {Cheng},
  \citenamefont {Ho}, \citenamefont {Yang}, \citenamefont {Chen}, \citenamefont
  {Hsieh},\ and\ \citenamefont {Cheng}}]{alkane-activate-2024}%
  \BibitemOpen
  \bibfield  {author} {\bibinfo {author} {\bibfnamefont {Y.-H.}\ \bibnamefont
  {Cheng}}, \bibinfo {author} {\bibfnamefont {Y.-S.}\ \bibnamefont {Ho}},
  \bibinfo {author} {\bibfnamefont {C.-J.}\ \bibnamefont {Yang}}, \bibinfo
  {author} {\bibfnamefont {C.-Y.}\ \bibnamefont {Chen}}, \bibinfo {author}
  {\bibfnamefont {C.-T.}\ \bibnamefont {Hsieh}},\ and\ \bibinfo {author}
  {\bibfnamefont {M.-J.}\ \bibnamefont {Cheng}},\ }\bibfield  {title} {\enquote
  {\bibinfo {title} {{Electron Dynamics in Alkane C–H Activation Mediated by
  Transition Metal Complexes}},}\ }\href
  {https://doi.org/10.1021/acs.jpca.4c01131} {\bibfield  {journal} {\bibinfo
  {journal} {J. Phys. Chem. A}\ }\textbf {\bibinfo {volume} {128}},\ \bibinfo
  {pages} {4638--4650} (\bibinfo {year} {2024})}\BibitemShut {NoStop}%
\bibitem [{\citenamefont {Klein}\ and\ \citenamefont
  {Knizia}(2018)}]{cpcet-vs-hat-2018}%
  \BibitemOpen
  \bibfield  {author} {\bibinfo {author} {\bibfnamefont {J.~E. M.~N.}\
  \bibnamefont {Klein}}\ and\ \bibinfo {author} {\bibfnamefont
  {G.}~\bibnamefont {Knizia}},\ }\bibfield  {title} {\enquote {\bibinfo {title}
  {{cPCET versus HAT: A Direct Theoretical Method for Distinguishing X–H
  Bond-Activation Mechanisms}},}\ }\href
  {https://doi.org/https://doi.org/10.1002/anie.201805511} {\bibfield
  {journal} {\bibinfo  {journal} {Angew. Chem., Int. Ed.}\ }\textbf {\bibinfo
  {volume} {57}},\ \bibinfo {pages} {11913--11917} (\bibinfo {year}
  {2018})}\BibitemShut {NoStop}%
\bibitem [{\citenamefont {Balakrishnan}\ \emph {et~al.}(2025)\citenamefont
  {Balakrishnan}, \citenamefont {Chen}, \citenamefont {Cheng}, \citenamefont
  {Wang},\ and\ \citenamefont {Cheng}}]{reduction-CO2-2025}%
  \BibitemOpen
  \bibfield  {author} {\bibinfo {author} {\bibfnamefont {A.}~\bibnamefont
  {Balakrishnan}}, \bibinfo {author} {\bibfnamefont {W.-S.}\ \bibnamefont
  {Chen}}, \bibinfo {author} {\bibfnamefont {Y.-H.}\ \bibnamefont {Cheng}},
  \bibinfo {author} {\bibfnamefont {K.-H.}\ \bibnamefont {Wang}},\ and\
  \bibinfo {author} {\bibfnamefont {M.-J.}\ \bibnamefont {Cheng}},\ }\bibfield
  {title} {\enquote {\bibinfo {title} {{Unveiling Electron Dynamics in the
  Electrochemical Reduction of CO$_2$ to Methane on Copper}},}\ }\href
  {https://doi.org/https://doi.org/10.1002/cplu.202500250} {\bibfield
  {journal} {\bibinfo  {journal} {ChemPlusChem}\ }\textbf {\bibinfo {volume}
  {90}},\ \bibinfo {pages} {e2500250} (\bibinfo {year} {2025})}\BibitemShut
  {NoStop}%
\bibitem [{\citenamefont {Hsieh}\ \emph {et~al.}(2023)\citenamefont {Hsieh},
  \citenamefont {Tang}, \citenamefont {Ho}, \citenamefont {Shao},\ and\
  \citenamefont {Cheng}}]{methane-to-methanol-2023}%
  \BibitemOpen
  \bibfield  {author} {\bibinfo {author} {\bibfnamefont {C.-T.}\ \bibnamefont
  {Hsieh}}, \bibinfo {author} {\bibfnamefont {Y.-T.}\ \bibnamefont {Tang}},
  \bibinfo {author} {\bibfnamefont {Y.-S.}\ \bibnamefont {Ho}}, \bibinfo
  {author} {\bibfnamefont {W.-K.}\ \bibnamefont {Shao}},\ and\ \bibinfo
  {author} {\bibfnamefont {M.-J.}\ \bibnamefont {Cheng}},\ }\bibfield  {title}
  {\enquote {\bibinfo {title} {{Methane to Methanol Conversion over N-Doped
  Graphene Facilitated by Electrochemical Oxygen Evolution: A First-Principles
  Study}},}\ }\href {https://doi.org/10.1021/acs.jpcc.2c07736} {\bibfield
  {journal} {\bibinfo  {journal} {J. Phys. Chem. C}\ }\textbf {\bibinfo
  {volume} {127}},\ \bibinfo {pages} {308--318} (\bibinfo {year}
  {2023})}\BibitemShut {NoStop}%
\bibitem [{\citenamefont {Derricotte}\ and\ \citenamefont
  {Evangelista}(2017)}]{livvo-core-excite-2017}%
  \BibitemOpen
  \bibfield  {author} {\bibinfo {author} {\bibfnamefont {W.~D.}\ \bibnamefont
  {Derricotte}}\ and\ \bibinfo {author} {\bibfnamefont {F.~A.}\ \bibnamefont
  {Evangelista}},\ }\bibfield  {title} {\enquote {\bibinfo {title} {Localized
  intrinsic valence virtual orbitals as a tool for the automatic classification
  of core excited states},}\ }\href {https://doi.org/10.1021/acs.jctc.7b00493}
  {\bibfield  {journal} {\bibinfo  {journal} {J. Chem. Theory Comput.}\
  }\textbf {\bibinfo {volume} {13}},\ \bibinfo {pages} {5984--5999} (\bibinfo
  {year} {2017})}\BibitemShut {NoStop}%
\bibitem [{\citenamefont {Corry}\ and\ \citenamefont
  {O’Malley}(2020)}]{loc-bond-o2-2020}%
  \BibitemOpen
  \bibfield  {author} {\bibinfo {author} {\bibfnamefont {T.~A.}\ \bibnamefont
  {Corry}}\ and\ \bibinfo {author} {\bibfnamefont {P.~J.}\ \bibnamefont
  {O’Malley}},\ }\bibfield  {title} {\enquote {\bibinfo {title} {{Localized
  Bond Orbital Analysis of the Bonds of O$_2$}},}\ }\href
  {https://doi.org/10.1021/acs.jpca.0c07836} {\bibfield  {journal} {\bibinfo
  {journal} {J. Phys. Chem. A}\ }\textbf {\bibinfo {volume} {124}},\ \bibinfo
  {pages} {9771--9776} (\bibinfo {year} {2020})}\BibitemShut {NoStop}%
\bibitem [{\citenamefont {Alabugin}, \citenamefont {dos Passos~Gomes},\ and\
  \citenamefont {Abdo}(2019)}]{hyperconjugation-2019}%
  \BibitemOpen
  \bibfield  {author} {\bibinfo {author} {\bibfnamefont {I.~V.}\ \bibnamefont
  {Alabugin}}, \bibinfo {author} {\bibfnamefont {G.}~\bibnamefont {dos
  Passos~Gomes}},\ and\ \bibinfo {author} {\bibfnamefont {M.~A.}\ \bibnamefont
  {Abdo}},\ }\bibfield  {title} {\enquote {\bibinfo {title}
  {Hyperconjugation},}\ }\href
  {https://doi.org/https://doi.org/10.1002/wcms.1389} {\bibfield  {journal}
  {\bibinfo  {journal} {Wiley Interdiscip. Rev.: Comput. Mol. Sci.}\ }\textbf
  {\bibinfo {volume} {9}},\ \bibinfo {pages} {e1389} (\bibinfo {year}
  {2019})}\BibitemShut {NoStop}%
\bibitem [{\citenamefont {Li}\ \emph {et~al.}(2020)\citenamefont {Li},
  \citenamefont {Govind}, \citenamefont {Isborn}, \citenamefont
  {DePrince~III},\ and\ \citenamefont {Lopata}}]{td-elstru-rev-2020}%
  \BibitemOpen
  \bibfield  {author} {\bibinfo {author} {\bibfnamefont {X.}~\bibnamefont
  {Li}}, \bibinfo {author} {\bibfnamefont {N.}~\bibnamefont {Govind}}, \bibinfo
  {author} {\bibfnamefont {C.}~\bibnamefont {Isborn}}, \bibinfo {author}
  {\bibfnamefont {A.~E.}\ \bibnamefont {DePrince~III}},\ and\ \bibinfo {author}
  {\bibfnamefont {K.}~\bibnamefont {Lopata}},\ }\bibfield  {title} {\enquote
  {\bibinfo {title} {Real-time time-dependent electronic structure theory},}\
  }\href {https://doi.org/10.1021/acs.chemrev.0c00223} {\bibfield  {journal}
  {\bibinfo  {journal} {Chem. Rev.}\ }\textbf {\bibinfo {volume} {120}},\
  \bibinfo {pages} {9951--9993} (\bibinfo {year} {2020})}\BibitemShut {NoStop}%
\bibitem [{\citenamefont {Chan}\ and\ \citenamefont
  {Head-Gordon}(2002)}]{dmrg-study-polynom-2002}%
  \BibitemOpen
  \bibfield  {author} {\bibinfo {author} {\bibfnamefont {G.~K.-L.}\
  \bibnamefont {Chan}}\ and\ \bibinfo {author} {\bibfnamefont {M.}~\bibnamefont
  {Head-Gordon}},\ }\bibfield  {title} {\enquote {\bibinfo {title} {{Highly
  correlated calculations with a polynomial cost algorithm: A study of the
  density matrix renormalization group}},}\ }\href
  {https://doi.org/10.1063/1.1449459} {\bibfield  {journal} {\bibinfo
  {journal} {J. Chem. Phys.}\ }\textbf {\bibinfo {volume} {116}},\ \bibinfo
  {pages} {4462--4476} (\bibinfo {year} {2002})}\BibitemShut {NoStop}%
\bibitem [{\citenamefont {Lubich}\ \emph {et~al.}(2013)\citenamefont {Lubich},
  \citenamefont {Rohwedder}, \citenamefont {Schneider},\ and\ \citenamefont
  {Vandereycken}}]{tdvp-tensor_train-2013}%
  \BibitemOpen
  \bibfield  {author} {\bibinfo {author} {\bibfnamefont {C.}~\bibnamefont
  {Lubich}}, \bibinfo {author} {\bibfnamefont {T.}~\bibnamefont {Rohwedder}},
  \bibinfo {author} {\bibfnamefont {R.}~\bibnamefont {Schneider}},\ and\
  \bibinfo {author} {\bibfnamefont {B.}~\bibnamefont {Vandereycken}},\
  }\bibfield  {title} {\enquote {\bibinfo {title} {{Dynamical Approximation by
  Hierarchical Tucker and Tensor-Train Tensors}},}\ }\href
  {https://doi.org/10.1137/120885723} {\bibfield  {journal} {\bibinfo
  {journal} {SIAM J. Matrix Anal. Appl.}\ }\textbf {\bibinfo {volume} {34}},\
  \bibinfo {pages} {470--494} (\bibinfo {year} {2013})}\BibitemShut {NoStop}%
\bibitem [{\citenamefont {Lubich}, \citenamefont {Oseledets},\ and\
  \citenamefont {Vandereycken}(2015)}]{tdvp-time_integration-2015}%
  \BibitemOpen
  \bibfield  {author} {\bibinfo {author} {\bibfnamefont {C.}~\bibnamefont
  {Lubich}}, \bibinfo {author} {\bibfnamefont {I.~V.}\ \bibnamefont
  {Oseledets}},\ and\ \bibinfo {author} {\bibfnamefont {B.}~\bibnamefont
  {Vandereycken}},\ }\bibfield  {title} {\enquote {\bibinfo {title} {Time
  integration of tensor trains},}\ }\href {https://doi.org/10.1137/140976546}
  {\bibfield  {journal} {\bibinfo  {journal} {SIAM J. Numer. Anal.}\ }\textbf
  {\bibinfo {volume} {53}},\ \bibinfo {pages} {917--941} (\bibinfo {year}
  {2015})}\BibitemShut {NoStop}%
\bibitem [{\citenamefont {Haegeman}\ \emph {et~al.}(2016)\citenamefont
  {Haegeman}, \citenamefont {Lubich}, \citenamefont {Oseledets}, \citenamefont
  {Vandereycken},\ and\ \citenamefont {Verstraete}}]{tdvp-unify-2016}%
  \BibitemOpen
  \bibfield  {author} {\bibinfo {author} {\bibfnamefont {J.}~\bibnamefont
  {Haegeman}}, \bibinfo {author} {\bibfnamefont {C.}~\bibnamefont {Lubich}},
  \bibinfo {author} {\bibfnamefont {I.}~\bibnamefont {Oseledets}}, \bibinfo
  {author} {\bibfnamefont {B.}~\bibnamefont {Vandereycken}},\ and\ \bibinfo
  {author} {\bibfnamefont {F.}~\bibnamefont {Verstraete}},\ }\bibfield  {title}
  {\enquote {\bibinfo {title} {Unifying time evolution and optimization with
  matrix product states},}\ }\href {https://doi.org/10.1103/PhysRevB.94.165116}
  {\bibfield  {journal} {\bibinfo  {journal} {Phys. Rev. B}\ }\textbf {\bibinfo
  {volume} {94}},\ \bibinfo {pages} {165116} (\bibinfo {year}
  {2016})}\BibitemShut {NoStop}%
\bibitem [{\citenamefont {Paeckel}\ \emph {et~al.}(2019)\citenamefont
  {Paeckel}, \citenamefont {Köhler}, \citenamefont {Swoboda}, \citenamefont
  {Manmana}, \citenamefont {Schollwöck},\ and\ \citenamefont
  {Hubig}}]{PAECKEL_tddmrg_review-2019}%
  \BibitemOpen
  \bibfield  {author} {\bibinfo {author} {\bibfnamefont {S.}~\bibnamefont
  {Paeckel}}, \bibinfo {author} {\bibfnamefont {T.}~\bibnamefont {Köhler}},
  \bibinfo {author} {\bibfnamefont {A.}~\bibnamefont {Swoboda}}, \bibinfo
  {author} {\bibfnamefont {S.~R.}\ \bibnamefont {Manmana}}, \bibinfo {author}
  {\bibfnamefont {U.}~\bibnamefont {Schollwöck}},\ and\ \bibinfo {author}
  {\bibfnamefont {C.}~\bibnamefont {Hubig}},\ }\bibfield  {title} {\enquote
  {\bibinfo {title} {Time-evolution methods for matrix-product states},}\
  }\href {https://doi.org/https://doi.org/10.1016/j.aop.2019.167998} {\bibfield
   {journal} {\bibinfo  {journal} {Ann. Phys.}\ }\textbf {\bibinfo {volume}
  {411}},\ \bibinfo {pages} {167998} (\bibinfo {year} {2019})}\BibitemShut
  {NoStop}%
\bibitem [{\citenamefont {Frahm}\ and\ \citenamefont
  {Pfannkuche}(2019)}]{tddmrg-pfannkuche-2019}%
  \BibitemOpen
  \bibfield  {author} {\bibinfo {author} {\bibfnamefont {L.-H.}\ \bibnamefont
  {Frahm}}\ and\ \bibinfo {author} {\bibfnamefont {D.}~\bibnamefont
  {Pfannkuche}},\ }\bibfield  {title} {\enquote {\bibinfo {title} {Ultrafast
  \textit{ab initio} quantum chemistry using matrix product states},}\ }\href
  {https://doi.org/10.1021/acs.jctc.8b01291} {\bibfield  {journal} {\bibinfo
  {journal} {J. Chem. Theory Comput.}\ }\textbf {\bibinfo {volume} {15}},\
  \bibinfo {pages} {2154--2165} (\bibinfo {year} {2019})}\BibitemShut {NoStop}%
\bibitem [{\citenamefont {Baiardi}(2021)}]{tddmrg-baiardi-2021}%
  \BibitemOpen
  \bibfield  {author} {\bibinfo {author} {\bibfnamefont {A.}~\bibnamefont
  {Baiardi}},\ }\bibfield  {title} {\enquote {\bibinfo {title} {Electron
  dynamics with the time-dependent density matrix renormalization group},}\
  }\href {https://doi.org/10.1021/acs.jctc.0c01048} {\bibfield  {journal}
  {\bibinfo  {journal} {J. Chem. Theory Comput.}\ }\textbf {\bibinfo {volume}
  {17}},\ \bibinfo {pages} {3320--3334} (\bibinfo {year} {2021})}\BibitemShut
  {NoStop}%
\bibitem [{\citenamefont {Zhai}\ and\ \citenamefont
  {Chan}(2021)}]{low-comm-dmrg-2021}%
  \BibitemOpen
  \bibfield  {author} {\bibinfo {author} {\bibfnamefont {H.}~\bibnamefont
  {Zhai}}\ and\ \bibinfo {author} {\bibfnamefont {G.~K.-L.}\ \bibnamefont
  {Chan}},\ }\bibfield  {title} {\enquote {\bibinfo {title} {Low communication
  high performance \textit{ab initio} density matrix renormalization group
  algorithms},}\ }\href {https://doi.org/10.1063/5.0050902} {\bibfield
  {journal} {\bibinfo  {journal} {J. Chem. Phys.}\ }\textbf {\bibinfo {volume}
  {154}},\ \bibinfo {pages} {224116} (\bibinfo {year} {2021})}\BibitemShut
  {NoStop}%
\bibitem [{\citenamefont {Zhai}\ \emph {et~al.}(2023)\citenamefont {Zhai},
  \citenamefont {Larsson}, \citenamefont {Lee}, \citenamefont {Cui},
  \citenamefont {Zhu}, \citenamefont {Sun}, \citenamefont {Peng}, \citenamefont
  {Peng}, \citenamefont {Liao}, \citenamefont {Tölle}, \citenamefont {Yang},
  \citenamefont {Li},\ and\ \citenamefont {Chan}}]{block2-2023}%
  \BibitemOpen
  \bibfield  {author} {\bibinfo {author} {\bibfnamefont {H.}~\bibnamefont
  {Zhai}}, \bibinfo {author} {\bibfnamefont {H.~R.}\ \bibnamefont {Larsson}},
  \bibinfo {author} {\bibfnamefont {S.}~\bibnamefont {Lee}}, \bibinfo {author}
  {\bibfnamefont {Z.-H.}\ \bibnamefont {Cui}}, \bibinfo {author} {\bibfnamefont
  {T.}~\bibnamefont {Zhu}}, \bibinfo {author} {\bibfnamefont {C.}~\bibnamefont
  {Sun}}, \bibinfo {author} {\bibfnamefont {L.}~\bibnamefont {Peng}}, \bibinfo
  {author} {\bibfnamefont {R.}~\bibnamefont {Peng}}, \bibinfo {author}
  {\bibfnamefont {K.}~\bibnamefont {Liao}}, \bibinfo {author} {\bibfnamefont
  {J.}~\bibnamefont {Tölle}}, \bibinfo {author} {\bibfnamefont
  {J.}~\bibnamefont {Yang}}, \bibinfo {author} {\bibfnamefont {S.}~\bibnamefont
  {Li}},\ and\ \bibinfo {author} {\bibfnamefont {G.~K.-L.}\ \bibnamefont
  {Chan}},\ }\bibfield  {title} {\enquote {\bibinfo {title} {{Block2: A
  comprehensive open source framework to develop and apply state-of-the-art
  DMRG algorithms in electronic structure and beyond}},}\ }\href
  {https://doi.org/10.1063/5.0180424} {\bibfield  {journal} {\bibinfo
  {journal} {J. Chem. Phys.}\ }\textbf {\bibinfo {volume} {159}},\ \bibinfo
  {pages} {234801} (\bibinfo {year} {2023})}\BibitemShut {NoStop}%
\bibitem [{\citenamefont {Larsson}(2024)}]{hrl-mctdh-rev-2024}%
  \BibitemOpen
  \bibfield  {author} {\bibinfo {author} {\bibfnamefont {H.~R.}\ \bibnamefont
  {Larsson}},\ }\bibfield  {title} {\enquote {\bibinfo {title} {{A tensor
  network view of multilayer multiconfiguration time-dependent Hartree
  methods}},}\ }\href {https://doi.org/10.1080/00268976.2024.2306881}
  {\bibfield  {journal} {\bibinfo  {journal} {Mol. Phys.}\ }\textbf {\bibinfo
  {volume} {122}},\ \bibinfo {pages} {e2306881} (\bibinfo {year}
  {2024})}\BibitemShut {NoStop}%
\bibitem [{\citenamefont {Martin}(2003)}]{nto-2003}%
  \BibitemOpen
  \bibfield  {author} {\bibinfo {author} {\bibfnamefont {R.~L.}\ \bibnamefont
  {Martin}},\ }\bibfield  {title} {\enquote {\bibinfo {title} {Natural
  transition orbitals},}\ }\href {https://doi.org/10.1063/1.1558471} {\bibfield
   {journal} {\bibinfo  {journal} {J. Chem. Phys.}\ }\textbf {\bibinfo {volume}
  {118}},\ \bibinfo {pages} {4775--4777} (\bibinfo {year} {2003})}\BibitemShut
  {NoStop}%
\bibitem [{\citenamefont {Krylov}(2020)}]{orb-to-obs-2020}%
  \BibitemOpen
  \bibfield  {author} {\bibinfo {author} {\bibfnamefont {A.~I.}\ \bibnamefont
  {Krylov}},\ }\bibfield  {title} {\enquote {\bibinfo {title} {From orbitals to
  observables and back},}\ }\href {https://doi.org/10.1063/5.0018597}
  {\bibfield  {journal} {\bibinfo  {journal} {J. Chem. Phys.}\ }\textbf
  {\bibinfo {volume} {153}},\ \bibinfo {pages} {080901} (\bibinfo {year}
  {2020})}\BibitemShut {NoStop}%
\bibitem [{\citenamefont {Herbert}(2024)}]{visualize-tddft-excite-2024}%
  \BibitemOpen
  \bibfield  {author} {\bibinfo {author} {\bibfnamefont {J.~M.}\ \bibnamefont
  {Herbert}},\ }\bibfield  {title} {\enquote {\bibinfo {title} {Visualizing and
  characterizing excited states from time-dependent density functional
  theory},}\ }\href {https://doi.org/10.1039/D3CP04226J} {\bibfield  {journal}
  {\bibinfo  {journal} {Phys. Chem. Chem. Phys.}\ }\textbf {\bibinfo {volume}
  {26}},\ \bibinfo {pages} {3755--3794} (\bibinfo {year} {2024})}\BibitemShut
  {NoStop}%
\bibitem [{\citenamefont {Bovill}\ \emph {et~al.}(2026)\citenamefont {Bovill},
  \citenamefont {Abou~Taka}, \citenamefont {Harb},\ and\ \citenamefont
  {Hratchian}}]{exc-relax-diffdens-2026}%
  \BibitemOpen
  \bibfield  {author} {\bibinfo {author} {\bibfnamefont {A.~J.}\ \bibnamefont
  {Bovill}}, \bibinfo {author} {\bibfnamefont {A.}~\bibnamefont {Abou~Taka}},
  \bibinfo {author} {\bibfnamefont {H.}~\bibnamefont {Harb}},\ and\ \bibinfo
  {author} {\bibfnamefont {H.~P.}\ \bibnamefont {Hratchian}},\ }\bibfield
  {title} {\enquote {\bibinfo {title} {Excitation/relaxation analysis of
  electronic transitions using difference density natural orbitals},}\ }\href
  {https://doi.org/10.1021/acs.jctc.5c01792} {\bibfield  {journal} {\bibinfo
  {journal} {J. Chem. Theory Comput.}\ }\textbf {\bibinfo {volume} {22}},\
  \bibinfo {pages} {930--939} (\bibinfo {year} {2026})}\BibitemShut {NoStop}%
\bibitem [{\citenamefont {Dutoi}, \citenamefont {Wormit},\ and\ \citenamefont
  {Cederbaum}(2011)}]{charge-sep-holmob-2011}%
  \BibitemOpen
  \bibfield  {author} {\bibinfo {author} {\bibfnamefont {A.~D.}\ \bibnamefont
  {Dutoi}}, \bibinfo {author} {\bibfnamefont {M.}~\bibnamefont {Wormit}},\ and\
  \bibinfo {author} {\bibfnamefont {L.~S.}\ \bibnamefont {Cederbaum}},\
  }\bibfield  {title} {\enquote {\bibinfo {title} {Ultrafast charge separation
  driven by differential particle and hole mobilities},}\ }\href
  {https://doi.org/10.1063/1.3506617} {\bibfield  {journal} {\bibinfo
  {journal} {J. Chem. Phys.}\ }\textbf {\bibinfo {volume} {134}},\ \bibinfo
  {pages} {024303} (\bibinfo {year} {2011})}\BibitemShut {NoStop}%
\bibitem [{\citenamefont {Langkabel}\ \emph {et~al.}(2022)\citenamefont
  {Langkabel}, \citenamefont {Albrecht}, \citenamefont {Bande},\ and\
  \citenamefont {Krause}}]{visible-optical-excitation-2022}%
  \BibitemOpen
  \bibfield  {author} {\bibinfo {author} {\bibfnamefont {F.}~\bibnamefont
  {Langkabel}}, \bibinfo {author} {\bibfnamefont {P.~A.}\ \bibnamefont
  {Albrecht}}, \bibinfo {author} {\bibfnamefont {A.}~\bibnamefont {Bande}},\
  and\ \bibinfo {author} {\bibfnamefont {P.}~\bibnamefont {Krause}},\
  }\bibfield  {title} {\enquote {\bibinfo {title} {Making optical excitations
  visible – an exciton wavefunction extension to the time-dependent
  configuration interaction method},}\ }\href
  {https://doi.org/https://doi.org/10.1016/j.chemphys.2022.111502} {\bibfield
  {journal} {\bibinfo  {journal} {Chem. Phys.}\ }\textbf {\bibinfo {volume}
  {557}},\ \bibinfo {pages} {111502} (\bibinfo {year} {2022})}\BibitemShut
  {NoStop}%
\bibitem [{\citenamefont {Moitra}\ \emph {et~al.}(2023)\citenamefont {Moitra},
  \citenamefont {Konecny}, \citenamefont {Kadek}, \citenamefont {Rubio},\ and\
  \citenamefont {Repisky}}]{rel-rtddft-valcor-atas-2023}%
  \BibitemOpen
  \bibfield  {author} {\bibinfo {author} {\bibfnamefont {T.}~\bibnamefont
  {Moitra}}, \bibinfo {author} {\bibfnamefont {L.}~\bibnamefont {Konecny}},
  \bibinfo {author} {\bibfnamefont {M.}~\bibnamefont {Kadek}}, \bibinfo
  {author} {\bibfnamefont {A.}~\bibnamefont {Rubio}},\ and\ \bibinfo {author}
  {\bibfnamefont {M.}~\bibnamefont {Repisky}},\ }\bibfield  {title} {\enquote
  {\bibinfo {title} {Accurate relativistic real-time time-dependent density
  functional theory for valence and core attosecond transient absorption
  spectroscopy},}\ }\href {https://doi.org/10.1021/acs.jpclett.2c03599}
  {\bibfield  {journal} {\bibinfo  {journal} {J. Phys. Chem. Lett.}\ }\textbf
  {\bibinfo {volume} {14}},\ \bibinfo {pages} {1714--1724} (\bibinfo {year}
  {2023})}\BibitemShut {NoStop}%
\bibitem [{\citenamefont {Upadhyay}\ \emph {et~al.}(2026)\citenamefont
  {Upadhyay}, \citenamefont {Zheng}, \citenamefont {Wang}, \citenamefont
  {Shayit}, \citenamefont {Liu}, \citenamefont {Sun},\ and\ \citenamefont
  {Li}}]{chiral-magnet-from-4curr-2026}%
  \BibitemOpen
  \bibfield  {author} {\bibinfo {author} {\bibfnamefont {S.}~\bibnamefont
  {Upadhyay}}, \bibinfo {author} {\bibfnamefont {X.}~\bibnamefont {Zheng}},
  \bibinfo {author} {\bibfnamefont {T.}~\bibnamefont {Wang}}, \bibinfo {author}
  {\bibfnamefont {A.}~\bibnamefont {Shayit}}, \bibinfo {author} {\bibfnamefont
  {J.}~\bibnamefont {Liu}}, \bibinfo {author} {\bibfnamefont {D.}~\bibnamefont
  {Sun}},\ and\ \bibinfo {author} {\bibfnamefont {X.}~\bibnamefont {Li}},\
  }\bibfield  {title} {\enquote {\bibinfo {title} {Chirality-driven
  magnetization emerges from relativistic four-current dynamics},}\ }\href
  {https://doi.org/10.1063/5.0313445} {\bibfield  {journal} {\bibinfo
  {journal} {APL Comput. Phys.}\ }\textbf {\bibinfo {volume} {2}},\ \bibinfo
  {pages} {016103} (\bibinfo {year} {2026})}\BibitemShut {NoStop}%
\bibitem [{\citenamefont {Wang}\ \emph
  {et~al.}(2025{\natexlab{b}})\citenamefont {Wang}, \citenamefont {Driver},
  \citenamefont {Franz}, \citenamefont {Koloren\ifmmode~\check{c}\else
  \v{c}\fi{}}, \citenamefont {Thierstein}, \citenamefont {Robles},
  \citenamefont {Isele}, \citenamefont {Guo}, \citenamefont {Cesar},
  \citenamefont {Alexander}, \citenamefont {Beauvarlet}, \citenamefont {Borne},
  \citenamefont {Cheng}, \citenamefont {DiMauro}, \citenamefont {Duris},
  \citenamefont {Glownia}, \citenamefont {Gra\ss{}l}, \citenamefont {Hockett},
  \citenamefont {Hoffman}, \citenamefont {Kamalov}, \citenamefont {Larsen},
  \citenamefont {Li}, \citenamefont {Li}, \citenamefont {Lin}, \citenamefont
  {Obaid}, \citenamefont {Rosenberger}, \citenamefont {Walter}, \citenamefont
  {Wolf}, \citenamefont {Marangos}, \citenamefont {Kling}, \citenamefont
  {Bucksbaum}, \citenamefont {Marinelli},\ and\ \citenamefont
  {Cryan}}]{probe-chr-core-vacant-2025}%
  \BibitemOpen
  \bibfield  {author} {\bibinfo {author} {\bibfnamefont {J.}~\bibnamefont
  {Wang}}, \bibinfo {author} {\bibfnamefont {T.}~\bibnamefont {Driver}},
  \bibinfo {author} {\bibfnamefont {P.~L.}\ \bibnamefont {Franz}}, \bibinfo
  {author} {\bibfnamefont {P.~c.~v.}\ \bibnamefont
  {Koloren\ifmmode~\check{c}\else \v{c}\fi{}}}, \bibinfo {author}
  {\bibfnamefont {E.}~\bibnamefont {Thierstein}}, \bibinfo {author}
  {\bibfnamefont {R.~R.}\ \bibnamefont {Robles}}, \bibinfo {author}
  {\bibfnamefont {E.}~\bibnamefont {Isele}}, \bibinfo {author} {\bibfnamefont
  {Z.}~\bibnamefont {Guo}}, \bibinfo {author} {\bibfnamefont {D.}~\bibnamefont
  {Cesar}}, \bibinfo {author} {\bibfnamefont {O.}~\bibnamefont {Alexander}},
  \bibinfo {author} {\bibfnamefont {S.}~\bibnamefont {Beauvarlet}}, \bibinfo
  {author} {\bibfnamefont {K.}~\bibnamefont {Borne}}, \bibinfo {author}
  {\bibfnamefont {X.}~\bibnamefont {Cheng}}, \bibinfo {author} {\bibfnamefont
  {L.~F.}\ \bibnamefont {DiMauro}}, \bibinfo {author} {\bibfnamefont
  {J.}~\bibnamefont {Duris}}, \bibinfo {author} {\bibfnamefont {J.~M.}\
  \bibnamefont {Glownia}}, \bibinfo {author} {\bibfnamefont {M.}~\bibnamefont
  {Gra\ss{}l}}, \bibinfo {author} {\bibfnamefont {P.}~\bibnamefont {Hockett}},
  \bibinfo {author} {\bibfnamefont {M.}~\bibnamefont {Hoffman}}, \bibinfo
  {author} {\bibfnamefont {A.}~\bibnamefont {Kamalov}}, \bibinfo {author}
  {\bibfnamefont {K.~A.}\ \bibnamefont {Larsen}}, \bibinfo {author}
  {\bibfnamefont {S.}~\bibnamefont {Li}}, \bibinfo {author} {\bibfnamefont
  {X.}~\bibnamefont {Li}}, \bibinfo {author} {\bibfnamefont {M.-F.}\
  \bibnamefont {Lin}}, \bibinfo {author} {\bibfnamefont {R.}~\bibnamefont
  {Obaid}}, \bibinfo {author} {\bibfnamefont {P.}~\bibnamefont {Rosenberger}},
  \bibinfo {author} {\bibfnamefont {P.}~\bibnamefont {Walter}}, \bibinfo
  {author} {\bibfnamefont {T.~J.~A.}\ \bibnamefont {Wolf}}, \bibinfo {author}
  {\bibfnamefont {J.~P.}\ \bibnamefont {Marangos}}, \bibinfo {author}
  {\bibfnamefont {M.~F.}\ \bibnamefont {Kling}}, \bibinfo {author}
  {\bibfnamefont {P.~H.}\ \bibnamefont {Bucksbaum}}, \bibinfo {author}
  {\bibfnamefont {A.}~\bibnamefont {Marinelli}},\ and\ \bibinfo {author}
  {\bibfnamefont {J.~P.}\ \bibnamefont {Cryan}},\ }\bibfield  {title} {\enquote
  {\bibinfo {title} {Probing electronic coherence between core-level vacancies
  at different atomic sites},}\ }\href
  {https://doi.org/10.1103/PhysRevX.15.011008} {\bibfield  {journal} {\bibinfo
  {journal} {Phys. Rev. X}\ }\textbf {\bibinfo {volume} {15}},\ \bibinfo
  {pages} {011008} (\bibinfo {year} {2025}{\natexlab{b}})}\BibitemShut
  {NoStop}%
\bibitem [{\citenamefont {Gordon}(1968)}]{mol-correl-func-1968}%
  \BibitemOpen
  \bibfield  {author} {\bibinfo {author} {\bibfnamefont {R.}~\bibnamefont
  {Gordon}},\ }\bibfield  {title} {\enquote {\bibinfo {title} {Correlation
  functions for molecular motion},}\ }in\ \href
  {https://doi.org/https://doi.org/10.1016/B978-1-4832-3116-7.50008-4} {\emph
  {\bibinfo {booktitle} {Advances in Magnetic Resonance}}},\ \bibinfo {series}
  {Advances in Magnetic and Optical Resonance}, Vol.~\bibinfo {volume} {3},\
  \bibinfo {editor} {edited by\ \bibinfo {editor} {\bibfnamefont {J.~S.}\
  \bibnamefont {Waugh}}}\ (\bibinfo  {publisher} {Academic Press},\ \bibinfo
  {year} {1968})\ pp.\ \bibinfo {pages} {1--42}\BibitemShut {NoStop}%
\bibitem [{\citenamefont {{Tannor}}(2007)}]{tannor-book-2007}%
  \BibitemOpen
  \bibfield  {author} {\bibinfo {author} {\bibfnamefont {D.~J.}\ \bibnamefont
  {{Tannor}}},\ }\href
  {https://uscibooks.aip.org/books/introduction-to-quantum-mechanics-a-time-dependent-perspective/}
  {\emph {\bibinfo {title} {{Introduction to Quantum Mechanics: A
  Time-Dependent Perspective}}}}\ (\bibinfo  {publisher} {University Science
  Books},\ \bibinfo {year} {2007})\BibitemShut {NoStop}%
\bibitem [{\citenamefont {Jeckelmann}(2002)}]{dyn-dmrg-2002}%
  \BibitemOpen
  \bibfield  {author} {\bibinfo {author} {\bibfnamefont {E.}~\bibnamefont
  {Jeckelmann}},\ }\bibfield  {title} {\enquote {\bibinfo {title} {Dynamical
  density-matrix renormalization-group method},}\ }\href
  {https://doi.org/10.1103/PhysRevB.66.045114} {\bibfield  {journal} {\bibinfo
  {journal} {Phys. Rev. B}\ }\textbf {\bibinfo {volume} {66}},\ \bibinfo
  {pages} {045114} (\bibinfo {year} {2002})}\BibitemShut {NoStop}%
\bibitem [{\citenamefont {Ronca}\ \emph {et~al.}(2017)\citenamefont {Ronca},
  \citenamefont {Li}, \citenamefont {Jimenez-Hoyos},\ and\ \citenamefont
  {Chan}}]{ddmrg++-2017}%
  \BibitemOpen
  \bibfield  {author} {\bibinfo {author} {\bibfnamefont {E.}~\bibnamefont
  {Ronca}}, \bibinfo {author} {\bibfnamefont {Z.}~\bibnamefont {Li}}, \bibinfo
  {author} {\bibfnamefont {C.~A.}\ \bibnamefont {Jimenez-Hoyos}},\ and\
  \bibinfo {author} {\bibfnamefont {G.~K.-L.}\ \bibnamefont {Chan}},\
  }\bibfield  {title} {\enquote {\bibinfo {title} {Time-step targeting
  time-dependent and dynamical density matrix renormalization group algorithms
  with \textit{ab initio} hamiltonians},}\ }\href
  {https://doi.org/10.1021/acs.jctc.7b00682} {\bibfield  {journal} {\bibinfo
  {journal} {J. Chem. Theory Comput.}\ }\textbf {\bibinfo {volume} {13}},\
  \bibinfo {pages} {5560--5571} (\bibinfo {year} {2017})}\BibitemShut {NoStop}%
\bibitem [{\citenamefont {Rano}\ and\ \citenamefont
  {Larsson}(2025)}]{comp-eigenst-ttns-2025}%
  \BibitemOpen
  \bibfield  {author} {\bibinfo {author} {\bibfnamefont {M.}~\bibnamefont
  {Rano}}\ and\ \bibinfo {author} {\bibfnamefont {H.~R.}\ \bibnamefont
  {Larsson}},\ }\bibfield  {title} {\enquote {\bibinfo {title} {{Computing
  excited eigenstates using inexact Lanczos methods and tree tensor network
  states}},}\ }\href {https://doi.org/10.1063/5.0301263} {\bibfield  {journal}
  {\bibinfo  {journal} {J. Chem. Phys.}\ }\textbf {\bibinfo {volume} {163}},\
  \bibinfo {pages} {164110} (\bibinfo {year} {2025})}\BibitemShut {NoStop}%
\bibitem [{\citenamefont {Manne}\ and\ \citenamefont
  {Åberg}(1970)}]{koopman-inner-ion-1970}%
  \BibitemOpen
  \bibfield  {author} {\bibinfo {author} {\bibfnamefont {R.}~\bibnamefont
  {Manne}}\ and\ \bibinfo {author} {\bibfnamefont {T.}~\bibnamefont {Åberg}},\
  }\bibfield  {title} {\enquote {\bibinfo {title} {Koopmans' theorem for
  inner-shell ionization},}\ }\href
  {https://doi.org/https://doi.org/10.1016/0009-2614(70)80309-8} {\bibfield
  {journal} {\bibinfo  {journal} {Chem. Phys. Lett.}\ }\textbf {\bibinfo
  {volume} {7}},\ \bibinfo {pages} {282--284} (\bibinfo {year}
  {1970})}\BibitemShut {NoStop}%
\bibitem [{\citenamefont {Mignolet}, \citenamefont {Levine},\ and\
  \citenamefont {Remacle}(2012)}]{loc-dyn-dens-sudden-2012}%
  \BibitemOpen
  \bibfield  {author} {\bibinfo {author} {\bibfnamefont {B.}~\bibnamefont
  {Mignolet}}, \bibinfo {author} {\bibfnamefont {R.~D.}\ \bibnamefont
  {Levine}},\ and\ \bibinfo {author} {\bibfnamefont {F.}~\bibnamefont
  {Remacle}},\ }\bibfield  {title} {\enquote {\bibinfo {title} {Localized
  electron dynamics in attosecond-pulse-excited molecular systems: Probing the
  time-dependent electron density by sudden photoionization},}\ }\href
  {https://doi.org/10.1103/PhysRevA.86.053429} {\bibfield  {journal} {\bibinfo
  {journal} {Phys. Rev. A}\ }\textbf {\bibinfo {volume} {86}},\ \bibinfo
  {pages} {053429} (\bibinfo {year} {2012})}\BibitemShut {NoStop}%
\bibitem [{\citenamefont {Grell}\ \emph {et~al.}(2015)\citenamefont {Grell},
  \citenamefont {Bokarev}, \citenamefont {Winter}, \citenamefont {Seidel},
  \citenamefont {Aziz}, \citenamefont {Aziz},\ and\ \citenamefont
  {Kühn}}]{mr-photoel-so-2015}%
  \BibitemOpen
  \bibfield  {author} {\bibinfo {author} {\bibfnamefont {G.}~\bibnamefont
  {Grell}}, \bibinfo {author} {\bibfnamefont {S.~I.}\ \bibnamefont {Bokarev}},
  \bibinfo {author} {\bibfnamefont {B.}~\bibnamefont {Winter}}, \bibinfo
  {author} {\bibfnamefont {R.}~\bibnamefont {Seidel}}, \bibinfo {author}
  {\bibfnamefont {E.~F.}\ \bibnamefont {Aziz}}, \bibinfo {author}
  {\bibfnamefont {S.~G.}\ \bibnamefont {Aziz}},\ and\ \bibinfo {author}
  {\bibfnamefont {O.}~\bibnamefont {Kühn}},\ }\bibfield  {title} {\enquote
  {\bibinfo {title} {Multi-reference approach to the calculation of
  photoelectron spectra including spin-orbit coupling},}\ }\href
  {https://doi.org/10.1063/1.4928511} {\bibfield  {journal} {\bibinfo
  {journal} {J. Chem. Phys.}\ }\textbf {\bibinfo {volume} {143}},\ \bibinfo
  {pages} {074104} (\bibinfo {year} {2015})}\BibitemShut {NoStop}%
\bibitem [{\citenamefont {Kochetov}\ and\ \citenamefont
  {Bokarev}(2022)}]{rhodyn-2022}%
  \BibitemOpen
  \bibfield  {author} {\bibinfo {author} {\bibfnamefont {V.}~\bibnamefont
  {Kochetov}}\ and\ \bibinfo {author} {\bibfnamefont {S.~I.}\ \bibnamefont
  {Bokarev}},\ }\bibfield  {title} {\enquote {\bibinfo {title} {{RhoDyn: A
  $\rho$-TD-RASCI Framework to Study Ultrafast Electron Dynamics in
  Molecules}},}\ }\href {https://doi.org/10.1021/acs.jctc.1c01097} {\bibfield
  {journal} {\bibinfo  {journal} {J. Chem. Theory Comput.}\ }\textbf {\bibinfo
  {volume} {18}},\ \bibinfo {pages} {46--58} (\bibinfo {year}
  {2022})}\BibitemShut {NoStop}%
\bibitem [{\citenamefont {Pickup}(1977)}]{fast-photoion-1977}%
  \BibitemOpen
  \bibfield  {author} {\bibinfo {author} {\bibfnamefont {B.~T.}\ \bibnamefont
  {Pickup}},\ }\bibfield  {title} {\enquote {\bibinfo {title} {On the theory of
  fast photoionization processes},}\ }\href
  {https://doi.org/https://doi.org/10.1016/0301-0104(77)85131-8} {\bibfield
  {journal} {\bibinfo  {journal} {Chem. Phys.}\ }\textbf {\bibinfo {volume}
  {19}},\ \bibinfo {pages} {193--208} (\bibinfo {year} {1977})}\BibitemShut
  {NoStop}%
\bibitem [{\citenamefont {Cederbaum}\ \emph {et~al.}(1986)\citenamefont
  {Cederbaum}, \citenamefont {Domcke}, \citenamefont {Schirmer},\ and\
  \citenamefont {Niessen}}]{mo-breakdown-advchem-1986}%
  \BibitemOpen
  \bibfield  {author} {\bibinfo {author} {\bibfnamefont {L.~S.}\ \bibnamefont
  {Cederbaum}}, \bibinfo {author} {\bibfnamefont {W.}~\bibnamefont {Domcke}},
  \bibinfo {author} {\bibfnamefont {J.}~\bibnamefont {Schirmer}},\ and\
  \bibinfo {author} {\bibfnamefont {W.~V.}\ \bibnamefont {Niessen}},\ }\enquote
  {\bibinfo {title} {Correlation effects in the ionization of molecules:
  Breakdown of the molecular orbital picture},}\ in\ \href
  {https://doi.org/https://doi.org/10.1002/9780470142899.ch3} {\emph {\bibinfo
  {booktitle} {Advances in Chemical Physics}}}\ (\bibinfo  {publisher} {John
  Wiley \& Sons, Ltd},\ \bibinfo {year} {1986})\ pp.\ \bibinfo {pages}
  {115--159}\BibitemShut {NoStop}%
\bibitem [{Note1()}]{Note1}%
  \BibitemOpen
  \bibinfo {note} {Note that, in the following, we will use the terms ``2h1p
  configurations'' and ``2h1p states'' interchangeably. In our case, the
  operators act on a correlated state, giving a ``2h1p state.`` However, the
  term ``2h1p configuration'' is more commonly used.}\BibitemShut {Stop}%
\bibitem [{\citenamefont {Lu}\ \emph {et~al.}(2004)\citenamefont {Lu},
  \citenamefont {Wang}, \citenamefont {Schmidt}, \citenamefont {Bytautas},
  \citenamefont {Ho},\ and\ \citenamefont
  {Ruedenberg}}]{molintrsc_minbas-i-2004}%
  \BibitemOpen
  \bibfield  {author} {\bibinfo {author} {\bibfnamefont {W.~C.}\ \bibnamefont
  {Lu}}, \bibinfo {author} {\bibfnamefont {C.~Z.}\ \bibnamefont {Wang}},
  \bibinfo {author} {\bibfnamefont {M.~W.}\ \bibnamefont {Schmidt}}, \bibinfo
  {author} {\bibfnamefont {L.}~\bibnamefont {Bytautas}}, \bibinfo {author}
  {\bibfnamefont {K.~M.}\ \bibnamefont {Ho}},\ and\ \bibinfo {author}
  {\bibfnamefont {K.}~\bibnamefont {Ruedenberg}},\ }\bibfield  {title}
  {\enquote {\bibinfo {title} {{Molecule intrinsic minimal basis sets. I. Exact
  resolution of \textit{ab initio} optimized molecular orbitals in terms of
  deformed atomic minimal-basis orbitals}},}\ }\href
  {https://doi.org/10.1063/1.1638731} {\bibfield  {journal} {\bibinfo
  {journal} {J. Chem. Phys.}\ }\textbf {\bibinfo {volume} {120}},\ \bibinfo
  {pages} {2629--2637} (\bibinfo {year} {2004})}\BibitemShut {NoStop}%
\bibitem [{\citenamefont {Steen}, \citenamefont {Knizia},\ and\ \citenamefont
  {Klein}(2019)}]{masked-phenyl-niiv-2019}%
  \BibitemOpen
  \bibfield  {author} {\bibinfo {author} {\bibfnamefont {J.~S.}\ \bibnamefont
  {Steen}}, \bibinfo {author} {\bibfnamefont {G.}~\bibnamefont {Knizia}},\ and\
  \bibinfo {author} {\bibfnamefont {J.~E. M.~N.}\ \bibnamefont {Klein}},\
  }\bibfield  {title} {\enquote {\bibinfo {title} {{$\sigma$-Noninnocence:
  Masked Phenyl-Cation Transfer at Formal Ni$^\text{IV}$}},}\ }\href
  {https://doi.org/https://doi.org/10.1002/anie.201906658} {\bibfield
  {journal} {\bibinfo  {journal} {Angew. Chem., Int. Ed.}\ }\textbf {\bibinfo
  {volume} {58}},\ \bibinfo {pages} {13133--13139} (\bibinfo {year}
  {2019})}\BibitemShut {NoStop}%
\bibitem [{\citenamefont {Senjean}\ \emph {et~al.}(2021)\citenamefont
  {Senjean}, \citenamefont {Sen}, \citenamefont {Repisky}, \citenamefont
  {Knizia},\ and\ \citenamefont {Visscher}}]{gen-intrn-orb-quart-2021}%
  \BibitemOpen
  \bibfield  {author} {\bibinfo {author} {\bibfnamefont {B.}~\bibnamefont
  {Senjean}}, \bibinfo {author} {\bibfnamefont {S.}~\bibnamefont {Sen}},
  \bibinfo {author} {\bibfnamefont {M.}~\bibnamefont {Repisky}}, \bibinfo
  {author} {\bibfnamefont {G.}~\bibnamefont {Knizia}},\ and\ \bibinfo {author}
  {\bibfnamefont {L.}~\bibnamefont {Visscher}},\ }\bibfield  {title} {\enquote
  {\bibinfo {title} {{Generalization of Intrinsic Orbitals to Kramers-Paired
  Quaternion Spinors, Molecular Fragments, and Valence Virtual Spinors}},}\
  }\href {https://doi.org/10.1021/acs.jctc.0c00964} {\bibfield  {journal}
  {\bibinfo  {journal} {J. Chem. Theory Comput.}\ }\textbf {\bibinfo {volume}
  {17}},\ \bibinfo {pages} {1337--1354} (\bibinfo {year} {2021})}\BibitemShut
  {NoStop}%
\bibitem [{\citenamefont {Kalozoumis}\ \emph {et~al.}(2013)\citenamefont
  {Kalozoumis}, \citenamefont {Morfonios}, \citenamefont {Diakonos},\ and\
  \citenamefont {Schmelcher}}]{local-symm-1d-2013}%
  \BibitemOpen
  \bibfield  {author} {\bibinfo {author} {\bibfnamefont {P.~A.}\ \bibnamefont
  {Kalozoumis}}, \bibinfo {author} {\bibfnamefont {C.}~\bibnamefont
  {Morfonios}}, \bibinfo {author} {\bibfnamefont {F.~K.}\ \bibnamefont
  {Diakonos}},\ and\ \bibinfo {author} {\bibfnamefont {P.}~\bibnamefont
  {Schmelcher}},\ }\bibfield  {title} {\enquote {\bibinfo {title} {Local
  symmetries in one-dimensional quantum scattering},}\ }\href
  {https://doi.org/10.1103/PhysRevA.87.032113} {\bibfield  {journal} {\bibinfo
  {journal} {Phys. Rev. A}\ }\textbf {\bibinfo {volume} {87}},\ \bibinfo
  {pages} {032113} (\bibinfo {year} {2013})}\BibitemShut {NoStop}%
\bibitem [{\citenamefont {Schmelcher}, \citenamefont {Krönke},\ and\
  \citenamefont {Diakonos}(2017)}]{dyn-local-symm-2017}%
  \BibitemOpen
  \bibfield  {author} {\bibinfo {author} {\bibfnamefont {P.}~\bibnamefont
  {Schmelcher}}, \bibinfo {author} {\bibfnamefont {S.}~\bibnamefont
  {Krönke}},\ and\ \bibinfo {author} {\bibfnamefont {F.~K.}\ \bibnamefont
  {Diakonos}},\ }\bibfield  {title} {\enquote {\bibinfo {title} {Dynamics of
  local symmetry correlators for interacting many-particle systems},}\ }\href
  {https://doi.org/10.1063/1.4974096} {\bibfield  {journal} {\bibinfo
  {journal} {J. Chem. Phys.}\ }\textbf {\bibinfo {volume} {146}},\ \bibinfo
  {pages} {044116} (\bibinfo {year} {2017})}\BibitemShut {NoStop}%
\bibitem [{Note4()}]{Note4}%
  \BibitemOpen
  \bibinfo {note} {This assumes that the first excited state is approximately
  described by an orbital excitation.}\BibitemShut {Stop}%
\bibitem [{Note5()}]{Note5}%
  \BibitemOpen
  \bibinfo {note} {For simplicity, we avoid here the introduction of additional
  arrows that would depict the motion of a hole/positive charge.}\BibitemShut
  {Stop}%
\bibitem [{Note6()}]{Note6}%
  \BibitemOpen
  \bibinfo {note} {This mechanism is independent from a negative hole density
  appearing in the phenyl ring, which appears for all three dynamics and is due
  to 2h1p configurations and global symmetry changes discussed
  later.}\BibitemShut {Stop}%
\bibitem [{Note7()}]{Note7}%
  \BibitemOpen
  \bibinfo {note} {Note that quantifying these interactions is beyond the scope
  of this work.}\BibitemShut {Stop}%
\bibitem [{\citenamefont {Hermann}\ \emph {et~al.}(2016)\citenamefont
  {Hermann}, \citenamefont {Liu}, \citenamefont {Manz}, \citenamefont {Paulus},
  \citenamefont {P\'erez-Torres}, \citenamefont {Pohl},\ and\ \citenamefont
  {Tremblay}}]{elflux-benzene-cm-2016}%
  \BibitemOpen
  \bibfield  {author} {\bibinfo {author} {\bibfnamefont {G.}~\bibnamefont
  {Hermann}}, \bibinfo {author} {\bibfnamefont {C.}~\bibnamefont {Liu}},
  \bibinfo {author} {\bibfnamefont {J.}~\bibnamefont {Manz}}, \bibinfo {author}
  {\bibfnamefont {B.}~\bibnamefont {Paulus}}, \bibinfo {author} {\bibfnamefont
  {J.~F.}\ \bibnamefont {P\'erez-Torres}}, \bibinfo {author} {\bibfnamefont
  {V.}~\bibnamefont {Pohl}},\ and\ \bibinfo {author} {\bibfnamefont {J.~C.}\
  \bibnamefont {Tremblay}},\ }\bibfield  {title} {\enquote {\bibinfo {title}
  {Multidirectional angular electronic flux during adiabatic attosecond charge
  migration in excited benzene},}\ }\href
  {https://doi.org/10.1021/acs.jpca.6b01948} {\bibfield  {journal} {\bibinfo
  {journal} {J. Phys. Chem. A}\ }\textbf {\bibinfo {volume} {120}},\ \bibinfo
  {pages} {5360--5369} (\bibinfo {year} {2016})}\BibitemShut {NoStop}%
\bibitem [{Note8()}]{Note8}%
  \BibitemOpen
  \bibinfo {note} {The unprojected hole densities are semi-quantitatively
  identical to the projected ones and are shown in Supplementary
  Section~S{7}\protect \xspace .}\BibitemShut {Stop}%
\bibitem [{\citenamefont {Scheidegger}, \citenamefont {Golubev},\ and\
  \citenamefont {Vaníček}(2025)}]{cm-increase-size-flexlty-2025}%
  \BibitemOpen
  \bibfield  {author} {\bibinfo {author} {\bibfnamefont {A.}~\bibnamefont
  {Scheidegger}}, \bibinfo {author} {\bibfnamefont {N.~V.}\ \bibnamefont
  {Golubev}},\ and\ \bibinfo {author} {\bibfnamefont {J.~J.~L.}\ \bibnamefont
  {Vaníček}},\ }\bibfield  {title} {\enquote {\bibinfo {title} {Can
  increasing the size and flexibility of a molecule reduce decoherence and
  prolong charge migration?}}\ }\href {https://doi.org/10.1073/pnas.2501319122}
  {\bibfield  {journal} {\bibinfo  {journal} {Proc. Natl. Acad. Sci. USA}\
  }\textbf {\bibinfo {volume} {122}},\ \bibinfo {pages} {e2501319122} (\bibinfo
  {year} {2025})}\BibitemShut {NoStop}%
\bibitem [{blo(2024)}]{block2-gh-2024}%
  \BibitemOpen
  \href@noop {} {\enquote {\bibinfo {title} {block2},}\ } (\bibinfo {year}
  {2024}),\ \bibinfo {note}
  {{https://github.com/block-hczhai/block2-preview}}\BibitemShut {NoStop}%
\bibitem [{\citenamefont {Sun}\ \emph {et~al.}(2018)\citenamefont {Sun},
  \citenamefont {Berkelbach}, \citenamefont {Blunt}, \citenamefont {Booth},
  \citenamefont {Guo}, \citenamefont {Li}, \citenamefont {Liu}, \citenamefont
  {McClain}, \citenamefont {Sayfutyarova}, \citenamefont {Sharma},
  \citenamefont {Wouters},\ and\ \citenamefont {Chan}}]{pyscf-wires-2018}%
  \BibitemOpen
  \bibfield  {author} {\bibinfo {author} {\bibfnamefont {Q.}~\bibnamefont
  {Sun}}, \bibinfo {author} {\bibfnamefont {T.~C.}\ \bibnamefont {Berkelbach}},
  \bibinfo {author} {\bibfnamefont {N.~S.}\ \bibnamefont {Blunt}}, \bibinfo
  {author} {\bibfnamefont {G.~H.}\ \bibnamefont {Booth}}, \bibinfo {author}
  {\bibfnamefont {S.}~\bibnamefont {Guo}}, \bibinfo {author} {\bibfnamefont
  {Z.}~\bibnamefont {Li}}, \bibinfo {author} {\bibfnamefont {J.}~\bibnamefont
  {Liu}}, \bibinfo {author} {\bibfnamefont {J.~D.}\ \bibnamefont {McClain}},
  \bibinfo {author} {\bibfnamefont {E.~R.}\ \bibnamefont {Sayfutyarova}},
  \bibinfo {author} {\bibfnamefont {S.}~\bibnamefont {Sharma}}, \bibinfo
  {author} {\bibfnamefont {S.}~\bibnamefont {Wouters}},\ and\ \bibinfo {author}
  {\bibfnamefont {G.~K.-L.}\ \bibnamefont {Chan}},\ }\bibfield  {title}
  {\enquote {\bibinfo {title} {{PySCF: the Python-based simulations of
  chemistry framework}},}\ }\href
  {https://doi.org/https://doi.org/10.1002/wcms.1340} {\bibfield  {journal}
  {\bibinfo  {journal} {Wiley Interdiscip. Rev.: Comput. Mol. Sci.}\ }\textbf
  {\bibinfo {volume} {8}},\ \bibinfo {pages} {e1340} (\bibinfo {year}
  {2018})}\BibitemShut {NoStop}%
\bibitem [{\citenamefont {Sun}\ \emph {et~al.}(2020)\citenamefont {Sun},
  \citenamefont {Zhang}, \citenamefont {Banerjee}, \citenamefont {Bao},
  \citenamefont {Barbry}, \citenamefont {Blunt}, \citenamefont {Bogdanov},
  \citenamefont {Booth}, \citenamefont {Chen}, \citenamefont {Cui},
  \citenamefont {Eriksen}, \citenamefont {Gao}, \citenamefont {Guo},
  \citenamefont {Hermann}, \citenamefont {Hermes}, \citenamefont {Koh},
  \citenamefont {Koval}, \citenamefont {Lehtola}, \citenamefont {Li},
  \citenamefont {Liu}, \citenamefont {Mardirossian}, \citenamefont {McClain},
  \citenamefont {Motta}, \citenamefont {Mussard}, \citenamefont {Pham},
  \citenamefont {Pulkin}, \citenamefont {Purwanto}, \citenamefont {Robinson},
  \citenamefont {Ronca}, \citenamefont {Sayfutyarova}, \citenamefont
  {Scheurer}, \citenamefont {Schurkus}, \citenamefont {Smith}, \citenamefont
  {Sun}, \citenamefont {Sun}, \citenamefont {Upadhyay}, \citenamefont {Wagner},
  \citenamefont {Wang}, \citenamefont {White}, \citenamefont {Whitfield},
  \citenamefont {Williamson}, \citenamefont {Wouters}, \citenamefont {Yang},
  \citenamefont {Yu}, \citenamefont {Zhu}, \citenamefont {Berkelbach},
  \citenamefont {Sharma}, \citenamefont {Sokolov},\ and\ \citenamefont
  {Chan}}]{pyscf-jcp-2020}%
  \BibitemOpen
  \bibfield  {author} {\bibinfo {author} {\bibfnamefont {Q.}~\bibnamefont
  {Sun}}, \bibinfo {author} {\bibfnamefont {X.}~\bibnamefont {Zhang}}, \bibinfo
  {author} {\bibfnamefont {S.}~\bibnamefont {Banerjee}}, \bibinfo {author}
  {\bibfnamefont {P.}~\bibnamefont {Bao}}, \bibinfo {author} {\bibfnamefont
  {M.}~\bibnamefont {Barbry}}, \bibinfo {author} {\bibfnamefont {N.~S.}\
  \bibnamefont {Blunt}}, \bibinfo {author} {\bibfnamefont {N.~A.}\ \bibnamefont
  {Bogdanov}}, \bibinfo {author} {\bibfnamefont {G.~H.}\ \bibnamefont {Booth}},
  \bibinfo {author} {\bibfnamefont {J.}~\bibnamefont {Chen}}, \bibinfo {author}
  {\bibfnamefont {Z.-H.}\ \bibnamefont {Cui}}, \bibinfo {author} {\bibfnamefont
  {J.~J.}\ \bibnamefont {Eriksen}}, \bibinfo {author} {\bibfnamefont
  {Y.}~\bibnamefont {Gao}}, \bibinfo {author} {\bibfnamefont {S.}~\bibnamefont
  {Guo}}, \bibinfo {author} {\bibfnamefont {J.}~\bibnamefont {Hermann}},
  \bibinfo {author} {\bibfnamefont {M.~R.}\ \bibnamefont {Hermes}}, \bibinfo
  {author} {\bibfnamefont {K.}~\bibnamefont {Koh}}, \bibinfo {author}
  {\bibfnamefont {P.}~\bibnamefont {Koval}}, \bibinfo {author} {\bibfnamefont
  {S.}~\bibnamefont {Lehtola}}, \bibinfo {author} {\bibfnamefont
  {Z.}~\bibnamefont {Li}}, \bibinfo {author} {\bibfnamefont {J.}~\bibnamefont
  {Liu}}, \bibinfo {author} {\bibfnamefont {N.}~\bibnamefont {Mardirossian}},
  \bibinfo {author} {\bibfnamefont {J.~D.}\ \bibnamefont {McClain}}, \bibinfo
  {author} {\bibfnamefont {M.}~\bibnamefont {Motta}}, \bibinfo {author}
  {\bibfnamefont {B.}~\bibnamefont {Mussard}}, \bibinfo {author} {\bibfnamefont
  {H.~Q.}\ \bibnamefont {Pham}}, \bibinfo {author} {\bibfnamefont
  {A.}~\bibnamefont {Pulkin}}, \bibinfo {author} {\bibfnamefont
  {W.}~\bibnamefont {Purwanto}}, \bibinfo {author} {\bibfnamefont {P.~J.}\
  \bibnamefont {Robinson}}, \bibinfo {author} {\bibfnamefont {E.}~\bibnamefont
  {Ronca}}, \bibinfo {author} {\bibfnamefont {E.~R.}\ \bibnamefont
  {Sayfutyarova}}, \bibinfo {author} {\bibfnamefont {M.}~\bibnamefont
  {Scheurer}}, \bibinfo {author} {\bibfnamefont {H.~F.}\ \bibnamefont
  {Schurkus}}, \bibinfo {author} {\bibfnamefont {J.~E.~T.}\ \bibnamefont
  {Smith}}, \bibinfo {author} {\bibfnamefont {C.}~\bibnamefont {Sun}}, \bibinfo
  {author} {\bibfnamefont {S.-N.}\ \bibnamefont {Sun}}, \bibinfo {author}
  {\bibfnamefont {S.}~\bibnamefont {Upadhyay}}, \bibinfo {author}
  {\bibfnamefont {L.~K.}\ \bibnamefont {Wagner}}, \bibinfo {author}
  {\bibfnamefont {X.}~\bibnamefont {Wang}}, \bibinfo {author} {\bibfnamefont
  {A.}~\bibnamefont {White}}, \bibinfo {author} {\bibfnamefont {J.~D.}\
  \bibnamefont {Whitfield}}, \bibinfo {author} {\bibfnamefont {M.~J.}\
  \bibnamefont {Williamson}}, \bibinfo {author} {\bibfnamefont
  {S.}~\bibnamefont {Wouters}}, \bibinfo {author} {\bibfnamefont
  {J.}~\bibnamefont {Yang}}, \bibinfo {author} {\bibfnamefont {J.~M.}\
  \bibnamefont {Yu}}, \bibinfo {author} {\bibfnamefont {T.}~\bibnamefont
  {Zhu}}, \bibinfo {author} {\bibfnamefont {T.~C.}\ \bibnamefont {Berkelbach}},
  \bibinfo {author} {\bibfnamefont {S.}~\bibnamefont {Sharma}}, \bibinfo
  {author} {\bibfnamefont {A.~Y.}\ \bibnamefont {Sokolov}},\ and\ \bibinfo
  {author} {\bibfnamefont {G.~K.-L.}\ \bibnamefont {Chan}},\ }\bibfield
  {title} {\enquote {\bibinfo {title} {{Recent developments in the PySCF
  program package}},}\ }\href {https://doi.org/10.1063/5.0006074} {\bibfield
  {journal} {\bibinfo  {journal} {J. Chem. Phys.}\ }\textbf {\bibinfo {volume}
  {153}},\ \bibinfo {pages} {024109} (\bibinfo {year} {2020})}\BibitemShut
  {NoStop}%
\bibitem [{pys(2024)}]{pyscf-gh-2024}%
  \BibitemOpen
  \href@noop {} {\enquote {\bibinfo {title} {{PySCF}},}\ } (\bibinfo {year}
  {2024}),\ \bibinfo {note} {{https://github.com/pyscf/pyscf}}\BibitemShut
  {NoStop}%
\bibitem [{\citenamefont {Sun}\ \emph {et~al.}(2026)\citenamefont {Sun},
  \citenamefont {Hermes}, \citenamefont {Wu}, \citenamefont {Zhai},
  \citenamefont {Zhang}, \citenamefont {Ahmed}, \citenamefont {Aucar},
  \citenamefont {Backhouse}, \citenamefont {Banerjee}, \citenamefont {Bao},
  \citenamefont {Bogdanov}, \citenamefont {Bystrom}, \citenamefont {Chapoton},
  \citenamefont {Chen}, \citenamefont {Chernyshov}, \citenamefont {Clifford},
  \citenamefont {{Cohen-Janes}}, \citenamefont {Cui}, \citenamefont {Damour},
  \citenamefont {Dattani}, \citenamefont {Dittmer}, \citenamefont {Ehlert},
  \citenamefont {Eriksen}, \citenamefont {Evangelista}, \citenamefont {Ewing},
  \citenamefont {Farahvash}, \citenamefont {Focke}, \citenamefont {Gao},
  \citenamefont {Gasperich}, \citenamefont {Gillispie}, \citenamefont
  {Greiner}, \citenamefont {Hennefarth}, \citenamefont {Hermann}, \citenamefont
  {Hillenbrand}, \citenamefont {Huhtasalo}, \citenamefont {Ibrahim},
  \citenamefont {Jangid}, \citenamefont {Javaremi}, \citenamefont {Jenkins},
  \citenamefont {Jin}, \citenamefont {King}, \citenamefont {Kooi},
  \citenamefont {Kurian}, \citenamefont {Larsson}, \citenamefont {Lau},
  \citenamefont {Lee}, \citenamefont {Lehtola}, \citenamefont {Li},
  \citenamefont {Li}, \citenamefont {Li}, \citenamefont {Li}, \citenamefont
  {Li}, \citenamefont {Lykhin}, \citenamefont {Mahajan}, \citenamefont
  {Mauger}, \citenamefont {del {Mazo-Sevillano}}, \citenamefont {Moussa},
  \citenamefont {Nakano}, \citenamefont {Neufeld}, \citenamefont {Peng},
  \citenamefont {Pham}, \citenamefont {Pinski}, \citenamefont {Pokhilko},
  \citenamefont {Pu}, \citenamefont {Qian}, \citenamefont {Quiton},
  \citenamefont {Schulze}, \citenamefont {Scott}, \citenamefont {Seal},
  \citenamefont {Serna}, \citenamefont {Smith}, \citenamefont {Smyser},
  \citenamefont {Stahl}, \citenamefont {Sun}, \citenamefont {Sung},
  \citenamefont {Trushin}, \citenamefont {Upadhyay}, \citenamefont {Vo},
  \citenamefont {Vogels}, \citenamefont {Wang}, \citenamefont {Wang},
  \citenamefont {Wang}, \citenamefont {Wang}, \citenamefont {Wang},
  \citenamefont {Williamson}, \citenamefont {Yang}, \citenamefont {Ye},
  \citenamefont {Yeh}, \citenamefont {Yu}, \citenamefont {Yu}, \citenamefont
  {Yu}, \citenamefont {Zhang}, \citenamefont {Zhang}, \citenamefont {Zhang},
  \citenamefont {Zhao}, \citenamefont {Zhou}, \citenamefont {Zhu},
  \citenamefont {Zhu}, \citenamefont {Berkelbach}, \citenamefont {Gagliardi},
  \citenamefont {Sharma}, \citenamefont {Sokolov},\ and\ \citenamefont
  {{Kin-Lic Chan}}}]{preprint-pyscf-2026}%
  \BibitemOpen
  \bibfield  {author} {\bibinfo {author} {\bibfnamefont {Q.}~\bibnamefont
  {Sun}}, \bibinfo {author} {\bibfnamefont {M.~R.}\ \bibnamefont {Hermes}},
  \bibinfo {author} {\bibfnamefont {X.}~\bibnamefont {Wu}}, \bibinfo {author}
  {\bibfnamefont {H.}~\bibnamefont {Zhai}}, \bibinfo {author} {\bibfnamefont
  {X.}~\bibnamefont {Zhang}}, \bibinfo {author} {\bibfnamefont {A.~M.}\
  \bibnamefont {Ahmed}}, \bibinfo {author} {\bibfnamefont {J.~J.}\ \bibnamefont
  {Aucar}}, \bibinfo {author} {\bibfnamefont {O.~J.}\ \bibnamefont
  {Backhouse}}, \bibinfo {author} {\bibfnamefont {S.}~\bibnamefont {Banerjee}},
  \bibinfo {author} {\bibfnamefont {P.}~\bibnamefont {Bao}}, \bibinfo {author}
  {\bibfnamefont {N.~A.}\ \bibnamefont {Bogdanov}}, \bibinfo {author}
  {\bibfnamefont {K.}~\bibnamefont {Bystrom}}, \bibinfo {author} {\bibfnamefont
  {F.}~\bibnamefont {Chapoton}}, \bibinfo {author} {\bibfnamefont {N.-Y.}\
  \bibnamefont {Chen}}, \bibinfo {author} {\bibfnamefont {I.~Y.}\ \bibnamefont
  {Chernyshov}}, \bibinfo {author} {\bibfnamefont {H.~S.}\ \bibnamefont
  {Clifford}}, \bibinfo {author} {\bibfnamefont {S.}~\bibnamefont
  {{Cohen-Janes}}}, \bibinfo {author} {\bibfnamefont {Z.-H.}\ \bibnamefont
  {Cui}}, \bibinfo {author} {\bibfnamefont {Y.~D.}\ \bibnamefont {Damour}},
  \bibinfo {author} {\bibfnamefont {N.}~\bibnamefont {Dattani}}, \bibinfo
  {author} {\bibfnamefont {L.~B.}\ \bibnamefont {Dittmer}}, \bibinfo {author}
  {\bibfnamefont {S.}~\bibnamefont {Ehlert}}, \bibinfo {author} {\bibfnamefont
  {J.~J.}\ \bibnamefont {Eriksen}}, \bibinfo {author} {\bibfnamefont {F.~A.}\
  \bibnamefont {Evangelista}}, \bibinfo {author} {\bibfnamefont {S.~A.}\
  \bibnamefont {Ewing}}, \bibinfo {author} {\bibfnamefont {A.}~\bibnamefont
  {Farahvash}}, \bibinfo {author} {\bibfnamefont {K.}~\bibnamefont {Focke}},
  \bibinfo {author} {\bibfnamefont {Y.}~\bibnamefont {Gao}}, \bibinfo {author}
  {\bibfnamefont {K.~E.}\ \bibnamefont {Gasperich}}, \bibinfo {author}
  {\bibfnamefont {N.}~\bibnamefont {Gillispie}}, \bibinfo {author}
  {\bibfnamefont {J.}~\bibnamefont {Greiner}}, \bibinfo {author} {\bibfnamefont
  {M.~R.}\ \bibnamefont {Hennefarth}}, \bibinfo {author} {\bibfnamefont
  {J.}~\bibnamefont {Hermann}}, \bibinfo {author} {\bibfnamefont
  {C.}~\bibnamefont {Hillenbrand}}, \bibinfo {author} {\bibfnamefont
  {J.}~\bibnamefont {Huhtasalo}}, \bibinfo {author} {\bibfnamefont
  {B.}~\bibnamefont {Ibrahim}}, \bibinfo {author} {\bibfnamefont
  {B.}~\bibnamefont {Jangid}}, \bibinfo {author} {\bibfnamefont {A.~N.}\
  \bibnamefont {Javaremi}}, \bibinfo {author} {\bibfnamefont {A.~J.}\
  \bibnamefont {Jenkins}}, \bibinfo {author} {\bibfnamefont {Y.}~\bibnamefont
  {Jin}}, \bibinfo {author} {\bibfnamefont {D.~S.}\ \bibnamefont {King}},
  \bibinfo {author} {\bibfnamefont {D.~P.}\ \bibnamefont {Kooi}}, \bibinfo
  {author} {\bibfnamefont {J.~S.}\ \bibnamefont {Kurian}}, \bibinfo {author}
  {\bibfnamefont {H.~R.}\ \bibnamefont {Larsson}}, \bibinfo {author}
  {\bibfnamefont {B.~T.~G.}\ \bibnamefont {Lau}}, \bibinfo {author}
  {\bibfnamefont {S.}~\bibnamefont {Lee}}, \bibinfo {author} {\bibfnamefont
  {S.}~\bibnamefont {Lehtola}}, \bibinfo {author} {\bibfnamefont
  {C.}~\bibnamefont {Li}}, \bibinfo {author} {\bibfnamefont {H.}~\bibnamefont
  {Li}}, \bibinfo {author} {\bibfnamefont {J.}~\bibnamefont {Li}}, \bibinfo
  {author} {\bibfnamefont {R.}~\bibnamefont {Li}}, \bibinfo {author}
  {\bibfnamefont {S.}~\bibnamefont {Li}}, \bibinfo {author} {\bibfnamefont
  {A.~O.}\ \bibnamefont {Lykhin}}, \bibinfo {author} {\bibfnamefont
  {A.}~\bibnamefont {Mahajan}}, \bibinfo {author} {\bibfnamefont
  {N.}~\bibnamefont {Mauger}}, \bibinfo {author} {\bibfnamefont
  {P.}~\bibnamefont {del {Mazo-Sevillano}}}, \bibinfo {author} {\bibfnamefont
  {J.}~\bibnamefont {Moussa}}, \bibinfo {author} {\bibfnamefont
  {K.}~\bibnamefont {Nakano}}, \bibinfo {author} {\bibfnamefont {V.~A.}\
  \bibnamefont {Neufeld}}, \bibinfo {author} {\bibfnamefont {L.}~\bibnamefont
  {Peng}}, \bibinfo {author} {\bibfnamefont {H.~Q.}\ \bibnamefont {Pham}},
  \bibinfo {author} {\bibfnamefont {P.}~\bibnamefont {Pinski}}, \bibinfo
  {author} {\bibfnamefont {P.}~\bibnamefont {Pokhilko}}, \bibinfo {author}
  {\bibfnamefont {Z.}~\bibnamefont {Pu}}, \bibinfo {author} {\bibfnamefont
  {Y.}~\bibnamefont {Qian}}, \bibinfo {author} {\bibfnamefont {S.~J.}\
  \bibnamefont {Quiton}}, \bibinfo {author} {\bibfnamefont {W.~T.}\
  \bibnamefont {Schulze}}, \bibinfo {author} {\bibfnamefont {T.~R.}\
  \bibnamefont {Scott}}, \bibinfo {author} {\bibfnamefont {A.}~\bibnamefont
  {Seal}}, \bibinfo {author} {\bibfnamefont {J.~D.}\ \bibnamefont {Serna}},
  \bibinfo {author} {\bibfnamefont {J.~E.~T.}\ \bibnamefont {Smith}}, \bibinfo
  {author} {\bibfnamefont {K.~E.}\ \bibnamefont {Smyser}}, \bibinfo {author}
  {\bibfnamefont {T.}~\bibnamefont {Stahl}}, \bibinfo {author} {\bibfnamefont
  {C.}~\bibnamefont {Sun}}, \bibinfo {author} {\bibfnamefont {K.~J.}\
  \bibnamefont {Sung}}, \bibinfo {author} {\bibfnamefont {E.}~\bibnamefont
  {Trushin}}, \bibinfo {author} {\bibfnamefont {S.}~\bibnamefont {Upadhyay}},
  \bibinfo {author} {\bibfnamefont {E.~A.}\ \bibnamefont {Vo}}, \bibinfo
  {author} {\bibfnamefont {T.}~\bibnamefont {Vogels}}, \bibinfo {author}
  {\bibfnamefont {S.}~\bibnamefont {Wang}}, \bibinfo {author} {\bibfnamefont
  {T.}~\bibnamefont {Wang}}, \bibinfo {author} {\bibfnamefont {X.}~\bibnamefont
  {Wang}}, \bibinfo {author} {\bibfnamefont {X.}~\bibnamefont {Wang}}, \bibinfo
  {author} {\bibfnamefont {Y.}~\bibnamefont {Wang}}, \bibinfo {author}
  {\bibfnamefont {M.}~\bibnamefont {Williamson}}, \bibinfo {author}
  {\bibfnamefont {J.}~\bibnamefont {Yang}}, \bibinfo {author} {\bibfnamefont
  {H.-Z.}\ \bibnamefont {Ye}}, \bibinfo {author} {\bibfnamefont {C.-N.}\
  \bibnamefont {Yeh}}, \bibinfo {author} {\bibfnamefont {H.}~\bibnamefont
  {Yu}}, \bibinfo {author} {\bibfnamefont {J.}~\bibnamefont {Yu}}, \bibinfo
  {author} {\bibfnamefont {V.~W.-z.}\ \bibnamefont {Yu}}, \bibinfo {author}
  {\bibfnamefont {C.}~\bibnamefont {Zhang}}, \bibinfo {author} {\bibfnamefont
  {D.}~\bibnamefont {Zhang}}, \bibinfo {author} {\bibfnamefont
  {Y.}~\bibnamefont {Zhang}}, \bibinfo {author} {\bibfnamefont
  {Z.}~\bibnamefont {Zhao}}, \bibinfo {author} {\bibfnamefont {Z.}~\bibnamefont
  {Zhou}}, \bibinfo {author} {\bibfnamefont {A.~J.}\ \bibnamefont {Zhu}},
  \bibinfo {author} {\bibfnamefont {T.}~\bibnamefont {Zhu}}, \bibinfo {author}
  {\bibfnamefont {T.~C.}\ \bibnamefont {Berkelbach}}, \bibinfo {author}
  {\bibfnamefont {L.}~\bibnamefont {Gagliardi}}, \bibinfo {author}
  {\bibfnamefont {S.}~\bibnamefont {Sharma}}, \bibinfo {author} {\bibfnamefont
  {A.~Y.}\ \bibnamefont {Sokolov}},\ and\ \bibinfo {author} {\bibfnamefont
  {G.}~\bibnamefont {{Kin-Lic Chan}}},\ }\bibfield  {title} {\enquote {\bibinfo
  {title} {The {{Python}} simulations of chemistry framework: 10 years of an
  open-source quantum chemistry project},}\ }\href
  {https://doi.org/10.1063/5.0337441} {\bibfield  {journal} {\bibinfo
  {journal} {J. Chem. Phys.}\ }\textbf {\bibinfo {volume} {165}},\ \bibinfo
  {pages} {102502} (\bibinfo {year} {2026})}\BibitemShut {NoStop}%
\bibitem [{\citenamefont {Tatsuaki}(2000)}]{tatsuaki-singlet-embed-2000}%
  \BibitemOpen
  \bibfield  {author} {\bibinfo {author} {\bibfnamefont {W.}~\bibnamefont
  {Tatsuaki}},\ }\bibfield  {title} {\enquote {\bibinfo {title}
  {Interaction-round-a-face density-matrix renormalization-group method applied
  to rotational-invariant quantum spin chains},}\ }\href
  {https://doi.org/10.1103/PhysRevE.61.3199} {\bibfield  {journal} {\bibinfo
  {journal} {Phys. Rev. E}\ }\textbf {\bibinfo {volume} {61}},\ \bibinfo
  {pages} {3199--3206} (\bibinfo {year} {2000})}\BibitemShut {NoStop}%
\bibitem [{\citenamefont {Sharma}\ and\ \citenamefont
  {Chan}(2012)}]{dmrg-spin-symm-2012}%
  \BibitemOpen
  \bibfield  {author} {\bibinfo {author} {\bibfnamefont {S.}~\bibnamefont
  {Sharma}}\ and\ \bibinfo {author} {\bibfnamefont {G.~K.-L.}\ \bibnamefont
  {Chan}},\ }\bibfield  {title} {\enquote {\bibinfo {title} {{Spin-adapted
  density matrix renormalization group algorithms for quantum chemistry}},}\
  }\href {https://doi.org/10.1063/1.3695642} {\bibfield  {journal} {\bibinfo
  {journal} {J. Chem. Phys.}\ }\textbf {\bibinfo {volume} {136}},\ \bibinfo
  {pages} {124121} (\bibinfo {year} {2012})}\BibitemShut {NoStop}%
\bibitem [{\citenamefont {Park}\ and\ \citenamefont
  {Light}(1986)}]{sil-light-1986}%
  \BibitemOpen
  \bibfield  {author} {\bibinfo {author} {\bibfnamefont {T.~J.}\ \bibnamefont
  {Park}}\ and\ \bibinfo {author} {\bibfnamefont {J.~C.}\ \bibnamefont
  {Light}},\ }\bibfield  {title} {\enquote {\bibinfo {title} {{Unitary quantum
  time evolution by iterative Lanczos reduction}},}\ }\href
  {https://doi.org/10.1063/1.451548} {\bibfield  {journal} {\bibinfo  {journal}
  {J. Chem. Phys.}\ }\textbf {\bibinfo {volume} {85}},\ \bibinfo {pages}
  {5870--5876} (\bibinfo {year} {1986})}\BibitemShut {NoStop}%
\bibitem [{\citenamefont {Ditchfield}, \citenamefont {Hehre},\ and\
  \citenamefont {Pople}(1971)}]{6-31g_1-1971}%
  \BibitemOpen
  \bibfield  {author} {\bibinfo {author} {\bibfnamefont {R.}~\bibnamefont
  {Ditchfield}}, \bibinfo {author} {\bibfnamefont {W.~J.}\ \bibnamefont
  {Hehre}},\ and\ \bibinfo {author} {\bibfnamefont {J.~A.}\ \bibnamefont
  {Pople}},\ }\bibfield  {title} {\enquote {\bibinfo {title}
  {{Self‐Consistent Molecular‐Orbital Methods. IX. An Extended
  Gaussian‐Type Basis for Molecular‐Orbital Studies of Organic
  Molecules}},}\ }\href {https://doi.org/10.1063/1.1674902} {\bibfield
  {journal} {\bibinfo  {journal} {J. Chem. Phys.}\ }\textbf {\bibinfo {volume}
  {54}},\ \bibinfo {pages} {724--728} (\bibinfo {year} {1971})}\BibitemShut
  {NoStop}%
\bibitem [{\citenamefont {Hehre}, \citenamefont {Ditchfield},\ and\
  \citenamefont {Pople}(1972)}]{6-31g_2-1972}%
  \BibitemOpen
  \bibfield  {author} {\bibinfo {author} {\bibfnamefont {W.~J.}\ \bibnamefont
  {Hehre}}, \bibinfo {author} {\bibfnamefont {R.}~\bibnamefont {Ditchfield}},\
  and\ \bibinfo {author} {\bibfnamefont {J.~A.}\ \bibnamefont {Pople}},\
  }\bibfield  {title} {\enquote {\bibinfo {title} {{Self-Consistent Molecular
  Orbital Methods. XII. Further Extensions of Gaussian-Type Basis Sets for Use
  in Molecular Orbital Studies of Organic Molecules}},}\ }\href
  {https://doi.org/10.1063/1.1677527} {\bibfield  {journal} {\bibinfo
  {journal} {J. Chem. Phys.}\ }\textbf {\bibinfo {volume} {56}},\ \bibinfo
  {pages} {2257--2261} (\bibinfo {year} {1972})}\BibitemShut {NoStop}%
\bibitem [{\citenamefont {Weigend}\ and\ \citenamefont
  {Ahlrichs}(2005)}]{def2-sv_p_-2005}%
  \BibitemOpen
  \bibfield  {author} {\bibinfo {author} {\bibfnamefont {F.}~\bibnamefont
  {Weigend}}\ and\ \bibinfo {author} {\bibfnamefont {R.}~\bibnamefont
  {Ahlrichs}},\ }\bibfield  {title} {\enquote {\bibinfo {title} {{Balanced
  basis sets of split valence{,} triple zeta valence and quadruple zeta valence
  quality for H to Rn: Design and assessment of accuracy}},}\ }\href
  {https://doi.org/10.1039/B508541A} {\bibfield  {journal} {\bibinfo  {journal}
  {Phys. Chem. Chem. Phys.}\ }\textbf {\bibinfo {volume} {7}},\ \bibinfo
  {pages} {3297--3305} (\bibinfo {year} {2005})}\BibitemShut {NoStop}%
\bibitem [{\citenamefont {Kuleff}\ \emph {et~al.}(2016)\citenamefont {Kuleff},
  \citenamefont {Kryzhevoi}, \citenamefont {Pernpointner},\ and\ \citenamefont
  {Cederbaum}}]{core-ionize-cm-2016}%
  \BibitemOpen
  \bibfield  {author} {\bibinfo {author} {\bibfnamefont {A.~I.}\ \bibnamefont
  {Kuleff}}, \bibinfo {author} {\bibfnamefont {N.~V.}\ \bibnamefont
  {Kryzhevoi}}, \bibinfo {author} {\bibfnamefont {M.}~\bibnamefont
  {Pernpointner}},\ and\ \bibinfo {author} {\bibfnamefont {L.~S.}\ \bibnamefont
  {Cederbaum}},\ }\bibfield  {title} {\enquote {\bibinfo {title} {Core
  ionization initiates subfemtosecond charge migration in the valence shell of
  molecules},}\ }\href {https://doi.org/10.1103/PhysRevLett.117.093002}
  {\bibfield  {journal} {\bibinfo  {journal} {Phys. Rev. Lett.}\ }\textbf
  {\bibinfo {volume} {117}},\ \bibinfo {pages} {093002} (\bibinfo {year}
  {2016})}\BibitemShut {NoStop}%
\end{thebibliography}
\end{document}